\documentclass[aps,pre,superscriptaddress, floatfix, 12pt]{revtex4-1}
\usepackage{amsmath,amsfonts,amssymb}
\usepackage{graphicx}
\usepackage{algorithm}
\usepackage{algpseudocode}
\usepackage{epstopdf}
\usepackage{float,placeins}
\usepackage[hyperfigures,breaklinks,colorlinks,citecolor=blue,linkcolor=black]{hyperref}
\usepackage[caption=false]{subfig}
\usepackage{mathbbol}

\def\s{\sigma}

\newcommand{\lasubsim}[3]{\lambda_{#1 \rightarrow (#1 #2)} (#3)}

\newcommand{\musub}[3]{    \mu_{#1 \rightarrow (#1 #2)} (X_{#1}(#3)|X_{#2}(#3))}

\usepackage{accents}
\newlength{\dhatheight}

\begin{document}
\title{From random point processes to hierarchical Cavity Master Equations for stochastic dynamics of disordered systems in Random Graphs: Ising models and epidemics}

\author{D. Machado}
\affiliation{Group of Complex Systems and Statistical Physics. Department of Theoretical Physics, Physics Faculty, University of Havana, Cuba}
\email{dmachado@fisica.uh.cu}
\author{R. Mulet}
\affiliation{Group of Complex Systems and Statistical Physics. Department of Theoretical Physics, Physics Faculty, University of Havana, Cuba}
\email{mulet@fisica.uh.cu}
\date{\today}

\begin{abstract}
\noindent
We start from the Theory of Random Point Processes to derive n-point coupled master equations describing the continuous dynamics of discrete variables in random graphs. These equations constitute a hierarchical set of approximations that generalize and improve the Cavity Master Equation (CME) previously obtained in \cite{CME-PRE}. Our derivation clarifies some of the hypotheses and approximations that originally lead to the CME, considered now as the first order of a more general technique. We tested the new scheme in the dynamics of three models defined over diluted graphs: the Ising ferromagnet, the Viana-Bray spin-glass and the susceptible-infectious-susceptible model for epidemics. In the first two, the new equations perform similarly to the best-known approaches in the literature. In the latter, they outperform the well-known Pair Quenched Mean-Field Approximation.
\end{abstract}
\maketitle

\section{Introduction}

Emissions from a radioactive source, time series of electrical energy in a nerve fiber, the instants of arrival of customers in a queue, and the flips of a spin due to thermal activation are all examples of random processes that occur in such short intervals of time that can be classified as punctual. Sometimes the distinction between point processes and other stochastic processes is hard to define. For example, any stochastic process in continuous time in which the sample paths are step functions is associated with a point process, also any process with a discrete state space where a time of entry into a state occurs randomly.

The Theory of Random Point Processes (TRPP) \cite{vanKampen, Daley2002RPP} provides the formal and practical background to study and understand these and similar random collections of point occurrences. Usually, it is a matter of taste of the researchers, or the convenience for the research to attain, or not, to the TRPP to approach a specific problem.

In this work we study the continuous dynamics of a system of discrete interacting variables. The model systems of interest can be viewed as a multivariate and multidimensional Random Point Process. Alternatively, the dynamics of each variable can be considered as a Random Point Process itself. We will show that starting from the TRPP it is possible to write hierarchical master-like equations for any group of variables in the system. 

The first attempt in this direction was proposed a few years ago through the adaptation of dynamic message-passing equations from discrete-time to continuous-time dynamics. It landed first in the field of epidemics propagation \cite{Karrer2010SIR, shrestha2015message} without a general formulation but with good results. Later, in \cite{CME-PRE} was proposed a new closure for the master equation that exploited the TRPP. The core of this method is a differential equation for the cavity conditional probability densities: the Cavity Master Equation (CME). This approach has proven to be very general as it has been applied to several models in graphs with finite connectivity like the Ising ferromagnet, the random field Ising model and the Viana-Bray spin-glass model \cite{CME-PRE}, the ferromagnetic $p$-spin under Glauber dynamics \cite{CME-Pspin} and more recently also the dynamics of a focused search algorithm to solve the three-SAT problem in a random graph \cite{CME-PRL}. In this work we generalize the CME derived in \cite{CME-PRE}, providing master equations for the probability densities of any group of connected variables.

We concentrate on the continuous-time dynamics of discrete-spin variables. They can be described by a master equation for the probability density of the states of the system \cite{Glauber63, coolen2005theory}. But to fully solve this master equation is a cumbersome task and results have been elusive, except for special cases. For example, the Sherrington-Kirkpatrick (SK) model in its non-symmetric version has received preferential attention among the fully-connected family \cite{Crisanti88, Rieger89}, and the exact solution for parallel and asynchronous dynamics of the dilute fully-asymmetric neural network model dates back to 1987 \cite{derrida1987exactly}.

We provide here a general alternative solution that improves the approach presented in \cite{CME-PRE}. Although it is devised for single instances, it
does not prevents its use to obtain global information. One can always numerically average the results of the integration of the single-instance equations. On the other hand, in \cite{CME-AvAndBP} the CME equations were use as the starting point in the derivation of an average case description for the dynamics of the Ising ferromagnet, in this case defined on Erdos-Renyi random graphs.

To compute averaged quantities, one of the greatest advances in the field is the Dynamical Replica Theory (DRT), first introduced in \cite{coolen1996DRT}. This approach permits the derivation of average case equations for the probabilities of some macroscopic observables in fully-connected \cite{coolen1996DRT, laughton1996order} and diluted \cite{mozeika2008dynamical, mozeika2009dynamical} graphs.
In practice DRT successfully reduces the dimensionality of the system, making the problem tractable. However, it also assumes that the microscopic probability distribution function is a constant within a subspace with a finite number of order parameters, something that has been put into question on specific models \cite{nishimori1996dynamical, delferraro2014pert}, and is not necessarily true for general non-equilibrium situations.

Another relevant approach to these problems is the scheme of analytical approximations introduced in \cite{BarthelwalkSAT2003, SemerjianwalkSAT2003} for the description of the dynamics of local search algorithms. In \cite{Semerjian_dyn_2004} this methodology is revisited and consolidated, showing also its equivalence to DRT under the assumption of replica symmetry. Due to its successful combination of accuracy and simplicity, we will come back to this scheme as a reference for evaluating our numerical results.

This work contributes to the settlement of the dynamic cavity method as a general tool to study the continuous-time dynamics of discrete-spin models. A brief description of the theory of RPP is given in subsection \ref{subsec:RPP}. Section \ref{sec:hCME} presents our hierarchical system of corrections to the original equations derived in \cite{CME-PRE} and used in \cite{CME-Pspin, CME-PRL, CME-AvAndBP}. They are numerically tested at section \ref{sec:numeric}. We do not intend either to compare these directly with DRT, whose numerics we consider appreciably harder in this kind of graphs, nor to give a better description of the dynamics of average quantities. Instead, we use the equivalent but simpler approximation schemes in \cite{Semerjian_dyn_2004} as a theoretical and numeric reference, and as a proxy to the comparison with DRT.

Unlike the two approaches previously described, our method gives information about local probability distributions and can be straightforwardly used to study problems that are intrinsically out of equilibrium \cite{CME-PRL}, i.e., where detailed balance is not conserved. As an example, the last subsection of \ref{sec:numeric} is devoted to the exploration of susceptible-infectious-susceptible model for epidemics. Finally, section \ref{sec:disc} contains the conclusions of our work.

\section{Random Point Processes}\label{subsec:RPP}

To facilitate the reading of section \ref{sec:hCME} we take a brief tour into the Theory of Random Point Processes, which we will use to parametrize probability distributions of spin histories. After some basic definitions and for completeness we will refer in subsection \ref{subsec:RPP_to_ME} to the simple case of an independent binary variable that randomly changes its state. There, we will use the random point processes formalism to derive a master equation for the variable's dynamics, and we will show how to extend this to the case of many interacting variables.

The core object of random point processes are spin trajectories or histories. A specific spin history that starts with $\sigma(t_0)=\sigma_0$ and ends with $\sigma(t)=\sigma$ is parametrized by the number of spin flips, the time in which they occur and the initial state of the system. The spin trajectory is then nothing but a random point process \cite{vanKampen, Daley2002RPP} and the probability measure in this sample space may be denoted as:

\begin{equation}
Q^{t}(X) = Q_{s}(t_0, t_1, ..., t_s, t \mid \sigma(t_0) = \sigma_0)
\label{eq:Q_notation}
\end{equation}
which represents the probability density of having a trajectory with $s$ jumps at $(t_1, t_1+dt_1), ..., (t_s, t_s+dt_s)$. When needed, we may write $X$ as $X(t)$ to emphasize that the final time of the spin history is precisely $t$. Seeking simplicity, in almost all equations we will not write the conditioning of all dynamic quantities on a given initial condition (for example, in (\ref{eq:Q_notation}) we have $\sigma(t_0) = \sigma_0$).

The probability density $Q(X)$ fulfills the following normalization relation:

\begin{equation}
1 = \sum_{X}^{t} Q^{t}(X) = \sum_{s=0}^{\infty} \int_{t_0}^{t} dt_1 \int_{t_1}^{t} dt_2 \, ... \, \int_{t_{s-1}}^{t} dt_s \, Q_{s}(t_0, t_1, ..., t_s, t \mid \sigma(t_0) = \sigma_0)
\label{eq:normalization_Q}
\end{equation}

The sum $\sum_{X}^{t}$ in equation (\ref{eq:normalization_Q}) goes over all histories that occur between $t_0$ and $t$, starting with $\sigma(t_0)=\sigma_0$. It is explicitly written in the right hand side as a sum $\sum_{s=0}^{\infty}$ of time integrals. Each term of that sum corresponds to histories that have exactly $s$ jumps, and the integrals account for all the possible jumping times $t_1, t_2, ..., t_s$.

Instantaneous magnitudes can be written as marginals of $Q^{t}(X)$. For example:

\begin{equation}
P^{t}(\sigma) = \sum_{X \mid \sigma(t) = \sigma}^{t} Q^{t}(X)
\label{eq:marginal_Q}
\end{equation}
where the sum $\sum_{X \mid \sigma(t) = \sigma}^{t}$ goes over all the histories $X(t)$ such that $\sigma(t)=\sigma$ and can be written similarly as $\sum_{X}^{t}$ (see (\ref{eq:normalization_Q})). In what follows we will use the symbol $\sum_{X \mid \sigma}^{t}$ as a shortening of $\sum_{X \mid \sigma(t) = \sigma}^{t}$.

As we will show, some properties of the time evolution of magnitudes like $P^{t}(\sigma)$ can be obtained by working directly over $Q^{t}(X)$.

\subsection{From random point processes to master equations} \label{subsec:RPP_to_ME}

Master equations are a common instrument in the study of continuous-time dynamics. As a proof of the validity of the random point process approach, we will show its equivalence with the formulation of master equations in two examples: the time evolution of a single non-interacting variable and the dynamics of N-interacting spins. Through the derivations, the reader will become familiar with properties and procedures that will be used in section \ref{sec:hCME}.

\subsubsection{Single variable}

The master equation for the time evolution a single spin without interactions reads:

\begin{equation}
\dfrac{dP^{t}(\sigma)}{dt} = -r(\sigma) P^{t}(\sigma) + r(-\sigma) P^{t}(\sigma)
\label{eq:master_equation_single}
\end{equation}
where $r(\sigma)$ is the transition rate between the states $\sigma$ and $-\sigma$.

In order to derive equation (\ref{eq:master_equation_full}) from the TRPP we will differentiate (\ref{eq:marginal_Q}). Differentiation in this context should be handled carefully since increasing $t$ means we are changing the sample space itself. Instead of using calculus rules to differentiate we use the definition of derivative as the limit of the incremental ratio:

\begin{equation}
\frac{dP^{t}(\sigma)}{dt} = \lim_{\Delta t \rightarrow 0} \dfrac{P^{t+\Delta t}(\sigma) - P^{t}(\sigma)}{\Delta t}
\label{eq:increment_ratio}
\end{equation}

For the sake of the discussion in the next sections let's write down the calculations in detail.

Let's say that the final state of the trajectory $X$ is $X(t)=\sigma$. Then the probability density $Q^{t}(X)$ can take the form \cite{vanKampen, Daley2002RPP}:

\begin{equation}
Q^{t}(X) = r(\s_0) e^{-r(\s_0)(t_1-t_0)} r(-\s_0)e^{-r(-\s_0)(t_2-t_1)}\ldots r(-\s)e^{-r(-\s)(t_{s}-t_{s-1})}e^{-r(\s)(t_{f}-t_{s})}
\label{eq:Q_single_param}
\end{equation}

The right hand side of equation (\ref{eq:Q_single_param}) is the probability density of the waiting times of the history $X:[t_0, t] \rightarrow \lbrace 1, -1 \rbrace$, whose jumps occur at $(t_1, t_2, ...\,, t_{s})$, with rates $r (\sigma)$.

Now we will write
$P^{t + \Delta t}(\s)$ as the marginalization of $Q^{t + \Delta t}(X)$ and then expand to first order in $\Delta t$. Explicitly writing that marginalization as in (\ref{eq:normalization_Q}):

\begin{eqnarray}
P^{t + \Delta t}(\s) = \sum_{s} \int_{t_0}^{t + \Delta t} dt_1 \int_{t_1}^{t + \Delta t} dt_2 \ldots \int_{t_{s-1}}^{t + \Delta t} dt_s \, Q^{t+\Delta t}(X)
\label{eq:marginalizing_Q_single_delta_t}
\end{eqnarray}

Here, the sum on the right-hand side goes over all the values of the number of jumps (denoted as $s$) such that $X(t)=\sigma$.

Now we must keep only the order $\Delta t$ term in (\ref{eq:marginalizing_Q_single_delta_t}). Let's expand first $Q^{t+\Delta t}(X)$ and leave the rest untouched. The equation (\ref{eq:Q_single_param}) can be rearranged as:

\begin{equation}
Q^{t + \Delta t}(X) = \Big[ \prod_{l=1}^{s} r(\sigma(t_l)) \Big] \exp \Big\lbrace - \int_{t_0}^{t+\Delta t} r(\sigma(\tau)) d\tau \Big\rbrace
\label{eq:Q_single_param_rew}
\end{equation}
where the jumps in the history $X$ occur at the times $t_l$, with $l=1, 2, \ldots, s$.

In (\ref{eq:Q_single_param_rew}), the probability of having the last jump in the interval $[t, t+\Delta t]$ is $r(-\sigma) \Delta t$ and, more generally, the probability of occurrence of $n$ jumps is proportional to $(\Delta t)^{n}$. Therefore, when $\Delta t$ goes to zero, with probability $1$ there are no jumps in $[t, t+\Delta t]$ and none of the times $t_l$ belongs to that interval.

We can then expand the exponential in (\ref{eq:Q_single_param_rew}) and write:

\begin{eqnarray}
Q^{t + \Delta t}(X) &=& \Big[\prod_{l=1}^{s} r(\sigma(t_l)) \Big] \exp \Big\lbrace - \int_{t_0}^{t} r(\sigma(\tau)) d\tau \Big\rbrace \big[ 1 - r(\sigma) \Delta t \big] + o(\Delta t) \nonumber \\
Q^{t + \Delta t}(X) &=& Q^{t}(X) \big[ 1 - r(\sigma) \Delta t \big] + o(\Delta t)
\label{eq:Q_single_param_expand}
\end{eqnarray}

On the other hand, the sum of iterated integrals in (\ref{eq:marginalizing_Q_single_delta_t}) is of order $(\Delta t)^{s}$ by itself. However, for each integral we have the property $\int_{t_0}^{t+\Delta t}d\tau=\int_{t_0}^{t}d\tau+\int_{t}^{t+\Delta t}d\tau$, and we can write:

\begin{eqnarray}
\sum_{s} \int_{t_0}^{t + \Delta t}dt_1\int_{t_1}^{t + \Delta t}dt_2\ldots\int_{t_{s-1}}^{t + \Delta t}dt_s
&=& \sum_{s} \int_{t_0}^{t }dt_1\int_{t_1}^{t}dt_2\ldots\int_{t_{s-1}}^{t}dt_s \;+ \nonumber \\
& & + \sum_{s} \int_{t_0}^{t }dt_1\int_{t_1}^{t}dt_2\ldots\int_{t}^{t + \Delta t}dt_s + o(\Delta t)
\label{eq:integral_expansion}
\end{eqnarray}

The first term of the right hand side of (\ref{eq:integral_expansion}) is an operator over the space of the histories that occur in the interval $[t_0, t]$. It can be safely applied to the expansion (\ref{eq:Q_single_param_expand}) of the probability density $Q^{t + \Delta t}(X)$ to give a contribution of order $O((\Delta t)^{0})\equiv O(1)$

\begin{equation}
I_0=\sum_{s}\int_{t_0}^{t }dt_1\int_{t_1}^{t}dt_2\ldots\int_{t_{s-1}}^{t}dt_s Q^{t}(X)=P^{t}(\sigma)
\label{eq:I_0_single}
\end{equation}
and a contribution of order $O(\Delta t)$

\begin{equation}
I_1=-\sum_{s}\int_{t_0}^{t }dt_1\int_{t_1}^{t}dt_2\ldots\int_{t_{s-1}}^{t}dt_s Q^{t}(X) \, r(\sigma) \Delta t=-P^{t}(\sigma) \, r(\sigma) \Delta t
\label{eq:I_1_single}
\end{equation}

Together, $I_0 + I_1$ represent the probability density of having $\sigma(t)=\sigma$ and no jumps in $[t, t+\Delta t]$. 

In the second sum of (\ref{eq:integral_expansion}) we have an operator that acts over the space where all jumps, except the last one, took place in $[t_0, t]$, and the last jump occurs in $[t, t+\Delta t]$. Therefore, we can obtain a second contribution of order $O(\Delta t)$ by applying this operator to the probability density $Q$ of having only one jump in $[t, t+\Delta t]$. Remembering the parametrization (\ref{eq:Q_single_param_rew}) and expanding in powers of $\Delta t$:

\begin{eqnarray}
I_2 &=& \sum_{s} \int_{t_0}^{t} dt_1 \ldots \int_{t_{s-2}}^{t} dt_{s-1} \Big[ \prod_{l=1}^{s-1} r(\sigma(t_l)) \Big] e^{ - \int_{t_0}^{t} r(\sigma(\tau)) d\tau } \int_{t}^{t + \Delta t} dt_s \, r(\sigma(t_s)) e^{ - \int_{t}^{t+\Delta t} r(\sigma(\tau)) d\tau } \nonumber \\
I_2 &=& \Big(\sum_{s} \int_{t_0}^{t} dt_1 \ldots \int_{t_{s-2}}^{t} dt_{s-1} \Big[ \prod_{l=1}^{s-1} r(\sigma(t_l)) \Big] e^{ - \int_{t_0}^{t} r(\sigma(\tau)) d\tau } \Big) \Big(r(-\sigma) \Delta t \Big) + o(\Delta t)
\label{eq:I_2_single_1}
\end{eqnarray}

In (\ref{eq:I_2_single_1}) we have used that, in order to fulfill the condition $\sigma(t)=\sigma$, the last jump must happen from $-\sigma$ to $\sigma$, with rate $r(-\sigma)$. The sum inside parenthesis in the equation now goes over all histories that end with $\sigma(t)=-\sigma$, and therefore is equal to $P^{t}(-\sigma)$. Thus $I_2 = P^{t}(-\sigma) \, r(-\sigma) \Delta t + o(\Delta t)$ represents the probability density of having $\sigma(t)=-\sigma$ and only one jump in $[t, t+\Delta t]$.

Putting all this together:

\begin{eqnarray}
P^{t + \Delta t}(\s) &=& I_0 + I_1 + I_2 + o(\Delta t) \nonumber \\
P^{t + \Delta t}(\s) &=& P^{t}(\sigma) - P^{t}(\sigma) \, r(\sigma) \Delta t + P^{t}(-\sigma) \, r(-\sigma) \Delta t + o(\Delta t)
\label{eq:P_single_order_Dt}
\end{eqnarray}

By subtracting $P^{t}(\sigma)$ from both sides of (\ref{eq:P_single_order_Dt}), dividing by $\Delta t$ and taking the limit $\Delta t \rightarrow 0$ we obtain the desired result (\ref{eq:master_equation_single}):

\begin{equation*}
\dfrac{dP^{t}(\s)}{dt} = -r(\s) P^{t}(\s) + r(-\s) P^{t}(-\s)
\end{equation*}

\subsubsection{Multiple variables}

In this case, where we have the time evolution of the N-interacting spins $\vec{\sigma}=\lbrace \sigma_1, ..., \sigma_N \rbrace$, the master equation is:

\begin{equation}
\dfrac{dP^{t}(\vec{\sigma})}{dt} = -\sum_{i =1}^{N} \big[ r_i(\vec{\sigma}) P^{t}(\vec{\sigma}) + r_i(F_i[\vec{\sigma}]) P^{t}(F_i[\vec{\sigma}]) \big]
\label{eq:master_equation_full}
\end{equation}
where $F_i[\vec{\sigma}]$ is an operator that transforms the state $\vec{\sigma}= \lbrace \sigma_1, ..., \sigma_i, ..., \sigma_N \rbrace$ into the state $F_i[\vec{\sigma}]= \lbrace \sigma_1, ..., -\sigma_i, ..., \sigma_N \rbrace$, and $r_i(\vec{\sigma})$ is the transition rate between $\vec{\sigma}$ and $F_i[\vec{\sigma}]$.

Similarly as in (\ref{eq:marginal_Q}), the probability density $P^{t}(\vec{\sigma})$ is given by:

\begin{equation}
P^{t}(\vec{\sigma}) = \sum_{\vec{X} \mid \vec{\sigma}}^{t} Q^{t}(\vec{X})
\label{eq:marginal_Q_vec}
\end{equation}
where $\vec{X}= \lbrace X_1, \ldots, X_N \rbrace$ is the vector of the histories of all spins.

Now we must differentiate (\ref{eq:marginal_Q_vec}) similarly as with (\ref{eq:marginal_Q}):

\begin{equation}
\frac{dP^{t}(\vec{\sigma})}{dt} = \lim_{\Delta t \rightarrow 0} \dfrac{P^{t+\Delta t}(\vec{\sigma}) - P^{t}(\vec{\sigma})}{\Delta t}
\label{eq:increment_ratio_vec}
\end{equation}

The set of individual histories of $N$-interacting spins, $Q^{t}(\vec{X})$, can be written as a product of probability densities $\Phi_a^{t}(X_a \mid \vec{X}_{\setminus a})$ of the history $X_a$ with the histories of all the other spins $\vec{X}_{\setminus a}$ fixed \cite{kipnis99}:

\begin{equation}
Q^{t}(X_1, X_2, ..., X_N) = \prod_{i=1}^{N} \Phi_i^{t}(X_i \mid \vec{X}_{\setminus i})
\label{eq:Q_prod}
\end{equation}

Each $\Phi_i^{t}(X_i \mid \vec{X}_{\setminus i})$ can be parametrized like in (\ref{eq:Q_single_param_rew}) \cite{vanKampen, Daley2002RPP}:

\begin{equation}
\Phi_i^{t} (X_i | \vec{X}_{\setminus i}) = \Big[ \prod_{l_i=1}^{s_i} r_{i}(\vec{\sigma}(t_{l_i})) \Big] \exp\Big\lbrace - \int_{t_0}^{t} r_{i}(\vec{\sigma}(\tau)) d\tau \Big\rbrace
\label{eq:Phi_param}
\end{equation}

Notice that when any $X_j$ with $j \in \partial i$ has a jump, the value of $r_{i} (\vec{\sigma}(\tau))$ changes in the integral inside the exponential function. As before, we will write
$P^{t + \Delta t}(\vec{\s})$ as the marginalization of $Q^{t + \Delta t}(\vec{X})$ and then expand to first order in $\Delta t$. Explicitly writing that marginalization as in (\ref{eq:marginalizing_Q_single_delta_t}):

\begin{eqnarray}
P^{t + \Delta t}(\vec{\s}) = \sum_{s_1=0}^{\infty} \sum_{s_2=0}^{\infty} \!\! \ldots \!\! \sum_{s_N=0}^{\infty} \Big[ \prod_{i =1}^{N} \int_{t_0}^{t + \Delta t} \!\!\!\!\! dt_1^{i}\int_{t_1^{i}}^{t + \Delta t} \!\!\!\!\! dt_2^{i}\ldots\int_{t_{s-1}^{i}}^{t + \Delta t} \!\!\!\!\! dt_s^{i} \Big] Q^{t+\Delta t}(X_1, X_2, \ldots, X_N)
\label{eq:marginalizing_Q_general_delta_t}
\end{eqnarray}

To expand $Q^{t+\Delta t}(\vec{X})$ requires expanding each $\Phi_i^{t + \Delta t}(X_i \mid \vec{X}_{\setminus i})$, which is almost the same we did in (\ref{eq:Q_single_param_expand}):

\begin{equation}
\Phi_i^{t+\Delta t} \big( X_i \mid \vec{X}_{\setminus i} \big) = \Phi_i^{t} \big(X_i \mid \vec{X}_{\setminus i} \big) \, \big[ 1 - r_{i}(\vec{\sigma}(t)) \, \Delta t \big] + o(\Delta t)
\label{eq:Phi_general_param_expand}
\end{equation}

Thus:
\begin{eqnarray}
Q^{t+\Delta t}(\vec{X}) &=& \prod_{i=1}^{N} \Phi_i^{t+\Delta t} \big( X_i \mid \vec{X}_{\setminus i} \big) \nonumber \\
Q^{t+\Delta t}(\vec{X}) &=& \prod_{i=1}^{N} \Phi_i^{t} \big( X_i \mid \vec{X}_{\setminus i} \big) \Big[1 - \Delta t \sum_{k=1}^{N} r_{k}(\vec{\sigma}(t)) \Big] + o(\Delta t) \nonumber \\
Q^{t+\Delta t}(\vec{X}) &=& Q^{t}(\vec{X}) \Big[1 - \Delta t \sum_{k=1}^{N} r_{k}(\vec{\sigma}(t)) \Big] + o(\Delta t)
\label{eq:Q_expand_gen}
\end{eqnarray}

Similarly as with (\ref{eq:I_0_single}) and (\ref{eq:I_1_single}) we will use (\ref{eq:Q_expand_gen}) to get two contributions of order $O(1)$ and $O(\Delta t)$, respectively:

\begin{eqnarray}
I_0&=&\sum_{s_1=0}^{\infty} \! \ldots \! \sum_{s_N=0}^{\infty} \Big[ \prod_{i =1}^{N} \int_{t_0}^{t} \!\!\! dt_1^{i}\ldots\int_{t_{s-1}^{i}}^{t} \!\!\! dt_s^{i} \Big] Q^{t}(\vec{X})=P^{t}(\vec{\sigma})
\label{eq:I_0_gen} \\
I_1&=& \Big( \sum_{s_1=0}^{\infty} \! \ldots \! \sum_{s_N=0}^{\infty} \Big[ \prod_{i =1}^{N} \int_{t_0}^{t} \!\!\! dt_1^{i}\ldots\int_{t_{s-1}^{i}}^{t} \!\!\! dt_s^{i} \Big] Q^{t}(\vec{X}) \Big) \Big(-\Delta t \sum_{k=1}^{N} r_{k}(\vec{\sigma}(t)) \Big) \nonumber \\
I_1&=&-P^{t}(\vec{\sigma}) \, \Delta t \, \sum_{k=1}^{N} r_{k}(\vec{\sigma}(t))
\label{eq:I_1_gen}
\end{eqnarray}

As in (\ref{eq:integral_expansion}) we can split the sum of iterated integrals in the right hand side of (\ref{eq:marginalizing_Q_general_delta_t}) to get $O(\Delta t)$ contributions. The latter is an operator that acts over the space of the histories where one and only one spin jumps in $[t, t + \Delta t]$. he rest is a calculation essentially equal to the one in (\ref{eq:I_2_single_1}):

\begin{eqnarray}
I_2 &=& \sum_{k=1}^{N}\, \sum_{s_1=0}^{\infty} \! \ldots \! \sum_{s_N=0}^{\infty} \Big[ \prod_{i \neq k} \int_{t_0}^{t} \!\!\! dt_1^{i}\ldots\int_{t_{s-1}^{i}}^{t} \!\!\! dt_s^{i} \Big] \int_{t_0}^{t} \!\!\! dt_1^{k}\ldots\int_{t_{s-2}}^{t} \!\!\! dt_{s-1}^{k} \Big[ \prod_{i \neq k} \Phi_{i}^{t}(X_{i} \mid \vec{X}_{\setminus i}) \Big] \times \nonumber \\
& & \times \Big[ \prod_{l_k=1}^{s_k-1} r_{k} (\vec{\sigma}(t_{l_k})) \Big] e^{ - \int_{t_0}^{t} r_{k} (\vec{\sigma}(\tau)) d\tau} r_{k} \big(F_{k}[\vec{\sigma}(t)] \big) \Delta t + o(\Delta t) \nonumber \\
I_2 &=& \sum_{k=1}^{N} P^{t}(F_{k}[\vec{\sigma}]) \, r_{k} \big(F_{k}[\vec{\sigma}(t)] \big) \, \Delta t
\label{eq:I_2_gen}
\end{eqnarray}

Finally, the expansion of $P^{t+\Delta t}(\vec{\sigma})$ gives:

\begin{eqnarray}
P^{t + \Delta t}(\vec{\s}) &=& I_0 + I_1 + I_2 + o(\Delta t) \nonumber \\
P^{t + \Delta t}(\vec{\s}) &=& P^{t}(\vec{\sigma}) - \Delta t \sum_{k=1}^{N} P^{t}(\vec{\sigma}) r_{k} (\vec{\sigma}) + \Delta t \sum_{k=1}^{N} P^{t}(F_k[\vec{\sigma}]) r_{k} (F_k[\vec{\sigma}]) + o(\Delta t)
\label{eq:P_gen_order_Dt}
\end{eqnarray}
and using (\ref{eq:increment_ratio_vec}) we obtain the usual Master Equation \cite{vanKampen} for a set of N-interacting spins (\ref{eq:master_equation_full}):

\begin{equation}
\dfrac{dP^{t}(\vec{\sigma})}{dt} = -\sum_{i =1}^{N} \big[ r_i(\vec{\sigma}) P^{t}(\vec{\sigma}) + r_i(F_i[\vec{\sigma}]) P^{t}(F_i[\vec{\sigma}]) \big] \label{eq:Gen_Mas_Eq}
\end{equation}

Unfortunately, to solve (\ref{eq:Gen_Mas_Eq}), even numerically, is in general a very difficult task because the densities $P^{t}(\vec{\sigma})$ are high dimensional objects.

\section{Hierarchical Cavity Master Equations} \label{sec:hCME}

This section contains the main analytical contribution of our work. We will exploit techniques similar to the ones presented above to write down a set of closed differential equations for the stochastic dynamics of discrete variables in a Random Graph. We generalize a closure presented in \cite{CME-PRE} substituting an uncontrolled approximation by new equations derived from first principles through the Theory of Random Point Processes.


To simplify the reading this section is divided in five subsections. At \ref{subsec:RPP_many_var_cav} we introduce the dynamic cavity method to study systems defined over tree-like graphs. In subsection \ref{subsec:writing_probs_local} we use that formulation to write the local probability densities sitting on any group of connected nodes in terms of dynamic cavity messages. Then we derive the known local master equations for those probability densities. In subsection \ref{subsec:writing_probs_cav} we repeat the same procedure to obtain analogous equations for cavity probability densities. A general approximated method for closing these equations is presented and discussed in subsection \ref{subsec:closure}. There, we show how these closed dynamic equations can be organized in a system of hierarchical approximations.

\subsection{Treelike architecture and cavity messages} \label{subsec:RPP_many_var_cav}


In this subsection we will introduce the dynamic cavity messages for tree-like graphs, where it is possible to separate the spins into two disconnected networks just by removing a single connection between two nodes.

We first select a spin, say $i$, and rewrite equation (\ref{eq:Q_prod}) expanding the tree around it and making use of its structure:

\begin{equation}
Q^{t}(X_1,\ldots,X_N) = \Phi_i^{t}(X_i|X_{\partial i}) \prod_{k \in \partial i} \Big[ \Phi_k^{t}(X_k|X_{\partial k}) \prod_{m \in \partial k \setminus i} \Big( \Phi_k^{t}(X_m|X_{\partial m}) \prod_{l \in \partial m \setminus k} \ldots \Big)\Big]
\label{eq:full_joint_prob_dist_tree}
\end{equation}

In equation (\ref{eq:full_joint_prob_dist_tree}) we used the symbol $\partial i$ for the set of spins in the neighborhood of $i$. Those are the only spins that directly interact with $i$ and therefore we adapted the previous notation of the $\Phi$ probability densities. The symbol $X_{\partial i}$ represents the set of histories of the nodes in $\partial i$. Furthermore, let $G_k^{(i)}$ be the sub-graph expanded from the site $k$ after removing the link $(ik)$. We define as $\{X\}_{ik}$ the set of histories of the spins included in $G_k^{(i)}$ except
$X_k$ itself. With these definitions we express (\ref{eq:full_joint_prob_dist_tree}) as:

\begin{equation}
Q^{t}(X_1,\ldots,X_N) = \Phi_i^{t}(X_i|X_{\partial i}) \prod_{k \in \partial i} M_{ki}^{t}(X_i,X_k,\{X\}_{ik})
\end{equation}

Here $M_{ki}^{t}$ is just a shortening for the expression inside brackets. Marginalizing $Q$ on all histories except $X_i,X_{\partial i}$ we get:

\begin{equation}
Q^{t}(X_i,X_{\partial i}) = \Phi_i^{t}(X_i|X_{\partial i}) \prod_{k \in \partial i} \mu_{k\rightarrow (ki)}^{t}(X_k|X_i)
\label{eqn:marginal_prob_dist_mess}
\end{equation}

The new functions $\mu_{k\rightarrow (ki)}^{t}(X_k|X_i)$, called dynamic cavity messages, are the marginals:
\begin{equation}
\mu_{k\rightarrow (ik)}^{t}(X_k|X_i) = \sum_{\{X\}_{ik}}^{t} M_{ki}^{t}(X_i,X_k,\{X\}_{ik})
\label{eq:cav_mess}
\end{equation}
and have the interpretation of the probability density of history $X_k$ given $X_i$ fixed.

If we take now two neighbors, $i$ and $j$, and make a similar reasoning we conclude that their marginal probability density can be written as:

\begin{equation}
Q^{t}(X_i,X_j) = \mu_{i\rightarrow (ij)}^{t}(X_i|X_j) \mu_{j\rightarrow (ji)}^{t}(X_j|X_i)
\label{eqn:pair_prob_dist_mess}
\end{equation}

These cavity messages can be parametrized as other similar dynamical quantities (see equations (\ref{eq:Q_single_param}), (\ref{eq:Phi_param})):

\begin{eqnarray}
\mu_{i \rightarrow (ij)}^{t}(X_i \mid X_j) = \lambda_{i \rightarrow (ij)} (X_i, X_j, t_0) \, e^{ -\int_{t_0}^{t_1} \lambda_{i}^{\tau} d\tau } \times \lambda_{i \rightarrow (ij)} (X_i, X_j, t_1) e^{ -\int_{t_1}^{t_2} \lambda_{i}^{\tau} d\tau } \times ... \label{eq:mu_param} \\
... \times \lambda_{i \rightarrow (ij)} (X_i, X_j, t_{s_i}) e^{ -\int_{t_{s_i}}^{t} \lambda_{i}^{\tau} d\tau } \nonumber
\end{eqnarray}
only that now the jumps occur with unknown rates $\lambda_{i \rightarrow (ij)} (X_i, X_j, t)$.

Then, as in (\ref{eq:Phi_general_param_expand}) we can expand $\Phi$ and $\mu$ to order $\Delta t$ (for the latter, see Appendix \ref{app:relation_mu_lambda}):

\begin{eqnarray}
\Phi_i^{t+\Delta t} \big( X_i \mid X_{\partial i} \big) &=& \Phi_i^{t} \big(X_i \mid X_{\partial i} \big) \, \big[ 1 - r_{i}(\sigma_i(t), \sigma_{\partial i} (t)) \, \Delta t \big] + o(\Delta t) \label{eq:Phi_tree_param_expand} \\
\mu_{i \rightarrow (ij)}^{t + \Delta t} (X_i | X_j) &=& \mu_{i \rightarrow (ij)}^t(X_i | X_j) \, [1-\lasubsim{i}{j}{X_i, X_j, t} \Delta t] + o(\Delta t) \label{eq:mu_tree_param_expand}
\end{eqnarray}

The equations (\ref{eq:cav_mess})-(\ref{eq:mu_tree_param_expand}) were already present in \cite{CME-PRE}, but they will be also at the basis of our derivation. They will be exploited below to write down a more general set of equations than the ones presented in \cite{CME-PRE}.


\subsection{Equations for local probability densities}\label{subsec:writing_probs_local}

\begin{figure}[htb]
\centering
\includegraphics[keepaspectratio=true,width=0.41\textwidth]{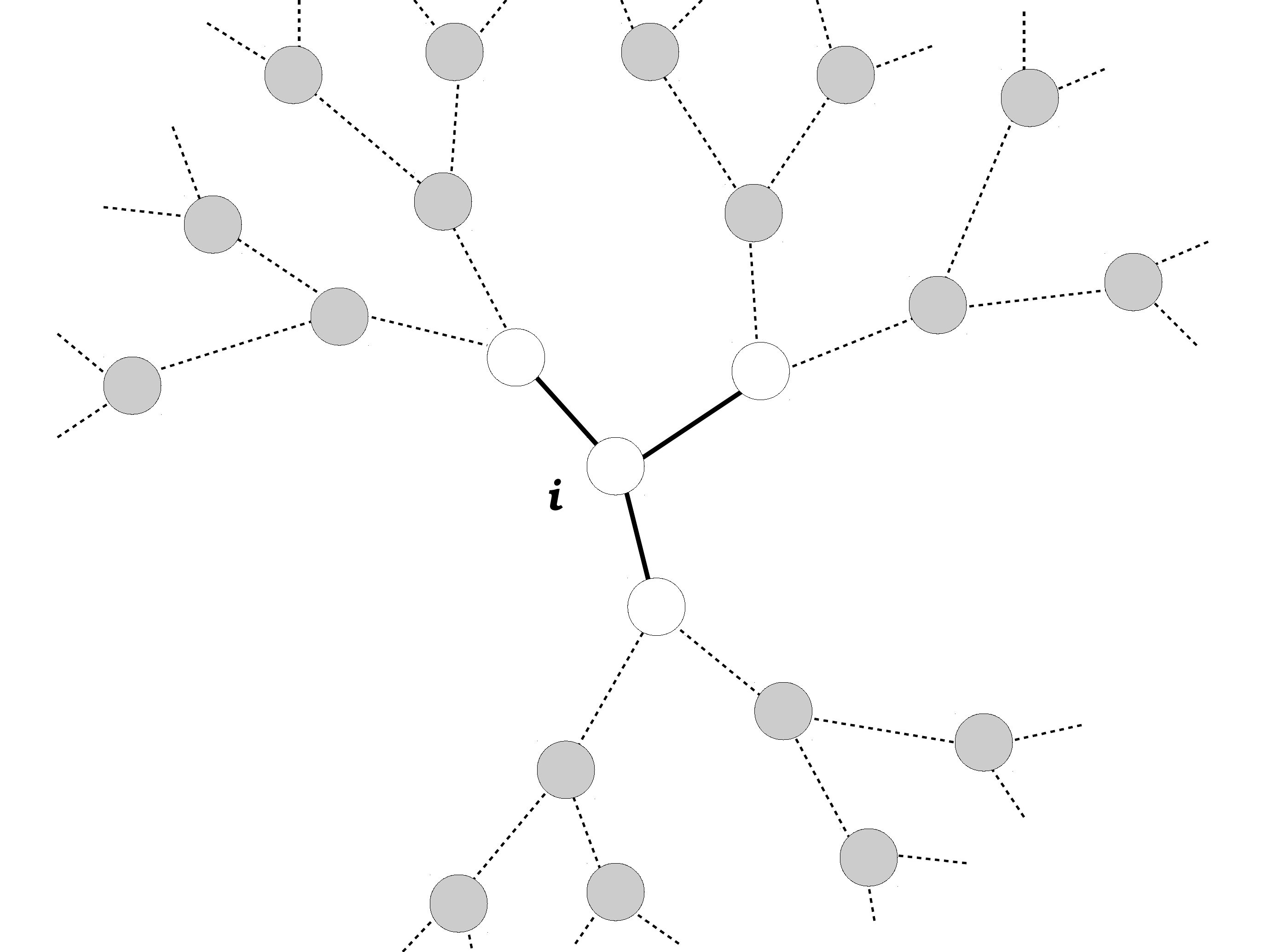}
\includegraphics[keepaspectratio=true,width=0.44\textwidth]{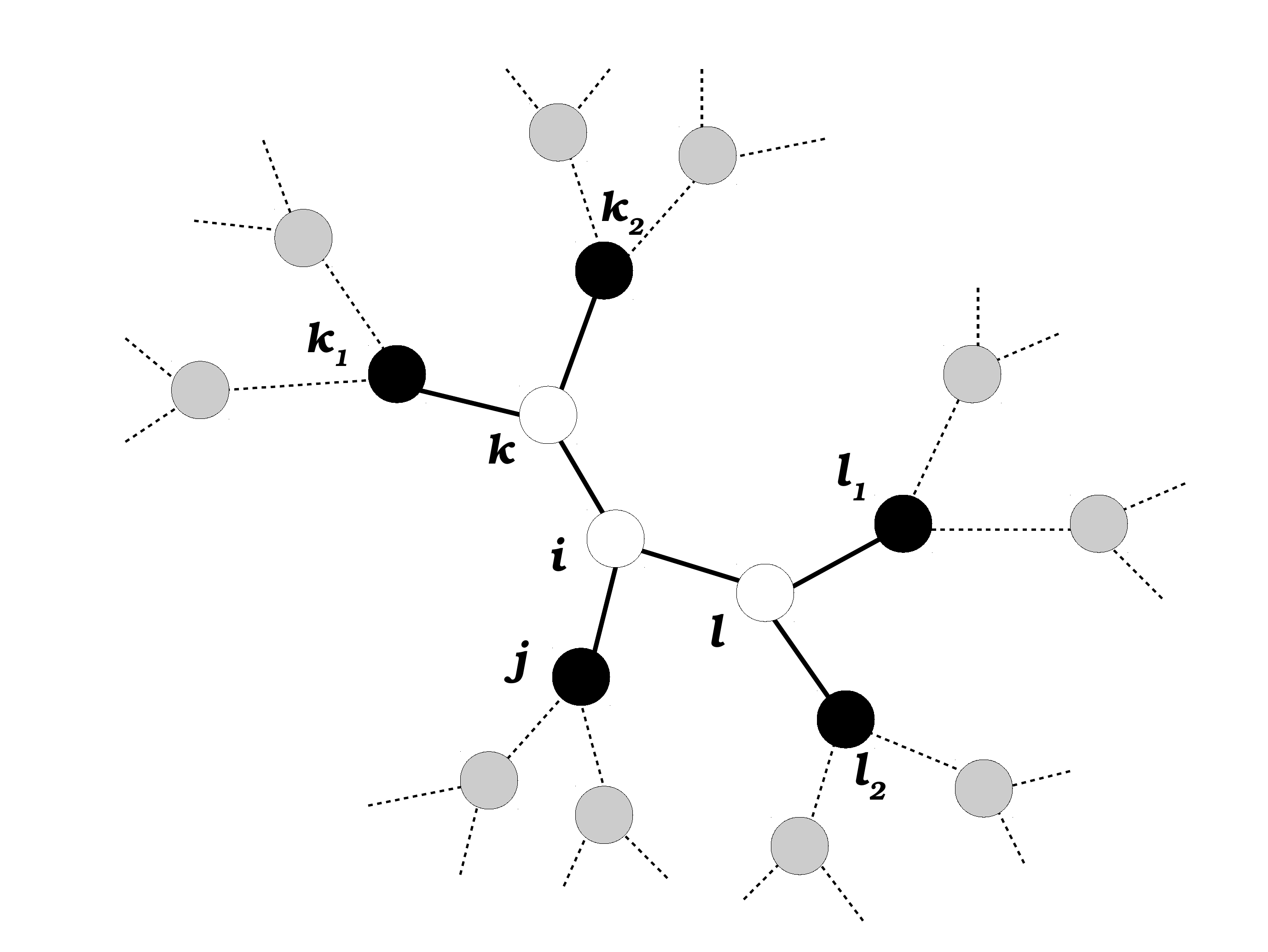}

\includegraphics[keepaspectratio=true,width=0.44\textwidth]{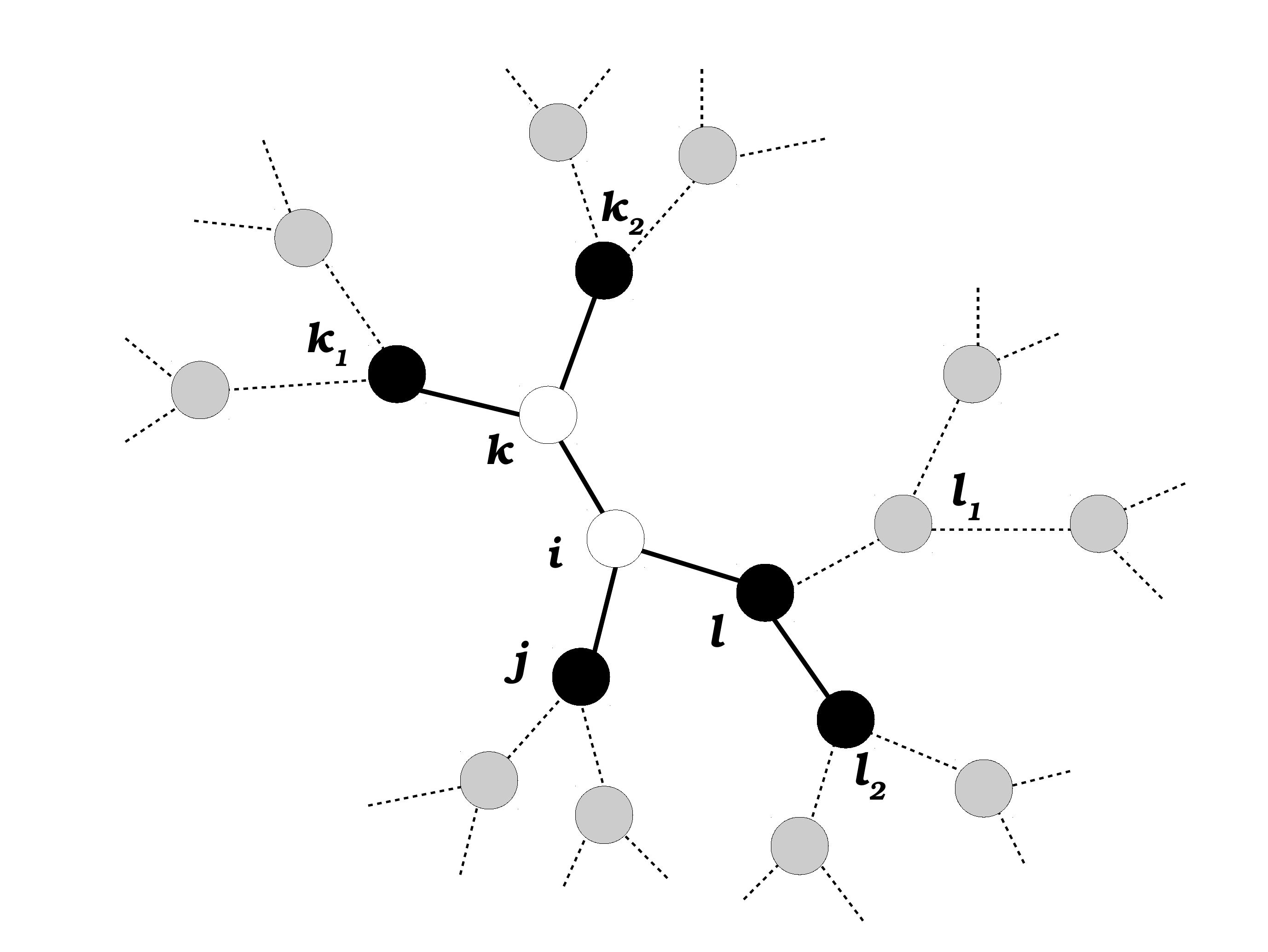}
\includegraphics[keepaspectratio=true,width=0.44\textwidth]{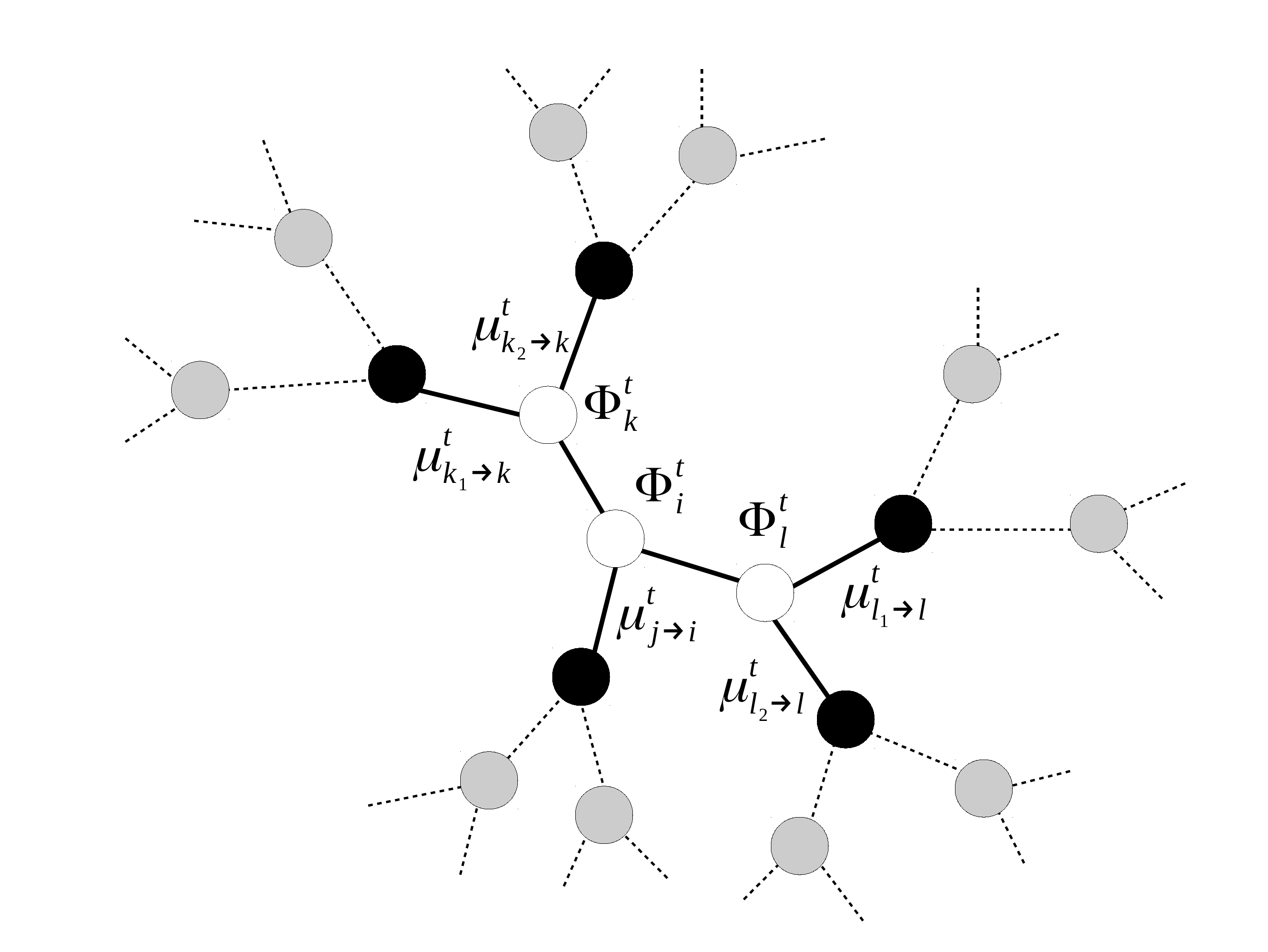}
\caption{\textbf{(Top-left panel):} Selection of a connected set with $4$ nodes (white) from a tree-like graph. The full graph is divided in $6$ tree-like sub-graphs (gray nodes). \textbf{(Top-right panel):} Dividing a connected set with $8$ nodes in two groups: inner (white) and outer nodes (black). This is a connected set of the first kind, because every outer node has only one neighbor inside the set. \textbf{(Bottom-left panel):} Connected set with 7 nodes. According to our classification the connected set is of the second kind because the node $l$ has two neighbors, $i$ and $l_2$, which belong to the set. \textbf{(Bottom-right panel):} Local weights and cavity messages present in the probability densities related to the connected set shown in top-right panel.}
\label{fig:illustration_connected_set}
\end{figure}

By eliminating a group of connected nodes from a tree-like graph, we always divide it into several tree-like sub-graphs. We illustrate this in the top-left panel of figure (\ref{fig:illustration_connected_set}). There, we colored in gray the sub-trees obtained after removing the white nodes in the center. As we learned in subsection \ref{subsec:RPP_many_var_cav}, if we marginalize the full distribution $Q
^{t}(\vec{X})$ over all the histories in these gray sub-graphs, we obtain the probability density of the set of histories corresponding to the connected group we have selected.

Therefore, if we denote the central node in the top-left panel of figure (\ref{fig:illustration_connected_set}) by the letter $i$, and remembering the steps from (\ref{eq:full_joint_prob_dist_tree}) to (\ref{eqn:marginal_prob_dist_mess}), we have:

\begin{eqnarray}
Q^{t}(\vec{X}_\text{connected set}) &=&\sum_{\vec{X}_{\text{gray nodes}}}^{t} Q^{t}(\vec{X}) \nonumber \\
Q^{t}(X_i,X_{\partial i}) &=& \Phi_i^{t}(X_i|X_{\partial i}) \prod_{k \in \partial i} \mu_{k\rightarrow (ki)}^{t}(X_k|X_i)
\label{eqn:marginal_prob_dist_mess_2}
\end{eqnarray}

This procedure can be easily generalized to write any local probability density of the histories of a connected set of nodes. We just have to recognize inner and outer nodes: the first ones have all their neighbors belonging to the connected set, and the second ones have at least one neighbor outside the connected set. The top-right and the bottom-left panels of figure (\ref{fig:illustration_connected_set}) show examples of connected sets. In each case, we have a group of inner nodes (in white), and a group of outer nodes (in black).

The probability density of a connected set is always a marginal of the full probability density. However, in order to avoid a more involved analysis, we will separate the connected sets into two categories. A connected set is \emph{of the first kind} when all its outer nodes have only one neighbor inside the set, and is \emph{of the second kind} when at least one outer node has more than one neighbor inside the set. This distinction is important because in the first case we can directly write the corresponding probability density as a product of local weights $\Phi$ and dynamic cavity messages $\mu$. We must simply include a local weight for each inner node and a cavity message for each outer node.

For example, the probability density corresponding to the connected set in the top-right panel of figure (\ref{fig:illustration_connected_set}) is:

\begin{eqnarray}
Q^{t}(\vec{X}_\text{connected set}) &=&\sum_{\vec{X}_{\text{gray nodes}}}^{t} Q^{t}(\vec{X}) \nonumber \\
Q^{t}(X_i, X_{\partial i}, X_{\partial k \setminus i}, X_{\partial l \setminus i}) &=& \Phi_i^{t}(X_i|X_{\partial i}) \Big[ \prod_{m \in \partial i \setminus j} \Phi_m^{t}(X_m|X_{\partial m}) \Big] \mu_{j\rightarrow (ji)}^{t}(X_j|X_i) \times \nonumber \\
& & \:\:\:\:\: \times\Big[ \prod_{n \in \partial k \setminus i} \mu_{n\rightarrow (nk)}^{t}(X_n|X_k) \Big] \Big[ \prod_{n \in \partial l \setminus i} \mu_{n\rightarrow (nl)}^{t}(X_n|X_l) \Big]
\label{eq:connected_set_example}
\end{eqnarray}
where $j, k, l \in \partial i$. A representation of this product of $\Phi$ and $\mu$ functions is shown in the bottom-right panel of (\ref{fig:illustration_connected_set}).

Connected sets of the second kind do not require a different treatment. Fortunately, we can always add nodes to convert these into sets of the first kind. For example, we can transform the bottom-left panel of figure (\ref{fig:illustration_connected_set}) into the top-right panel just by adding the node $l_1$ to the connected set. Then we can simply compute:

\begin{equation}
Q^{t}(X_i, X_{\partial i}, X_{\partial k \setminus i}, X_{l_2}) = \sum_{X_{l_1}}^{t} Q^{t}(X_i, X_{\partial i}, X_{\partial k \setminus i}, X_{l_1}, X_{l_2})
\label{eq:marginal_conn_set}
\end{equation}

Thus, the probability density of any set of the second kind can be written as a marginal of the probability density of a set of the first kind just like (\ref{eq:marginal_conn_set}) is written in terms of (\ref{eq:connected_set_example}).

In order to derive a differential equation for the time evolution of any instantaneous local probability density $P^{t}$, we should concentrate on connected sets of the first kind. To fix ideas, let's work with such a connected set containing $n$ inner nodes and $m$ outer nodes. As each outer node has one neighbor which is an inner node, we will have $n \geq m$.

Let's construct two subsets of nodes: $\mathcal{O}$ and $\mathcal{I}$. First, we fill $\mathcal{O}$ with all outer nodes in an arbitrary order: $\mathcal{O} = \lbrace o_1, o_2, \ldots, o_m \rbrace$. Then we build the set $\mathcal{I}=\lbrace i_1, i_2, \ldots, i_m, i_{m+1}, \ldots, i_n \rbrace$ in such a way that each node $i_{k} \in \mathcal{I}$ is a neighbor of the node $o_k \in \mathcal{O}$, with $k \leq m$.

The instantaneous probability density can be written as:

\begin{equation}
P^{t}(\vec{\sigma}_{\mathcal{O}}, \vec{\sigma}_{\mathcal{I}}) = \sum_{\vec{X}_{\mathcal{I}} | \vec{\sigma}_{\mathcal{I}}}^{t} \sum_{\vec{X}_{\mathcal{O}} | \vec{\sigma}_{\mathcal{O}}}^{t} Q^{t}(\vec{X}_{\mathcal{O}}, \vec{X}_{\mathcal{I}})
\label{eq:P_from_Q_tree}
\end{equation}
where $\vec{\sigma}_{\mathcal{I}}$ and $\vec{\sigma}_{\mathcal{O}}$ are the vectors of the instantaneous states of the inner and outer nodes, respectively, and $\vec{X}_{\mathcal{I}}$ and $\vec{X}_{\mathcal{O}}$ are the corresponding vectors of the histories of those nodes. 
On the other hand:

\begin{equation}
Q^{t}(\vec{X}_{\mathcal{O}}, \vec{X}_{\mathcal{I}}) = \Big[ \prod_{k=1}^{m} \mu_{o_k \rightarrow (o_k i_k)}^{t} (X_{o_k} | X_{i_k}) \Big] \Big[\prod_{k = 1}^{n} \Phi_{i_k}^{t}(X_{i_k}, X_{\partial i_k}) \Big]
\label{eq:general_Q_tree}
\end{equation}

Now we need to expand (\ref{eq:P_from_Q_tree}) to order $\Delta t$. As in subsection \ref{subsec:RPP_to_ME} we can write an order $\Delta t$ expression for $Q^{t+\Delta t}(\vec{X}_{\mathcal{O}}, \vec{X}_{\mathcal{I}})$. Remembering (\ref{eq:Phi_tree_param_expand}) and (\ref{eq:mu_tree_param_expand}) we have:

\begin{eqnarray}
Q^{t+\Delta t}(\vec{X}_{\mathcal{O}}, \vec{X}_{\mathcal{I}}) &=& \Big[ \prod_{k=1}^{m} \mu_{o_k \rightarrow i_k}^{t} \, [1-\lambda_{o_k \rightarrow i_k}^{t} \Delta t] + o(\Delta t) \Big] \Big[\prod_{k = 1}^{n} \Phi_{i_k}^{t} \, \big[ 1 - r_{i_k}^{t} \, \Delta t \big] + o(\Delta t) \Big] \nonumber \\
Q^{t+\Delta t}(\vec{X}_{\mathcal{O}}, \vec{X}_{\mathcal{I}}) &=& Q^{t}(\vec{X}_{\mathcal{O}}, \vec{X}_{\mathcal{I}}) - \Delta t \sum_{l = 1}^{n} r_{i_l}^{t} \; \Big[ \prod_{k=1}^{m} \mu_{o_k \rightarrow i_k}^{t} \Big] \Big[\prod_{k = 1}^{n} \Phi_{i_k}^{t} \Big] - \nonumber \\
& & - \Delta t \sum_{l = 1}^{m} \lambda_{o_l \rightarrow i_l}^{t} \; \Big[ \prod_{k=1}^{m} \mu_{o_k \rightarrow i_k}^{t} \Big] \Big[\prod_{k = 1}^{n} \Phi_{i_k}^{t} \Big]
\label{eq:Q_tree_expand}
\end{eqnarray}
where we have shortened our notation as follows:

\begin{eqnarray}
\Phi_{i_k}^{t} (X_{i_k} | X_{\partial i_k}) &\equiv& \Phi_{i_k}^{t} \nonumber \\
\mu_{o_k \rightarrow (o_k i_k)}^{t}(X_{o_k} | X_{i_k}) &\equiv& \mu_{o_k \rightarrow i_k}^{t} \label{eq:shortenings} \\
r_{i_k}(\sigma_{i_k}(t), \sigma_{\partial i_k} (t)) &\equiv& r_{i_k}^{t} \nonumber \\
\lasubsim{o_k}{i_k}{X_{o_k}, X_{i_k}, t} &\equiv& \lambda_{o_k \rightarrow i_k}^{t}\nonumber
\end{eqnarray}

Similarly as before, there are three contributions to the expansion of (\ref{eq:P_from_Q_tree}) to order $\Delta t$:

\begin{eqnarray}
I_0 &=& \sum_{\vec{X}_{\mathcal{O}} | \vec{\sigma}_{\mathcal{O}}}^{t} \sum_{\vec{X}_{\mathcal{I}} | \vec{\sigma}_{\mathcal{I}}}^{t} Q^{t}(\vec{X}_{\mathcal{O}}, \vec{X}_{\mathcal{I}}) = P^{t}(\vec{\sigma}_\mathcal{O}, \vec{\sigma}_{\mathcal{I}}) \label{eq:I0_gen_tree_P} \\
I_1 &=& -\Delta t \sum_{\vec{X}_{\mathcal{O}} | \vec{\sigma}_{\mathcal{O}}}^{t} \sum_{\vec{X}_{\mathcal{I}} |\vec{\sigma}_{\mathcal{I}}}^{t} \Big[ \prod_{k=1}^{m} \mu_{o_k \rightarrow i_k}^{t} \Big] \Big[\prod_{k = 1}^{n} \Phi_{i_k}^{t} \Big] \Big[ \sum_{l = 1}^{n} r_{i_l}^{t} + \sum_{l = 1}^{m} \lambda_{o_l \rightarrow i_l}^{t} \Big]
\label{eq:I1_gen_tree_P} \\
I_2 &=& \Delta t \sum_{l=1}^{n} \sum_{\vec{X}_{\mathcal{O}} | \vec{\sigma}_{\mathcal{O}}}^{t} \sum_{\vec{X}_{\mathcal{I}} | F_{i_l}[\vec{\sigma}_{\mathcal{I}}]}^{t} \Big[ \prod_{k=1}^{m} \mu_{o_k \rightarrow i_k}^{t} \Big] \Big[\prod_{k = 1}^{n} \Phi_{i_k}^{t} \Big] \: r_{i_l}^{t} + \nonumber \\
& & + \Delta t \sum_{l=1}^{m} \sum_{\vec{X}_{\mathcal{O}} | F_{o_l}[\vec{\sigma}_{\mathcal{O}}]}^{t} \sum_{\vec{X}_{\mathcal{I}} | \vec{\sigma}_{\mathcal{I}}}^{t} \Big[ \prod_{k=1}^{m} \mu_{o_k \rightarrow i_k}^{t} \Big] \Big[\prod_{k = 1}^{n} \Phi_{i_k}^{t} \Big] \: \lambda_{o_l \rightarrow i_l}^{t}
\label{eq:I2_gen_tree_P}
\end{eqnarray}

In Appendix \ref{app:relation_mu_lambda} (see equation (\ref{eq:lambda_mu_app_F})) we derive the identity:

\begin{equation}
\lambda_{i \rightarrow j}^{t}\;\;\mu_{i \rightarrow j}^{t} = \displaystyle \sum_{X_{\partial i\setminus j}}^{t}
r_{i}^{t} \, \Phi_i^{t} \prod_{k\in\partial i\setminus j} \mu_{k \rightarrow i}^{t} \label{eq:lambda_mu_app_full}
\end{equation}
which leads to:

\begin{eqnarray}
\Big[ \prod_{k=1}^{m} \mu_{o_k \rightarrow i_k}^{t} \Big] \Big[\prod_{k = 1}^{n} \Phi_{i_k}^{t} \Big] \: \lambda_{o_l \rightarrow i_l}^{t} &=& \Big[ \prod_{\substack{k=1 \\ k \neq l}}^{m} \mu_{o_k \rightarrow i_k}^{t} \Big] \Big[\prod_{k = 1}^{n} \Phi_{i_k}^{t} \Big] \: \lambda_{o_l \rightarrow i_l}^{t} \; \mu_{o_l \rightarrow i_l}^{t} \label{eq:product_with_lambda} \\
\Big[ \prod_{k=1}^{m} \mu_{o_k \rightarrow i_k}^{t} \Big] \Big[\prod_{k = 1}^{n} \Phi_{i_k}^{t} \Big] \: \lambda_{o_l \rightarrow i_l}^{t} &=& \Big[ \prod_{\substack{k=1 \\ k \neq l}}^{m} \mu_{o_k \rightarrow i_k}^{t} \Big] \Big[\prod_{k = 1}^{n} \Phi_{i_k}^{t} \Big] \sum_{X_{\partial o_l \setminus i_l}}^{t}
r_{o_l}^{t} \, \Phi_{o_l}^{t} \Big[\prod_{k\in\partial o_l \setminus i_l} \mu_{k \rightarrow o_l}^{t} \Big] \nonumber \\
\Big[ \prod_{k=1}^{m} \mu_{o_k \rightarrow i_k}^{t} \Big] \Big[\prod_{k = 1}^{n} \Phi_{i_k}^{t} \Big] \: \lambda_{o_l \rightarrow i_l}^{t} &=& \sum_{X_{\partial o_l \setminus i_l}}^{t} r_{o_l}^{t} \Big[ \prod_{\substack{k=1 \\ k \neq l}}^{m} \mu_{o_k \rightarrow i_k}^{t} \Big] \Big[\prod_{k = 1}^{n} \Phi_{i_k}^{t} \Big]
\Phi_{o_l}^{t} \Big[\prod_{k\in\partial o_l \setminus i_l} \mu_{k \rightarrow o_l}^{t} \Big]
\label{eq:bigger_probs_appears}
\end{eqnarray}

\begin{figure}[htb]
\centering
\includegraphics[keepaspectratio=true,width=0.45\textwidth]{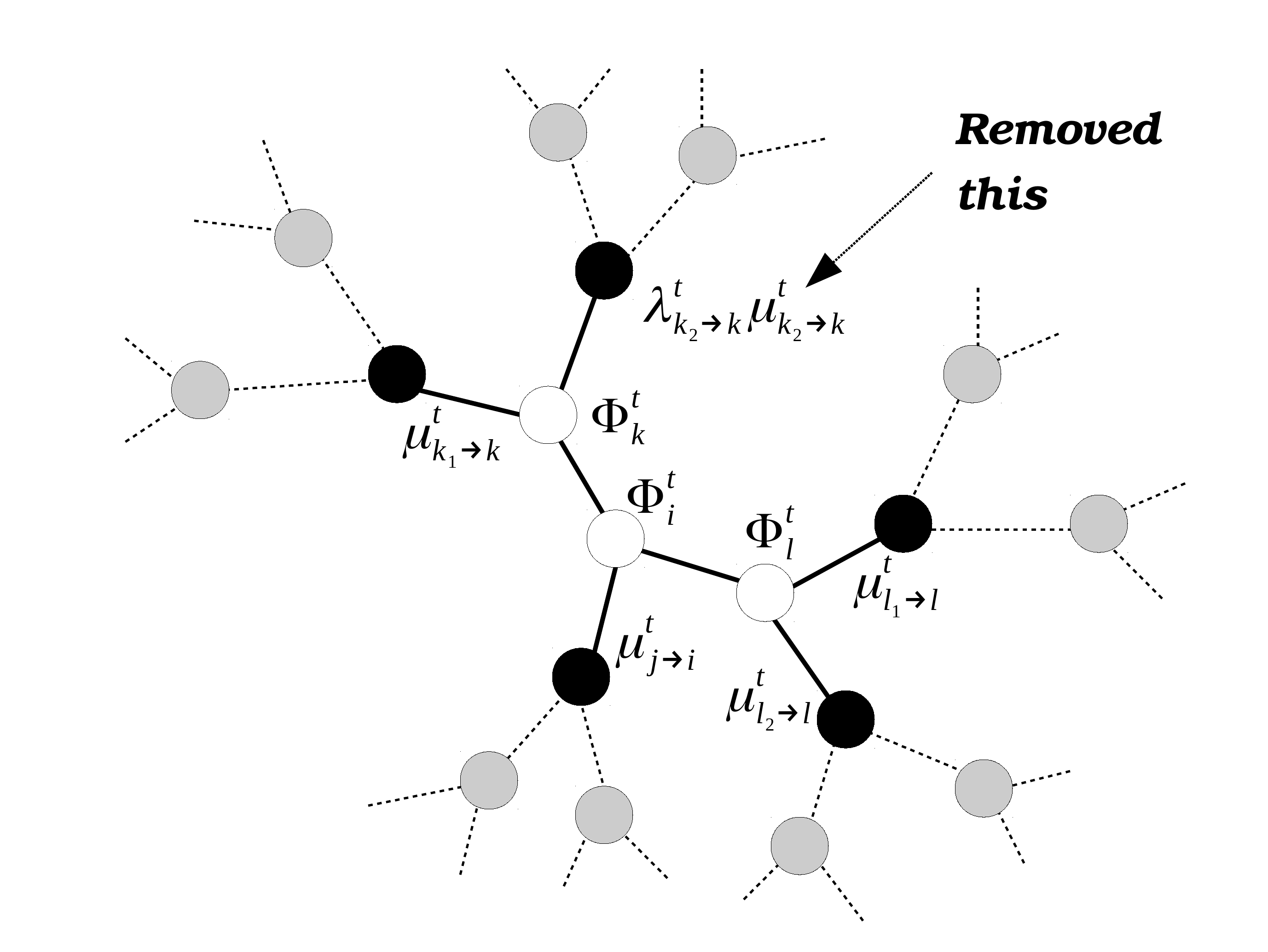}\includegraphics[keepaspectratio=true,width=0.09\textwidth]{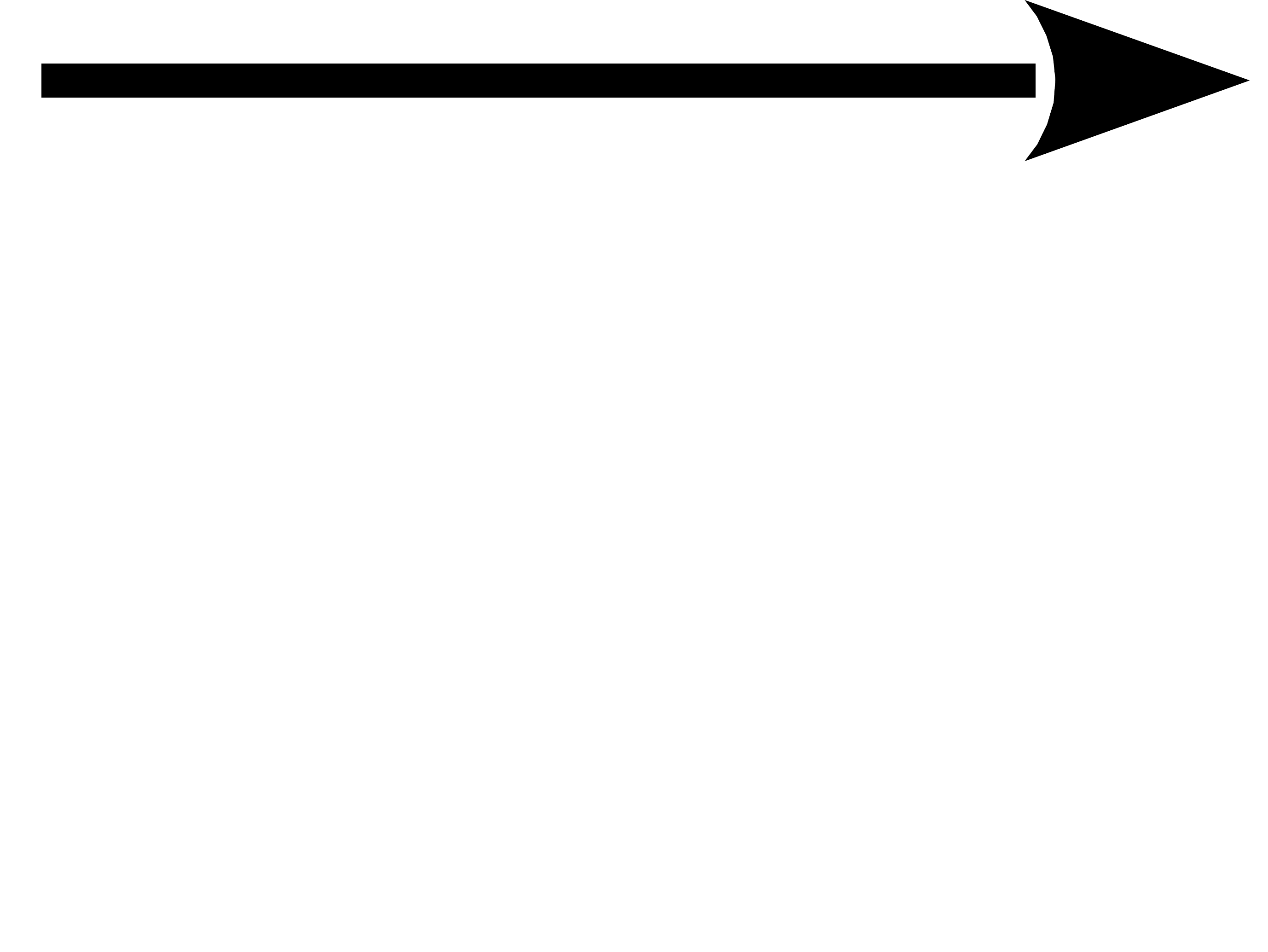}
\includegraphics[keepaspectratio=true,width=0.45\textwidth]{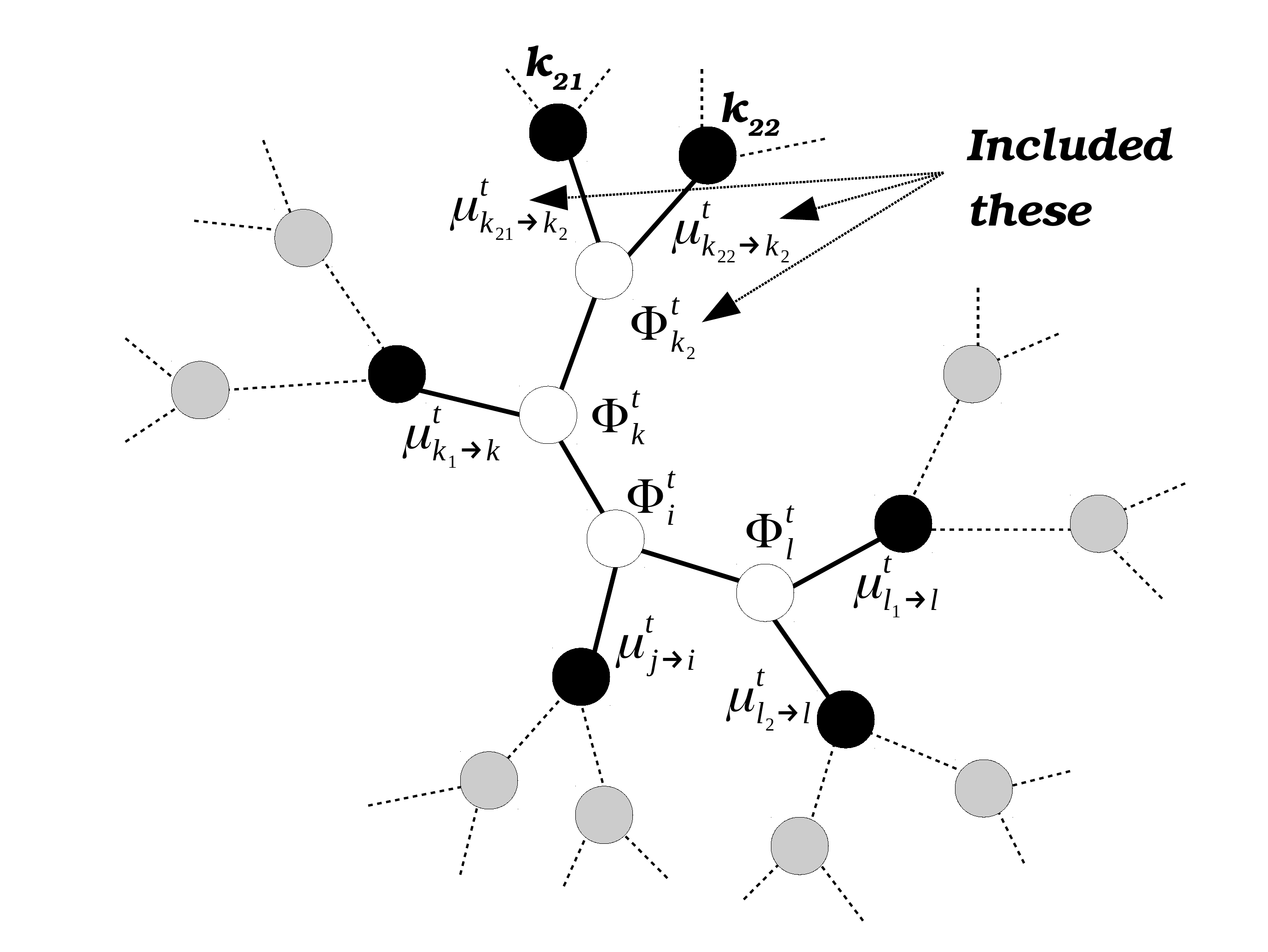}
\caption{This figure illustrates what happens when we apply the relation (\ref{eq:lambda_mu_app_full}) to equation (\ref{eq:product_with_lambda}) in the particular case of the connected set of the right panels of figure (\ref{fig:illustration_connected_set}). We took the outer node $k_2$ (see figure (\ref{fig:illustration_connected_set})) as $o_l$. After including a new local weight $\Phi_{k_2}^{t}$ and two new messages $\mu_{k_{21} \rightarrow k_2}^{t}$ and $\mu_{k_{21} \rightarrow k_2}^{t}$, we end up having the probability density of a bigger connected set that includes all neighbors of $k_2$.}
\label{fig:illustration_lambda_replacing}
\end{figure}

Now, what does the product $ \Big[ \prod_{\substack{k=1 \\ k \neq l}}^{m} \mu_{o_k \rightarrow i_k}^{t} \Big] \Big[\prod_{k = 1}^{n} \Phi_{i_k}^{t} \Big]
\Phi_{o_l}^{t} \Big[\prod_{k\in\partial o_l \setminus i_l} \mu_{k \rightarrow o_l}^{t} \Big]$ in the right hand side of (\ref{eq:bigger_probs_appears}) stands for? As in the example of figure (\ref{fig:illustration_lambda_replacing}), we took (\ref{eq:product_with_lambda}) and removed the factor $\lambda_{o_l \rightarrow i_l}^{t} \; \mu_{o_l \rightarrow i_l}^{t}$. Then, we included a local weight, $\Phi_{o_l}^{t}$, and a group of cavity messages $\mu_{k \rightarrow o_l}^{t}$. Therefore, in (\ref{eq:bigger_probs_appears}) we have all the local weights and cavity messages that appear in the probability density of a bigger connected set, which now includes all the neighbors of $o_l$.

Equation (\ref{eq:bigger_probs_appears}) transforms into:

\begin{eqnarray}
\Big[ \prod_{k=1}^{m} \mu_{o_k \rightarrow i_k}^{t} \Big] \Big[\prod_{k = 1}^{n} \Phi_{i_k}^{t} \Big] \: \lambda_{o_l \rightarrow i_l}^{t} &=& \sum_{X_{\partial o_l \setminus i_l}}^{t} r_{o_l}^{t} \, Q^{t}(X_{\partial o_l \setminus i_l}, \vec{X}_{\mathcal{O}}, \vec{X}_{\mathcal{I}})
\label{eq:bigger_probs}
\end{eqnarray}

Thus, we can rewrite (\ref{eq:I1_gen_tree_P}) and (\ref{eq:I2_gen_tree_P}) to get:

\begin{eqnarray}
I_1 &=& -\Delta t \sum_{\vec{X}_{\mathcal{O}} | \vec{\sigma}_{\mathcal{O}}}^{t} \sum_{\vec{X}_{\mathcal{I}} |\vec{\sigma}_{\mathcal{I}}}^{t} \Big[ \sum_{l = 1}^{n} r_{i_l}^{t} Q^{t}(\vec{X}_{\mathcal{O}}, \vec{X}_{\mathcal{I}}) + \sum_{l = 1}^{m} \sum_{X_{\partial o_l \setminus i_l}}^{t} r_{o_l}^{t} \, Q(X_{\partial o_l \setminus i_l}, \vec{X}_{\mathcal{O}}, \vec{X}_{\mathcal{I}}) \Big] \nonumber \\
I_1 &=& -\Delta t \Big[ \sum_{l=1}^{n} r_{i_l}(\sigma_{i_l}, \sigma_{\partial i_l}) P^{t}(\vec{\sigma}_{\mathcal{O}}, \vec{\sigma}_{\mathcal{I}}) + \sum_{l=1}^{m} \sum_{\sigma_{\partial o_l \setminus i_l}} r_{o_l}(\sigma_{o_l}, \sigma_{\partial o_l}) P^{t}(\sigma_{\partial o_l \setminus i_l}, \vec{\sigma}_{\mathcal{O}}, \vec{\sigma}_{\mathcal{I}}) \Big]
\label{eq:I1_gen_tree_P_subs}
\end{eqnarray}
\begin{eqnarray}
I_2 &=& \Delta t \sum_{l=1}^{n} \sum_{\vec{X}_{\mathcal{O}} | \vec{\sigma}_{\mathcal{O}}}^{t} \sum_{\vec{X}_{\mathcal{I}} | F_{i_l}[\vec{\sigma}_{\mathcal{I}}]}^{t} r_{i_l}^{t} \, Q(\vec{X}_{\mathcal{O}}, \vec{X}_{\mathcal{I}}) + \Delta t \sum_{l=1}^{m} \sum_{\vec{X}_{\mathcal{O}} | F_{o_l}[\vec{\sigma}_{\mathcal{O}}]}^{t} \sum_{\vec{X}_{\mathcal{I}} | \vec{\sigma}_{\mathcal{I}}}^{t} r_{o_l}^{t} \, Q(X_{\partial o_l \setminus i_l}, \vec{X}_{\mathcal{O}}, \vec{X}_{\mathcal{I}}) \nonumber \\
I_2 &=& \Delta t \Big[ \sum_{l=1}^{n} r_{i_l}( - \sigma_{i_l}, \sigma_{\partial i_l}) P^{t}(\vec{\sigma}_{\mathcal{O}}, F_{i_l}[\vec{\sigma}_{\mathcal{I}}]) + \nonumber \\
& & \:\:\:\:\:\:\:\:\:\:\:\:\:\:\:\:\:\:\:\:\:\: + \sum_{l=1}^{m} \sum_{\sigma_{\partial o_l \setminus i_l}} r_{o_l}( -\sigma_{o_l}, \sigma_{\partial o_l}) P^{t}( \sigma_{\partial o_l \setminus i_l}, F_{o_l}[\vec{\sigma}_{\mathcal{O}}], \vec{\sigma}_{\mathcal{I}}) \Big]
\label{eq:I2_gen_tree_P_subs}
\end{eqnarray}
where we have returned to the longer notation for the rates $r_{i}^{t}$ (see equation (\ref{eq:shortenings})).

Putting (\ref{eq:I0_gen_tree_P}), (\ref{eq:I1_gen_tree_P_subs}) and (\ref{eq:I2_gen_tree_P_subs}) together:

\begin{eqnarray}
P^{t + \Delta t}(\vec{\sigma}_{\mathcal{O}}, \vec{\sigma}_{\mathcal{I}}) &=& I_0 + I_1 + I_2 + o(\Delta t) \nonumber \\
P^{t + \Delta t}(\vec{\sigma}_{\mathcal{O}}, \vec{\sigma}_{\mathcal{I}}) &=& P^{t}(\vec{\sigma}_{\mathcal{O}}, \vec{\sigma}_{\mathcal{I}}) \! - \! \Delta t \! \sum_{l=1}^{n} \! \Big[ r_{i_l}(\sigma_{i_l}, \sigma_{\partial i_l}) P^{t}(\vec{\sigma}_{\mathcal{O}}, \vec{\sigma}_{\mathcal{I}}) \! - \! r_{i_l}(\! -\sigma_{i_l}, \sigma_{\partial i_l}) P^{t}(\vec{\sigma}_{\mathcal{O}}, F_{i_l}[\vec{\sigma}_{\mathcal{I}}]) \Big] - \nonumber \\
& & - \Delta t \sum_{l=1}^{m} \sum_{\sigma_{\partial o_l \setminus i_l}} r_{o_l}(\sigma_{o_l}, \sigma_{\partial o_l}) P^{t}(\sigma_{\partial o_l \setminus i_l}, \vec{\sigma}_{\mathcal{O}}, \vec{\sigma}_{\mathcal{I}}) + \nonumber \\
& &+\Delta t \sum_{l=1}^{m} \sum_{\sigma_{\partial o_l \setminus i_l}}r_{i_l}(-\sigma_{i_l}, \sigma_{\partial i_l}) P^{t}(\sigma_{\partial o_l \setminus i_l}, F_{o_l}[\vec{\sigma}_{\mathcal{O}}], \vec{\sigma}_{\mathcal{I}}) + o(\Delta t)
\label{eq:P_tree_expansion}
\end{eqnarray}

And finally:

\begin{eqnarray}
\frac{dP^{t}(\vec{\sigma}_{\mathcal{O}}, \vec{\sigma}_{\mathcal{I}})}{dt} &=& - \sum_{l=1}^{n} \Big[ r_{i_l}(\sigma_{i_l}, \sigma_{\partial i_l}) P^{t}(\vec{\sigma}_{\mathcal{O}}, \vec{\sigma}_{\mathcal{I}}) - r_{i_l}( -\sigma_{i_l}, \sigma_{\partial i_l}) P^{t}(\vec{\sigma}_{\mathcal{O}}, F_{i_l}[\vec{\sigma}_{\mathcal{I}}]) \Big] - \nonumber \\
& & - \sum_{l=1}^{m} \sum_{\sigma_{\partial o_l \setminus i_l}} \Big[ r_{o_l}(\sigma_{o_l}, \sigma_{\partial o_l}) P^{t}( \sigma_{\partial o_l \setminus i_l},\vec{\sigma}_{\mathcal{O}}, \vec{\sigma}_{\mathcal{I}}) - \nonumber \\
& & \:\:\:\:\:\:\:\:\:\:\:\:\:\:\:\:\:\:\:\:\:\:\:\:\: - r_{i_l}(-\sigma_{i_l}, \sigma_{\partial i_l}) P^{t}( \sigma_{\partial o_l \setminus i_l}, F_{o_l}[\vec{\sigma}_{\mathcal{O}}], \vec{\sigma}_{\mathcal{I}}) \Big]
\label{eq:P_derivative_tree_gen}
\end{eqnarray}

Equation (\ref{eq:P_derivative_tree_gen}) is like a master equation for the combined set of variables $\{\vec{\sigma}_{\mathcal{O}}, \vec{\sigma}_{\mathcal{I}}\}$, but it has a very peculiar structure. The first sum represents the contribution to the derivative due to flipping rates of spins sitting at inner nodes, and the second and third lines are the contribution related to outer nodes. Precisely there we can notice that the time derivative of $P^{t}(\vec{\sigma}_{\mathcal{O}}, \vec{\sigma}_{\mathcal{I}})$ depends on probability densities defined over bigger connected sets: $P^{t}( \sigma_{\partial o_l \setminus i_l},\vec{\sigma}_{\mathcal{O}}, \vec{\sigma}_{\mathcal{I}})$. This means that this is not a closed system of equations, and we need to complement it with other relations in order to obtain the time dependence of its variables.

In this subsection
we have presented a general method to obtain the exact differential equation of any local
probability density in a tree-like graph. Although we have worked with connected sets of
the first category (see figure (1)), the equation for connected sets of the second category is just a marginal
of (53).

\subsection{Equations for cavity probability densities}\label{subsec:writing_probs_cav}

To proceed further, it is necessary to find a way to close the set of equations (\ref{eq:P_derivative_tree_gen}). To do this, we need to introduce first a new set of master-like equations. They will represent Cavity Master Equations to be solved separately. With them we may write $P^{t}( \sigma_{\partial o_l \setminus i_l},\vec{\sigma}_{\mathcal{O}}, \vec{\sigma}_{\mathcal{I}})$ in terms of the quantities $P^{t}( \vec{\sigma}_{\mathcal{O}}, \vec{\sigma}_{\mathcal{I}})$ (see below), therefore closing the full system.




We start from the idea that, in a tree-like graph, any local probability density $Q$ of the histories of a connected set can be written as:

\begin{equation}
Q^{t}(\vec{X}_{\mathcal{O}}, \vec{X}_{\mathcal{I}}) = \Big[ \prod_{k=1}^{m} \mu_{o_k \rightarrow (o_k i_k)}^{t} (X_{o_k} | X_{i_k}) \Big] \Big[\prod_{k = 1}^{n} \Phi_{i_k}^{t}(X_{i_k}, X_{\partial i_k}) \Big]
\label{eq:general_Q_tree_2}
\end{equation}
or as a marginal of this kind of expressions. The densities $Q$ are defined over the space of all possible configurations of the $N$-vector $\vec{X}$ of nodes histories.

On the other hand, we may introduce a cavity probability density, $q$, defined over a reduced space where we partially or completely fix some of the histories. Thus, $Q$ and $q$ are fundamentally different.

Here, we will deal only with cavity probability densities which are defined over connected sets. As in previous subsection, we can focus on connected sets of the first kind without any loss in generality. The reduction of the configuration space that gives birth to a cavity treatment occurs in this case by fixing the history of one outer node. We can write one of those $q$ densities in terms of local weights and cavity messages from a probability density $Q$ just by removing the cavity messages $\mu$ corresponding to the fixed outer node. For example:

\begin{equation}
q^{t}(\vec{X}_{\mathcal{O} \setminus o_j}, \vec{X}_{\mathcal{I}} \: \big|\big| \: X_{o_j}) = \Big[ \prod_{\substack{k=1 \\ k \neq j}}^{m} \mu_{o_k \rightarrow (o_k i_k)}^{t} (X_{o_k} | X_{i_k}) \Big] \Big[\prod_{k = 1}^{n} \Phi_{i_k}^{t}(X_{i_k}, X_{\partial i_k}) \Big]
\label{eq:general_q_tree}
\end{equation}

\begin{figure}[htb]
\centering
\includegraphics[keepaspectratio=true,width=0.45\textwidth]{Figures/illustration_inner_outer_with_weights_and_messages.pdf}\includegraphics[keepaspectratio=true,width=0.09\textwidth]{Figures/arrow.pdf}
\includegraphics[keepaspectratio=true,width=0.45\textwidth]{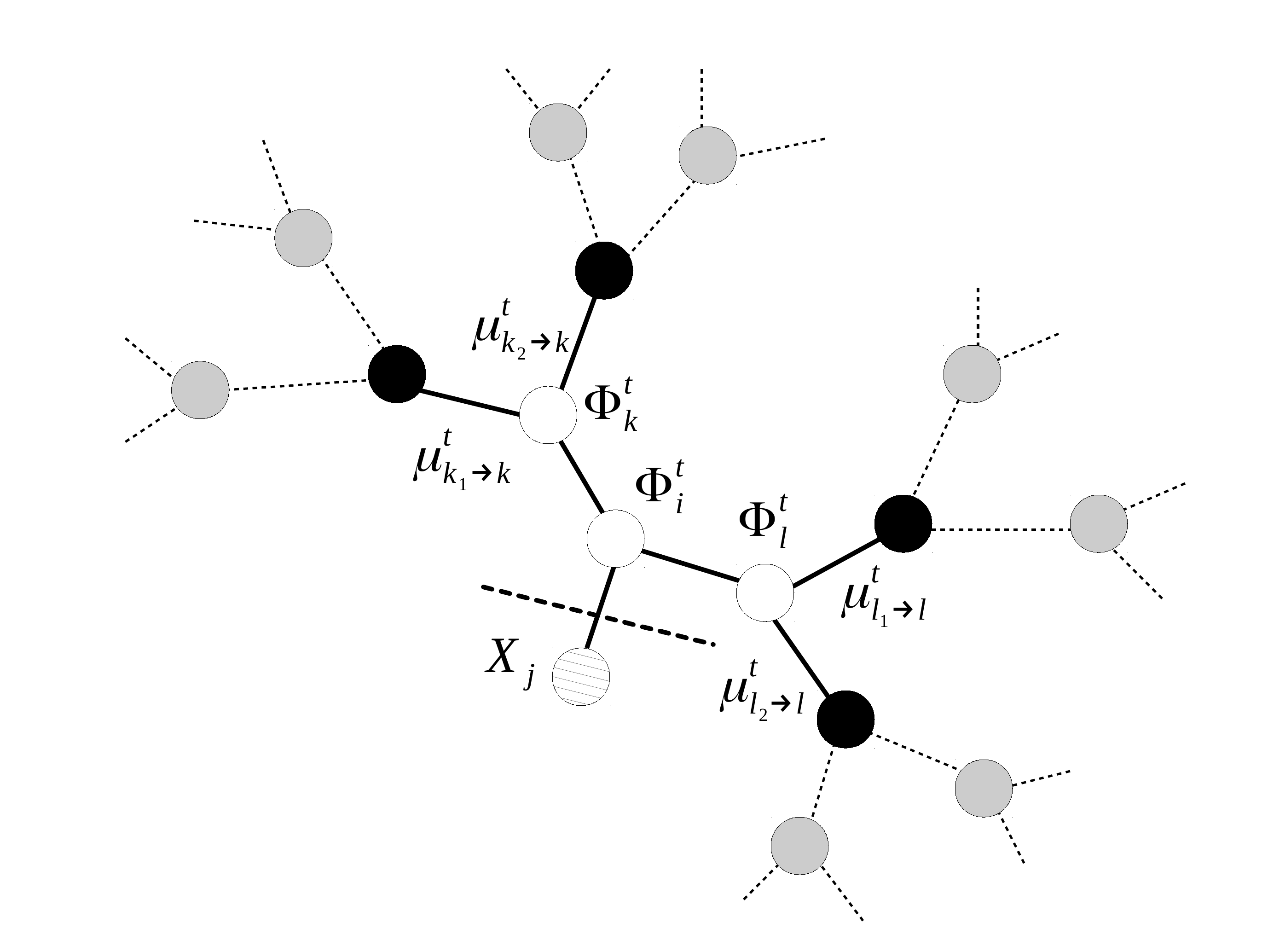}
\caption{This figure illustrates the relation between probability densities $Q$ (left panel) and cavity probability densities $q$ (right panel) defined over connected sets. In this case $q$ is obtained by removing the cavity message $\mu_{j \rightarrow i}^{t}$, and we say that $q$ is defined in the cavity where $X_{j}$ is fixed.}
\label{fig:illustration_cav_prob}
\end{figure}

We say that $q^{t}(\vec{X}_{\mathcal{O} \setminus o_j}, \vec{X}_{\mathcal{I}} \: \big|\big| \: X_{o_j})$ is defined in the cavity where $X_{o_j}$ is fixed, which is represented by the symbol $\big|\big|$. Figure (\ref{fig:illustration_cav_prob}) illustrates this idea by revisiting a particular example shown in previous subsection.

By marginalizing (\ref{eq:general_q_tree}) we get instantaneous cavity probability densities:

\begin{equation}
p^{t}(\vec{\sigma}_{\mathcal{O} \setminus o_j}, \vec{\sigma}_{\mathcal{I}} \: \big|\big| \: X_{o_j}) = \sum_{\vec{X}_{\mathcal{O} \setminus o_j } | \vec{\sigma}_{\mathcal{O} \setminus o_j} }^{t} \sum_{\vec{X}_{\mathcal{I}} | \vec{\sigma}_{\mathcal{I}}}^{t} q^{t}(\vec{X}_{\mathcal{O} \setminus o_j}, \vec{X}_{\mathcal{I}} \: \big|\big| \: X_{o_j})
\label{eq:general_p_tree}
\end{equation}
whose time differentiation is analogous to what we did in previous subsection. We will have $n$ terms like the ones in the first line of (\ref{eq:P_derivative_tree_gen}), and $m-1$ terms like the ones in the second and third lines of the same equation, only that now everything is defined in the cavity where $X_{o_j}$ is fixed.

Thus, proceeding as in \ref{subsec:writing_probs_local} it is easy to verify that:
\begin{eqnarray}
\frac{dp^{t}(\vec{\sigma}_{\mathcal{O} \setminus o_j}, \vec{\sigma}_{\mathcal{I}} \: \big|\big| \: X_{o_j})}{dt} &=& - \sum_{l=1}^{n} \! \Big[ r_{i_l}(\sigma_{i_l}, \sigma_{\partial i_l}) \, p^{t}( \vec{\sigma}_{\mathcal{O} \setminus o_j}, \vec{\sigma}_{\mathcal{I}} \: \big|\big| \: X_{o_j}) - \nonumber \\
& & \:\:\:\:\:\:\:\:\:\:\:\:\:\:\:\:\:\:\: - r_{i_l}(\! -\sigma_{i_l}, \sigma_{\partial i_l}) \, p^{t}( \vec{\sigma}_{\mathcal{O} \setminus o_j}, F_{i_l}[\vec{\sigma}_{\mathcal{I}}] \: \big|\big| \: X_{o_j}) \Big] - \nonumber \\
& & - \sum_{l=1}^{m} \sum_{\sigma_{\partial o_l \setminus i_l}} \Big[ r_{o_l}(\sigma_{o_l}, \sigma_{\partial o_l}) \, p^{t}(\sigma_{\partial o_l \setminus i_l}, \vec{\sigma}_{\mathcal{O} \setminus o_j }, \vec{\sigma}_{\mathcal{I}} \: \big|\big| \: X_{o_j}) - \nonumber \\
& & \:\:\:\:\:\:\:\:\:\:\:\:\:\:\:\:\:\:\:\:\:\: - r_{i_l}(-\sigma_{i_l}, \sigma_{\partial i_l}) \, p^{t}(\sigma_{\partial o_l \setminus i_l}, F_{o_l}[\vec{\sigma}_{\mathcal{O} \setminus o_j}] , \vec{\sigma}_{\mathcal{I}} \: \big|\big| \: X_{o_j}) \Big]
\label{eq:p_derivative_tree_gen}
\end{eqnarray}

As (\ref{eq:P_derivative_tree_gen}), equation (\ref{eq:p_derivative_tree_gen}) is exact in tree-like graphs. However, none of them can be analytically or numerically solved. Equations (\ref{eq:P_derivative_tree_gen}) and (\ref{eq:p_derivative_tree_gen}) contain expressions for the time derivatives of densities $P^{t}(\vec{\sigma}_{\mathcal{O}}, \vec{\sigma}_{\mathcal{I}})$ and $p^{t}(\vec{\sigma}_{\mathcal{O} \setminus o_j}, \vec{\sigma}_{\mathcal{I}} \: \big|\big| \: X_{o_j})$, but these expressions are not closed on the same variables. They both depend on probability densities defined over bigger sets of nodes.

\subsection{Closure and hierarchical approximations}\label{subsec:closure}

Although cavity probability densities are defined over configuration spaces which are fundamentally different than the full configuration space of all histories, we can use them to give a closure to equations like (\ref{eq:P_derivative_tree_gen}).

First off all, let's make a Markovian approximation in equation (\ref{eq:p_derivative_tree_gen}) buy substituting the full history $X_{o_j}$ by its final state $\sigma_{o_j}$:

\begin{eqnarray}
\frac{dp^{t}(\vec{\sigma}_{\mathcal{O} \setminus o_j}, \vec{\sigma}_{\mathcal{I}} \: \big|\big| \: \sigma_{o_j})}{dt} &=& - \sum_{l=1}^{n} \! \Big[ r_{i_l}(\sigma_{i_l}, \sigma_{\partial i_l}) \, p^{t}( \vec{\sigma}_{\mathcal{O} \setminus o_j}, \vec{\sigma}_{\mathcal{I}} \: \big|\big| \: \sigma_{o_j}) - \nonumber \\
& & \:\:\:\:\:\:\:\:\:\:\:\:\:\:\:\:\:\:\: - r_{i_l}(\! -\sigma_{i_l}, \sigma_{\partial i_l}) \, p^{t}( \vec{\sigma}_{\mathcal{O} \setminus o_j}, F_{i_l}[\vec{\sigma}_{\mathcal{I}}] \mid \sigma_{o_j}) \Big] - \nonumber \\
& & - \sum_{l=1}^{m} \sum_{\sigma_{\partial o_l \setminus i_l}} \Big[ r_{o_l}(\sigma_{o_l}, \sigma_{\partial o_l}) \, p^{t}(\sigma_{\partial o_l \setminus i_l}, \vec{\sigma}_{\mathcal{O} \setminus o_j }, \vec{\sigma}_{\mathcal{I}} \: \big|\big| \: \sigma_{o_j}) - \nonumber \\
& & \:\:\:\:\:\:\:\:\:\:\:\:\:\:\:\:\:\:\:\:\:\: - r_{i_l}(-\sigma_{i_l}, \sigma_{\partial i_l}) \, p^{t}(\sigma_{\partial o_l \setminus i_l}, F_{o_l}[\vec{\sigma}_{\mathcal{O} \setminus o_j}] , \vec{\sigma}_{\mathcal{I}} \: \big|\big| \: \sigma_{o_j}) \Big]
\label{eq:p_derivative_tree_gen_markov}
\end{eqnarray}

Equation (\ref{eq:p_derivative_tree_gen_markov}) gives the time derivative of a probability density $p^{t}(\vec{\sigma}_{\mathcal{O} \setminus o_j}, \vec{\sigma}_{\mathcal{I}} \: \big|\big| \: \sigma_{o_j})$ that can be defined over any connected set in the graph. Its first two lines represent the contribution to the derivative due to flipping rates of spins sitting at inner nodes, and the last two lines correspond to the contribution related to outer nodes (except for $\sigma_j$). As in (\ref{eq:P_derivative_tree_gen}), this equation is not closed. Densities like $p^{t}(\sigma_{\partial o_l \setminus i_l}, \vec{\sigma}_{\mathcal{O} \setminus o_j }, \vec{\sigma}_{\mathcal{I}} \: \big|\big| \: \sigma_{o_j})$ are defined over connected sets which include more nodes than $p^{t}(\vec{\sigma}_{\mathcal{O} \setminus o_j}, \vec{\sigma}_{\mathcal{I}} \: \big|\big| \: \sigma_{o_j})$. Again, we need closure relations for (\ref{eq:p_derivative_tree_gen_markov}).

\subsubsection{First-order closures}

It is useful at this time to re-visit the closure presented in \cite{CME-PRE}, considered now as the simplest approximation we can make to equations like (\ref{eq:p_derivative_tree_gen_markov}). Indeed, the CME at \cite{CME-PRE} is written for the cavity densities $p^{t}(\sigma_i \: \big| \big| \: \sigma_j)$, each one involving only two spins. Following (\ref{eq:p_derivative_tree_gen_markov}), the equation for this variables reads:

\begin{eqnarray}
\frac{d p^{t}(\sigma_i \: \big| \big| \: \sigma_j)}{dt} &=& - \!\! \sum_{\sigma_{\partial i \setminus j} } \!\! \Big\lbrace r_{i}(\sigma_i, \sigma_{\partial i}) \: p^{t}(\sigma_{\partial i \setminus j}, \sigma_i \: \big| \big| \: \sigma_j) - r_{i}(-\sigma_i, \sigma_{\partial i}) \: p^{t}(\sigma_{\partial i \setminus j}, -\sigma_i \: \big| \big| \: \sigma_j) \Big\rbrace
\label{eq:CME1_exact}
\end{eqnarray}

To close (\ref{eq:CME1_exact}), we need to write $p^{t}(\sigma_{\partial i \setminus j}, \sigma_i \: \big| \big| \: \sigma_j)$ in terms of $p^{t}(\sigma_i \: \big| \big| \: \sigma_j)$. In order to do so in a transparent way, we make first a small detour.

Consider the simplest case of a stochastic process with three random variables $A$, $B$ and $C$. There, an example of conditional cavity probability density reads:

\begin{equation}
p^{t}(A=a \mid B=b \: \big|\big| \: C=c) = \frac{p^{t}(a , b \: \big|\big| \: c)}{p^{t}( b \: \big|\big| \: c)}
\label{eq:gen_cond_cav}
\end{equation}
where $a$, $b$ and $c$ are values of the variables $A$, $B$ and $C$.

Notice that we used two different symbols to represent conditional and cavity relations: $\mid$ and $\big| \big|$, respectively. We intend to remark that these are conceptually different. In this example the conventional conditional probability density $P^{t}(a \mid b)$ is a measure of the set of configurations with $A(t)=a$ restricted to the sub-space where $B(t)=b$. However, this magnitude is defined in a stochastic process where $A$, $B$ and $C$ can take any of its possible values in the full configuration space. On the other hand, a cavity condition $p^{t}(a \big|\big| \: c)$ is a modification to our stochastic problem whose realizations are now defined on a configuration space where $C(t)=c$ in all cases. If that is not clear, a look back to (\ref{eq:general_q_tree}) shows that a cavity density is a result of removing the probabilistic weight that corresponds to the fixed variable ($C$ in this case). The question about what might be happening with $C$ at time $t$ is simply not asked.

It is in that context that we wrote $p^{t}(a \mid b \: \big|\big| \: c)$. This density is a measure of the set of configurations with $A(t)=a$ restricted to the sub-space where $B(t)=b$ of a stochastic process whose realizations always have $C(t)=c$.

Now we can write the exact relation:

\begin{equation}
p^{t}(\sigma_{\partial i \setminus j}, \sigma_i \: \big| \big| \: \sigma_j) = p^{t}(\sigma_{\partial i \setminus j} \mid \sigma_i \: \big| \big| \: \sigma_j) \, p^{t}(\sigma_i \: \big| \big| \: \sigma_j)
\label{eq:cav_cond_1}
\end{equation}

In \cite{CME-PRE}, this relation is approximated by:

\begin{equation}
p^{t}(\sigma_{\partial i \setminus j}, \sigma_i \: \big| \big| \: \sigma_j) = p^{t}(\sigma_{\partial i \setminus j} \mid \sigma_i \: \big| \big| \: \sigma_j) \, p^{t}(\sigma_i \: \big| \big| \: \sigma_j) \approx \Big[ \prod_{k \in \partial i \setminus j} p^{t}(\sigma_k \: \big| \big| \sigma_i) \Big] \, p^{t}(\sigma_i \: \big| \big| \: \sigma_j)
\label{eq:factorization_p}
\end{equation}

The substitution in (\ref{eq:factorization_p}) can be interpreted as a series of changes made to $p^{t}(\sigma_{\partial i \setminus j} \mid \sigma_i \: \big| \big| \: \sigma_j)$. First, the cavity relation with $\sigma_j$ is dropped to get $p^{t}(\sigma_{\partial i \setminus j} \mid \sigma_i \: \big| \big| \: \sigma_j) \approx p^{t}(\sigma_{\partial i \setminus j} \mid \sigma_i)$. Then, the conditional relation with $\sigma_i$ is substituted by a cavity relation $p^{t}(\sigma_{\partial i \setminus j} \mid \sigma_i) \approx p^{t}(\sigma_{\partial i \setminus j} \: \big| \big| \: \sigma_i)$.
These approximations increase their accuracy when the correlations between $\sigma_{\partial i \setminus j}$, $\sigma_j$ and $\sigma_i$ decrease \cite{CME-PRE}.

Completely dropping the dependence on $\sigma_j$ looks more drastic than replacing $\mid \sigma_i$ by $\: \big| \big| \: \sigma_i$, but we must notice that node $j$ is farther apart from any node in $\partial i \setminus j$ than node $i$, thus, $\sigma_{\partial i \setminus j}$ should be less correlated with $\sigma_j$ than with $\sigma_i$. Finally, it is easy to see that $p^{t}(\sigma_{\partial i \setminus j} \: \big| \big| \: \sigma_i)$ can be exactly factorized to get (\ref{eq:factorization_p}), and we obtain the CME as presented in \cite{CME-PRE}.

\begin{eqnarray}
\frac{d p^{t}(\sigma_i \: \big| \big| \: \sigma_j)}{dt} &=& - \sum_{\sigma_{\partial i \setminus j} } \Big\lbrace r_{i}(\sigma_i, \sigma_{\partial i}) \Big[ \prod_{k \in \partial i \setminus j} p^{t}(\sigma_k \: \big| \big| \: \sigma_i) \Big] p^{t}(\sigma_i \: \big| \big| \: \sigma_j) - \nonumber \\
& & \:\:\:\:\:\:\:\:\:\:\:\:\:\:\:\:\:\:- r_{i}(-\sigma_i, \sigma_{\partial i}) \Big[ \prod_{k \in \partial i \setminus j} p^{t}(\sigma_k \: \big| \big| -\sigma_i) \Big] p^{t}(-\sigma_i \: \big| \big| \: \sigma_j) \Big\rbrace
\label{eq:CME}
\end{eqnarray}

In \cite{CME-PRE, CME-Pspin, CME-PRL, CME-AvAndBP} these kind of equations are integrated together with:

\begin{eqnarray}
\frac{d P^{t}(\sigma_i)}{dt} &=& -
\sum_{\sigma_{\partial i } } \Big\lbrace r_{i}(\sigma_i, \sigma_{\partial i}) \, \Big[ \prod_{k \in \partial i} p^{t}(\sigma_k \: \big| \big| \sigma_i) \Big] \, P^{t}(\sigma_i) - \label{eq:ME} \\
& & \:\:\:\:\:\:\:\:\:\:\:\:\:\:\:\:\:\:- r_{i}(-\sigma_i, \sigma_{\partial i}) \, \Big[ \prod_{k \in \partial i} p^{t}(\sigma_k \: \big| \big| -\sigma_i) \Big] \, P^{t}(-\sigma_i) \Big\rbrace \nonumber
\end{eqnarray}
where a similar factorization is used.

\begin{equation}
P^{t}(\sigma_{\partial i}, \sigma_i) = P^{t}(\sigma_{\partial i} \mid \sigma_i ) P^{t}(\sigma_i) \approx p^{t}(\sigma_{\partial i} \: \big| \big| \: \sigma_i ) \, P^{t}(\sigma_i) = \Big[ \prod_{k \in \partial i \setminus j} p^{t}(\sigma_k \: \big| \big| \sigma_i) \Big] \, P^{t}(\sigma_i)
\label{eq:factorization_P}
\end{equation}

Although these equations had proven their utility in previous works, we can go beyond equation (\ref{eq:CME}) by directly obtaining the time dependence of the probability densities $p^{t}(\sigma_{\partial i \setminus j}, \sigma_i \: \big| \big| \: \sigma_j)$. These are defined over a larger connected sets, and according to the general equation (\ref{eq:p_derivative_tree_gen_markov}) its time derivative can be written as:

\begin{eqnarray}
\frac{d p^{t}(\sigma_{\partial i \setminus j}, \sigma_i \: \big| \big| \: \sigma_j)}{dt} &=& - r_{i}(\sigma_i, \sigma_{\partial i}) \, p^{t}(\sigma_{\partial i \setminus j}, \sigma_i \: \big| \big| \: \sigma_j) + r_{i}(-\sigma_i, \sigma_{\partial i}) \, p^{t}(\sigma_{\partial i \setminus j}, -\sigma_i \: \big| \big| \: \sigma_j) \nonumber \\
& & -
\sum_{l \in \partial i \setminus j} \sum_{\sigma_{\partial l \setminus i} } \Big\lbrace r_{l}(\sigma_l, \sigma_{\partial l}) \, p^{t}(\sigma_{\partial l \setminus i} , \sigma_{\partial i \setminus j}, \sigma_i \: \big| \big| \: \sigma_j) - \label{eq:CME2} \\
& & \:\:\:\:\:\:\:\:\:\:\:\:\:\:\:\:\:\:- r_{l}(-\sigma_l, \sigma_{\partial l}) \, p^{t}(\sigma_{\partial l \setminus i} , F_{l}[\sigma_{\partial i \setminus j}], \sigma_i \: \big| \big| \: \sigma_j) \Big\rbrace \nonumber
\end{eqnarray}

Again, we need to close these equations introducing some relation between the probability densities $p^{t}(\sigma_{\partial l \setminus i} , \sigma_{\partial i \setminus j}, \sigma_i \: \big| \big| \: \sigma_j)$ and $p^{t}(\sigma_{\partial i \setminus j}, \sigma_i \: \big| \big| \: \sigma_j)$. We can always concentrate on conditional cavity densities, which allow us to write:

\begin{equation}
p^{t}(\sigma_{\partial l \setminus i} , \sigma_{\partial i \setminus j}, \sigma_i \: \big| \big| \: \sigma_j) = p^{t}(\sigma_{\partial l \setminus i} \mid \sigma_{\partial i \setminus j}, \sigma_i \: \big| \big| \: \sigma_j) \, p^{t}(\sigma_{\partial i \setminus j}, \sigma_i \: \big| \big| \: \sigma_j)
\label{eq:cav_cond_2}
\end{equation}

As before, we can take $p^{t}(\sigma_{\partial l \setminus i} \mid \sigma_{\partial i \setminus j}, \sigma_i \: \big| \big| \: \sigma_j)$ and drop the dependence on $\sigma_j$. As all the nodes in $\partial i \setminus j$, except for $l$, are at the same distance from any node in $\partial l \setminus i$ than node $j$, the corresponding spins will be in average equally correlated with $\sigma_{\partial l \setminus i}$ than $\sigma_j$. Let's also drop the conventional conditional dependence on these nodes to get the relation $p^{t}(\sigma_{\partial l \setminus i} \mid \sigma_{\partial i \setminus j}, \sigma_i \: \big| \big| \: \sigma_j) \approx p^{t}(\sigma_{\partial l \setminus i} \mid \sigma_l , \sigma_i )$. This procedure is now more accurate than with (\ref{eq:CME}), because we are neglecting correlations associated with nodes which are farther apart in the graph.

Then, we replace the conventional conditional relation with $\sigma_i$ by a cavity relation to obtain:

\begin{equation}
p^{t}(\sigma_{\partial l \setminus i} \mid \sigma_{\partial i \setminus j}, \sigma_i \: \big| \big| \: \sigma_j) \approx p^{t}(\sigma_{\partial l \setminus i} \mid \sigma_l \: \big| \big| \: \sigma_i ) = \frac{p^{t}(\sigma_{\partial l \setminus i}, \sigma_l \: \big| \big| \: \sigma_i )}{\sum_{\sigma_{\partial l \setminus i}} p^{t}(\sigma_{\partial l \setminus i}, \sigma_l \: \big| \big| \: \sigma_i )} \label{eq:closure_CME2}
\end{equation}

With this, we have closed (\ref{eq:CME2}) through approximations for conditional cavity probability densities.

Equations (\ref{eq:CME2}) and (\ref{eq:closure_CME2}) must also be combined with differential equations for the probability densities $P^{t}$. In this case we can use, for example:

\begin{eqnarray}
\frac{d P^{t}(\sigma_{\partial i}, \sigma_i)}{dt} &=& - r_{i}(\sigma_i, \sigma_{\partial i}) \, P^{t}(\sigma_{\partial i}, \sigma_i ) + r_{i}(-\sigma_i, \sigma_{\partial i}) \, P^{t}(\sigma_{\partial i}, -\sigma_i) \nonumber \\
& & -
\sum_{l \in \partial i} \sum_{\sigma_{\partial l \setminus i} } \Big\lbrace r_{l}(\sigma_l, \sigma_{\partial l}) \, p^{t}(\sigma_{\partial l \setminus i} \mid \sigma_l \: \big| \big| \: \sigma_i ) \, P^{t}(\sigma_{\partial i}, \sigma_i) - \label{eq:ME2} \\
& & \:\:\:\:\:\:\:\:\:\:\:\:\:\:\:\:\:\:- r_{l}(-\sigma_l, \sigma_{\partial l}) \, p^{t}(\sigma_{\partial l \setminus i} \mid -\sigma_l \: \big| \big| \: \sigma_i ) \, P^{t}(F_{l}[\sigma_{\partial i}], \sigma_i) \Big\rbrace \nonumber
\end{eqnarray}
which involves essentially the same approximations than (\ref{eq:closure_CME2}):

\begin{equation}
P^{t}(\sigma_{\partial l \setminus i}, \sigma_{\partial i}, \sigma_i) = P^{t}(\sigma_{\partial l \setminus i} \mid \sigma_{\partial i}, \sigma_i) \, P^{t}(\sigma_{\partial i}, \sigma_i) \approx p^{t}(\sigma_{\partial l \setminus i} \mid \sigma_l \: \big| \big| \sigma_i \:) \, P^{t}(\sigma_{\partial i}, \sigma_i)
\label{eq:closure_P_CME2}
\end{equation}

The performance of the set of equations (\ref{eq:CME2})-(\ref{eq:closure_P_CME2}) that we call in the following \textit{CME-2} will outperform the CME as presented in \cite{CME-PRE} (see equations (\ref{eq:CME}) and (\ref{eq:ME})), fact that we illustrate in section \ref{sec:numeric}.

\subsubsection{General closure} \label{subsubsec:gen_clos}

Although we gave an initial idea of how to use the dynamic cavity method to obtain closed systems of differential equations for local probability densities, our goal is to provide a general method that works for any equation like (\ref{eq:p_derivative_tree_gen}). This requires some sense of structure and order that should facilitate the reader's comprehension. We will move among connected sets of the first kind (see figure (\ref{fig:illustration_connected_set})) to establish a hierarchy using the concept of distance within the graph.

We can find a path that connects every pair of nodes in (\ref{eq:cond_cav_1}). Furthermore, for each pair there is always a path with minimum length, and we can define a distance between nodes as the length of that minimal path. The distance $d(k, l)$ between nodes $k$ and $l$, with $k \neq l$, is then a positive integer.

\begin{figure}[htb]
\centering
\includegraphics[keepaspectratio=true,width=0.45\textwidth]{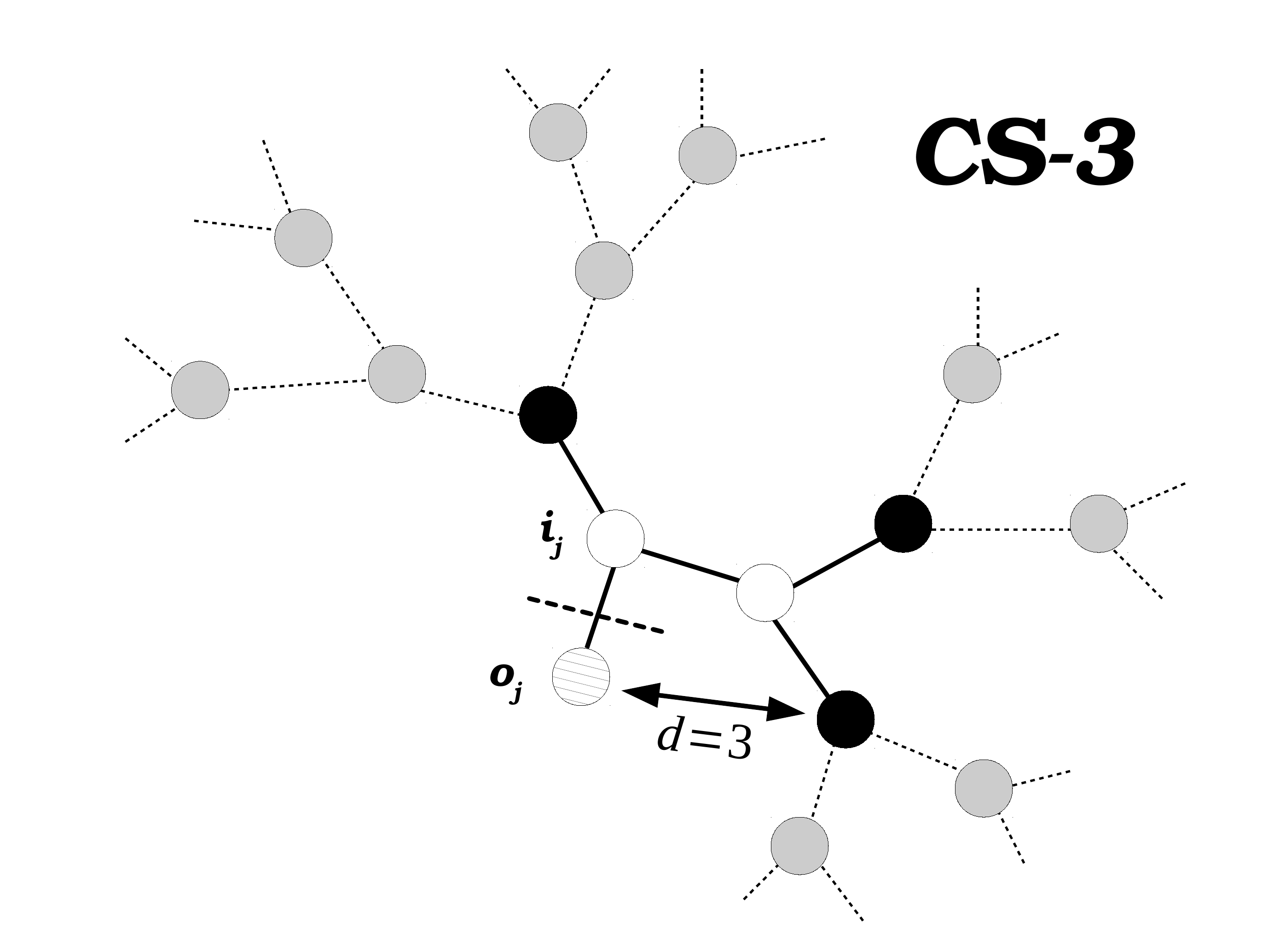}
\includegraphics[keepaspectratio=true,width=0.45\textwidth]{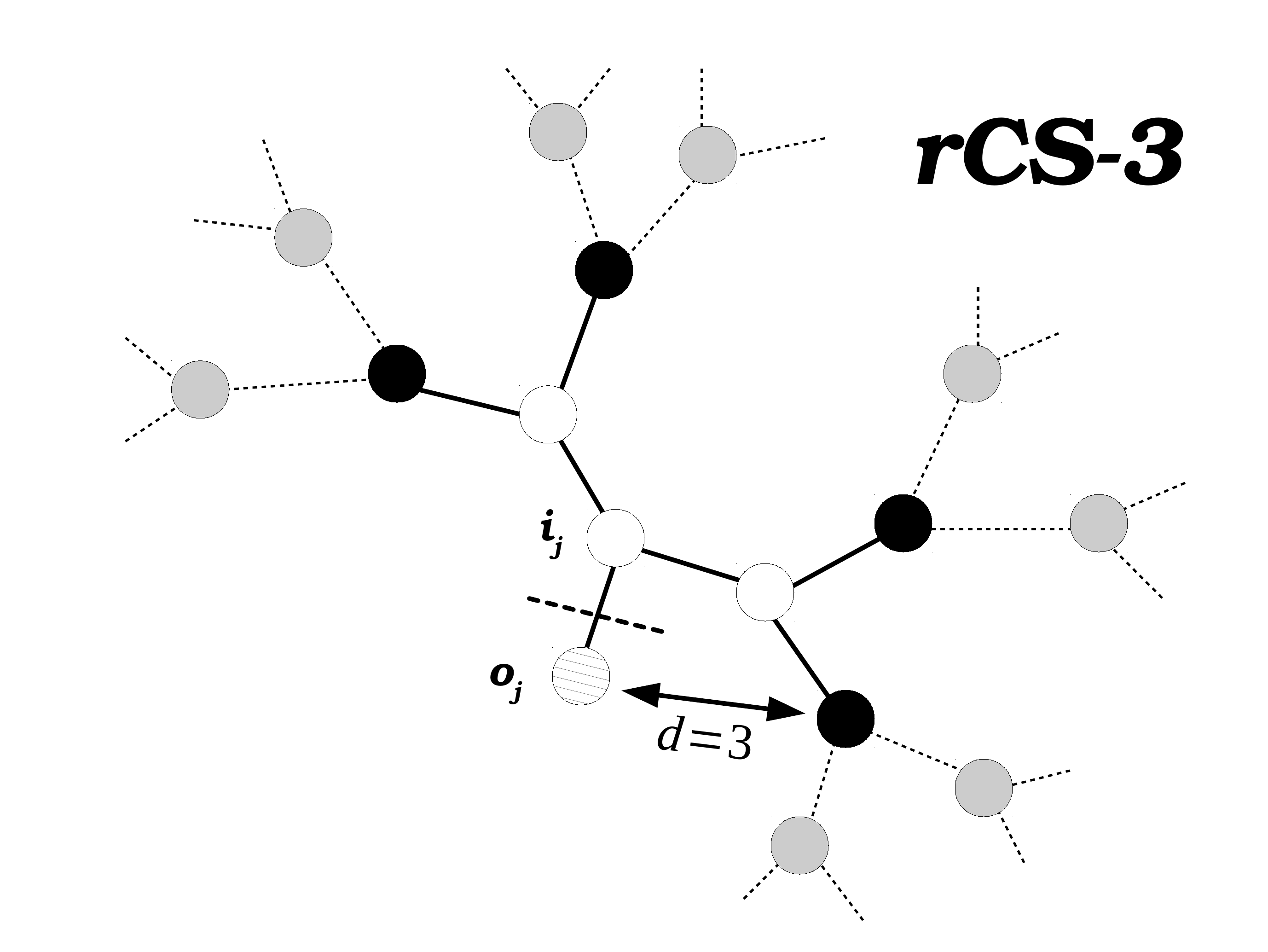}
\caption{A cavity probability density $p^{t}(\vec{\sigma}_{\mathcal{O} \setminus o_j}, \vec{\sigma}_{\mathcal{I}} \: \big|\big| \: \sigma_{o_j})$ is defined over a general connected set. This figure illustrates the classification of these sets according to the distance between the outer nodes and the origin $o_j$. \textbf{(Left panel):} The maximum distance is $d=3$. Thus, the set belongs to the class \textit{CS-3}. In this example all distances are not equal. \textbf{(Right panel):} Every outer node is at a distance $d=3$ from $o_j$. Thus, the set is \textit{rCS-3} and belongs to \textit{CS-3}.}
\label{fig:illustration_CS-N}
\end{figure}

From this, we can classify the connected set $\lbrace \mathcal{O} \setminus o_j,\mathcal{I}, o_j \rbrace$ according to the distances between the node $o_j$ (the origin) and all the outer nodes $\mathcal{O} \setminus o_j$. When the maximum distance $\max_{o_l \in \mathcal{O} \setminus o_j} \lbrace d(o_j, o_l) \rbrace$ is equal to the integer $Z$, we will say that the connected set is in the class \textit{CS-Z}. If all the outer nodes (except $o_j$) are at the same distance from $o_j$, \textit{i.e.}, if $d(o_j, o_l)=d(o_j, o_k)$ for all $o_k, o_l \in \mathcal{O} \setminus o_j$, we will say that the connected set is regular and use the notation \textit{rCS-Z} to refer to that specific set. Of course, every \textit{rCS-Z} belongs to the class \textit{CS-Z}. Figure (\ref{fig:illustration_CS-N}) illustrates this classification.

In this context can use the notation:
\begin{equation}
p^{t}(\vec{\sigma}_{\mathcal{O} \setminus o_j}, \vec{\sigma}_{\mathcal{I}} \: \big|\big| \: \sigma_{o_j}) \equiv p^{t}(\vec{\sigma}_Z, \vec{\sigma}_{Z-1},\ldots, \vec{\sigma}_2, \vec{\sigma}_1 \: \big|\big| \: \sigma_{o_j}) \equiv p^{t}(\vec{\sigma}_Z, \vec{\sigma}_{Z-1},\ldots, \vec{\sigma}_2, \sigma_{i_j} \: \big|\big| \: \sigma_{o_j})
\label{eq:notation_CS-Z}
\end{equation}
where $i_j$ is the inner node connected to $o_j$.

In (\ref{eq:notation_CS-Z}), the vector $\vec{\sigma}_z$ ($1 \leq z \leq Z$) contains the spins $\sigma_k$ such that $d(k, o_j)=z$, with the condition that $k \in \mathcal{O}$ or $k \in \mathcal{I}$. The cavity probability density in (\ref{eq:notation_CS-Z}) is defined over a connected set that belongs to the class \textit{CS-Z}, which we will represent with the symbol $p^{t}(\text{\textit{CS-Z}})$. In order to simplify our language, we will simply say that all $p^{t}(\text{\textit{CS-Z}})$ are \emph{in} \textit{CS-Z}.

Using our new notation (\ref{eq:notation_CS-Z}), the cavity probability densities in the last two lines of (\ref{eq:p_derivative_tree_gen_markov}), which are in \textit{CS-(Z+1)}, can be rewritten as:

\begin{equation}
p^{t}(\sigma_{\partial o_l \setminus i_l}, \vec{\sigma}_{\mathcal{O} \setminus o_j }, \vec{\sigma}_{\mathcal{I}} \: \big|\big| \: \sigma_{o_j}) \equiv p^{t}(\vec{\sigma}_{Z+1}, \vec{\sigma}_Z, \vec{\sigma}_{Z-1},\ldots, \vec{\sigma}_2, \sigma_{i_j} \: \big|\big| \: \sigma_{o_j})
\label{eq:redenote_big_probs}
\end{equation}
where $\vec{\sigma}_{Z+1}$ is exactly the same than $\sigma_{\partial o_l \setminus i_l}$.

Notice that every $p^{t}(\text{\textit{CS-Z}})$ is a marginal of a cavity probability density defined over the corresponding regular set \textit{rCS-Z} (see left panels of (\ref{fig:illustration_CS-N})). In view of all this, what we need to write is a system of equations like:

\begin{eqnarray}
\frac{d p^{t}(\text{\textit{rCS-Z}})}{dt} = F \big[ p^{t}(\text{\textit{rCS-Z}}) \big]
\label{eq:closure_formal}
\end{eqnarray}
which in our case is equivalent to find a relation $p^{t}(\text{\textit{CS-(Z+1)}}) = f\big[ p^{t}(\text{\textit{CS-Z}}) \big]$.

With this classification of connected sets according, essentially, to their size, we are ready to find that $f[\cdot]$ function. Let's write:

\begin{equation}
p^{t}(\vec{\sigma}_{Z+1} \mid \vec{\sigma}_{Z}, \ldots, \sigma_{i_j} \: \big|\big| \: \sigma_{o_j}) = \frac{p^{t}(\vec{\sigma}_{Z+1}, \vec{\sigma}_{Z}, \ldots, \sigma_{i_j} \: \big|\big| \: \sigma_{o_j})}{p^{t}(\vec{\sigma}_{Z}, \ldots, \sigma_{i_j} \: \big|\big| \: \sigma_{o_j})}
\label{eq:cond_cav_1}
\end{equation}

If the distance $Z+1$ is large compared to system's correlation length, we can assume that the fact of having fixed the value of $\sigma_{o_j}$ at time $t$ is irrelevant for the state $\vec{\sigma}_{Z+1}$ in the left hand side of (\ref{eq:cond_cav_1}):

\begin{equation}
p^{t}(\vec{\sigma}_{Z+1} \mid \vec{\sigma}_{Z}, \ldots, \sigma_{i_j} \: \big|\big| \: \sigma_{o_j}) \approx p^{t}(\vec{\sigma}_{Z+1} \mid \vec{\sigma}_{Z}, \ldots, \vec{\sigma}_2, \sigma_{i_j})
\label{eq:drop_sigma_oj}
\end{equation}

Following the same reasoning, we can drop all the spins in $\lbrace \vec{\sigma}_Z, \ldots, \vec{\sigma}_2 \rbrace$ whose distance to the nodes in $\vec{\sigma}_{Z+1}$ is larger than $Z$. We have already done this kind of approximations when writing the equations for $p^{t}(\sigma_i \: \big| \big| \sigma_j)$ and $p^{t}(\sigma_{\partial i \setminus j},\sigma_i \: \big| \big| \sigma_j)$, here we are just generalizing. The remaining spins $\lbrace \vec{\sigma}_{Z+1}, \vec{\sigma}'_Z, \ldots, \vec{\sigma}'_2, \sigma_{i_j} \rbrace$ are defined over a connected set of the class \textit{CS-Z}, with origin in $i_j$.

Furthermore, when the spatial correlation between $\vec{\sigma}_{Z+1}$ and $\sigma_{i_j}$ is low, we can make the approximation \cite{CME-PRE}:

\begin{equation}
p^{t}(\vec{\sigma}_{Z+1} \mid \vec{\sigma}'_{Z}, \ldots, \vec{\sigma}'_{2}, \sigma_{i_j}) \approx p^{t}(\vec{\sigma}_{Z+1} \mid \vec{\sigma}'_{Z}, \ldots, \vec{\sigma}'_{2} \: \big| \big| \: \sigma_{i_j}) = \frac{p^{t}(\vec{\sigma}_{Z+1}, \vec{\sigma}'_{Z}, \ldots, \vec{\sigma}'_{2} \: \big| \big| \: \sigma_{i_j})}{p^{t}(\vec{\sigma}'_{Z}, \ldots, \vec{\sigma}'_{2} \: \big| \big| \: \sigma_{i_j})}
\label{eq:drop_sigma_oj_2}
\end{equation}

Notice that the distance between $i_j$ and the nodes related to $\vec{\sigma}_{Z+1}$ is equal to $Z$, so we are assuming that the correlation decays noticeably at that distance.


From (\ref{eq:drop_sigma_oj}) and (\ref{eq:drop_sigma_oj_2}) we have:

\begin{equation}
p^{t}(\vec{\sigma}_{Z+1} \mid \vec{\sigma}_{Z}, \ldots, \sigma_{i_j} \: \big|\big| \: \sigma_{o_j}) \approx \frac{p^{t}(\vec{\sigma}_{Z+1}, \vec{\sigma}'_{Z}, \ldots, \vec{\sigma}'_{2} \: \big| \big| \: \sigma_{i_j})}{\sum_{\vec{\sigma}_{Z+1}} p^{t}(\vec{\sigma}_{Z+1}, \vec{\sigma}'_{Z}, \ldots, \vec{\sigma}'_{2} \: \big| \big| \: \sigma_{i_j})}
\label{eq:cond_cav_1_app}
\end{equation}
and the probability densities (\ref{eq:redenote_big_probs}) become:

\begin{eqnarray}
p^{t}(\vec{\sigma}_{Z+1}, \vec{\sigma}_{Z}, \ldots, \sigma_{i_j} \: \big|\big| \: \sigma_{o_j}) &=& p^{t}(\vec{\sigma}_{Z+1} \mid \vec{\sigma}_{Z}, \ldots, \sigma_{i_j} \: \big|\big| \: \sigma_{o_j}) \: p^{t}(\vec{\sigma}_{Z}, \ldots, \sigma_{i_j} \: \big|\big| \: \sigma_{o_j}) \nonumber \\
p^{t}(\vec{\sigma}_{Z+1}, \vec{\sigma}_{Z}, \ldots, \sigma_{i_j} \: \big|\big| \: \sigma_{o_j}) &\approx& \frac{p^{t}(\vec{\sigma}_{Z+1}, \vec{\sigma}'_{Z}, \ldots, \vec{\sigma}'_{2} \: \big| \big| \: \sigma_{i_j})}{\sum_{\vec{\sigma}_{Z+1}} p^{t}(\vec{\sigma}_{Z+1}, \vec{\sigma}'_{Z}, \ldots, \vec{\sigma}'_{2} \: \big| \big| \: \sigma_{i_j})} \: p^{t}(\vec{\sigma}_{Z}, \ldots, \sigma_{i_j} \: \big|\big| \: \sigma_{o_j})
\label{eq:app_final}
\end{eqnarray}

Thus, we have managed to write the function $f\big[ p^{t}(\text{\textit{CS-Z}}) \big]$ that gives any cavity probability density in \textit{CS-(Z+1)} in terms of several densities in \textit{CS-Z}. This constitutes a closure for (\ref{eq:p_derivative_tree_gen_markov}), and we can now numerically solve these equations.

It is important to say that after solving for $p^{t}(\text{\textit{CS-Z}})$, we can compute all the cavity probability densities in \textit{CS-1}, \textit{CS-2}, ..., \textit{CS-(Z-1)} as its marginals.

As for (\ref{eq:P_derivative_tree_gen}), we can write $P^{t}(\sigma_{\partial o_l \setminus i_l}, \vec{\sigma}_{\mathcal{O}}, \vec{\sigma}_{\mathcal{I}})$ in terms of $P^{t}( \vec{\sigma}_{\mathcal{O}}, \vec{\sigma}_{\mathcal{I}})$ and a cavity probability density whose value we can obtain from the numerical integration of (\ref{eq:p_derivative_tree_gen_markov}) with the relation (\ref{eq:app_final}). We do:

\begin{eqnarray}
P^{t}(\sigma_{\partial o_l \setminus i_l}, \vec{\sigma}_{\mathcal{O}}, \vec{\sigma}_{\mathcal{I}}) &=& P^{t}(\sigma_{\partial o_l \setminus i_l} \mid \vec{\sigma}_{\mathcal{O}}, \vec{\sigma}_{\mathcal{I}}) \, P^{t}(\vec{\sigma}_{\mathcal{O}}, \vec{\sigma}_{\mathcal{I}}) \nonumber \\
P^{t}(\sigma_{\partial o_l \setminus i_l}, \vec{\sigma}_{\mathcal{O}}, \vec{\sigma}_{\mathcal{I}}) &\approx& p^{t}(\sigma_{\partial o_l \setminus i_l} \mid \vec{\sigma}_{\mathcal{O} \setminus o_j}, \vec{\sigma}_{\mathcal{I}} \: \big| \big| \sigma_{o_j}) \, P^{t}(\vec{\sigma}_{\mathcal{O}}, \vec{\sigma}_{\mathcal{I}})
\label{eq:closure_P}
\end{eqnarray}

Here we changed the conditional probability density $P^{t}(\sigma_{\partial o_l \setminus i_l} \mid \vec{\sigma}_{\mathcal{O}}, \vec{\sigma}_{\mathcal{I}})$ by a conditional cavity density. Nevertheless, we introduced the cavity condition by fixing only the spin $\sigma_{o_j}$. Again, this substitution should be accurate when the distance between $o_j$ and the nodes in $\partial o_l \setminus i_l$ is large enough compared with correlation length. Finally, equations (\ref{eq:p_derivative_tree_gen_markov}), (\ref{eq:cond_cav_1}), (\ref{eq:app_final}) and (\ref{eq:closure_P}) give a closure to the differential equation (\ref{eq:P_derivative_tree_gen}) and we are ready to numerically obtain the time dependence of system's observables.

It is important to say that our closure relations could be used in several ways, but the key point will always be how to choose the distance $Z$ at which we neglect correlations and substitute conventional conditional relations by conditional cavity relations. This distance will be the parameter that defines what we will call a \emph{level of approximation}.

Summarizing, we define a level of approximation through the differential equations for the cavity probability densities (\ref{eq:p_derivative_tree_gen_markov}). If we write them for densities $p^{t}(\text{\textit{CS-Z}})$ we say that we are in the $Z^{\text{th}}$ level of approximation. Within this context, equations (\ref{eq:CME}) and (\ref{eq:ME}) constitute a first level approximation, and we will say they are \emph{in} the level \textit{CME-1}. On the other hand, equations (\ref{eq:CME2})-(\ref{eq:closure_P_CME2}) are \emph{in} the second level, which is \textit{CME-2}.

\section{Numeric results}
\label{sec:numeric}


The information about the nature of the interactions and the local dynamics goes into the spin-flip rates $r_{i}(\sigma_i, \sigma_{\partial i})$, which of course depend on the graph's structure. Here, we will work over two families of diluted random graphs which are locally tree-like: Erdos-Renyi \cite{erdHos1959random} and Random Regular Graphs.

In practice, once interactions are set, we select an initial condition for all the probability densities ($P^{t_0}$ and $p^{t_0}$), and the integration of the equations gives the full $P^{t}$ and $p^{t}$. In subsection \ref{subsec:1and2} we will explore the numerical differences obtained with first and second levels of approximation, \textit{CME-1} and \textit{CME-2}, into two well-known spin models from statistical mechanics, the Ising ferromagnet, and the Viana-Bray spin-glass, both defined over Erdos-Renyi graphs. In \ref{subsec:other_theo} we compare the \textit{CME-2} and \textit{CME-3} with the dynamical independent neighbor approximation presented in \cite{Semerjian_dyn_2004}, for the dynamics of the Ising ferromagnet. Finally, at \ref{subsec:SIS} we also contrast the results of \textit{CME-2}, with the ones of Pair Quenched Mean-Field Approximation for the dynamics of the susceptible-infectious-susceptible (SIS) model for epidemics \cite{cator2012second, mata2013pair}.

\subsection{First and second levels of approximation}\label{subsec:1and2}

One of the traditional forms that theorists choose for the spin flipping probabilities per time unit is Glauber's rule \cite{Glauber63}:

\begin{equation}
r_{i}(\sigma_i, \sigma_{\partial i}) = \frac{1}{2} \big[1 - \sigma_i \tanh \big( \beta \sum_{k \in \partial i} J_{ki} \sigma_k + \beta h_i \big) \big]
\label{eq:Glauber_rule}
\end{equation}

As usual, the spin variables $\sigma_i$ can take the values $\sigma_i= \pm 1$. The interaction between each pair of connected spins $(\sigma_i, \sigma_j)$ is encoded in the couplings parameters $J_{ij} = \pm 1$ and can be either satisfied ($J_{ij} \sigma_i \sigma_j = 1$) or unsatisfied ($J_{ij} \sigma_i \sigma_j = -1$). On the other hand, the $h_i$ are local fields that we will set here to zero. The parameter $\beta$ is the inverse of the temperature $T$.

We can then write:

\begin{equation}
r_{i}(\sigma_i, \sigma_{\partial i}) \equiv r_i(c_i, u_i) = \frac{1}{2} \big[1 - \tanh \big( \beta( 2 c_i - u_i) \big) \big]
\label{eq:Glauber_rule_sat_unsat}
\end{equation}
where $c_i$ is spin's connectivity and we can use the Kronecker's delta to write:

\begin{equation}
u_i \equiv u_i(\sigma_i, \sigma_{\partial i}) = \sum_{k \in \partial i} \delta_{J_{ki}, \sigma_k \sigma_i}
\label{eq:ui}
\end{equation}

When the number $u_i$ of unsatisfied interactions between $i$ and its neighbors is large, there is a high probability that $\sigma_i$ changes to $-\sigma_i$ (see equation (\ref{eq:Glauber_rule_sat_unsat})).

Let's assume we have an Erdos-Renyi graph with $N$ nodes, and in every node a spin variable. The corresponding ferromagnetic Ising model is the result of setting $J_{ij}=1$ for all the connected pairs $(i, j)$. With this definition and with the rules (\ref{eq:Glauber_rule_sat_unsat}) we can make Kinetic Monte Carlo simulations of the system's dynamics.

\begin{figure}[htb]
\centering
\includegraphics[keepaspectratio=true,width=0.45\textwidth]{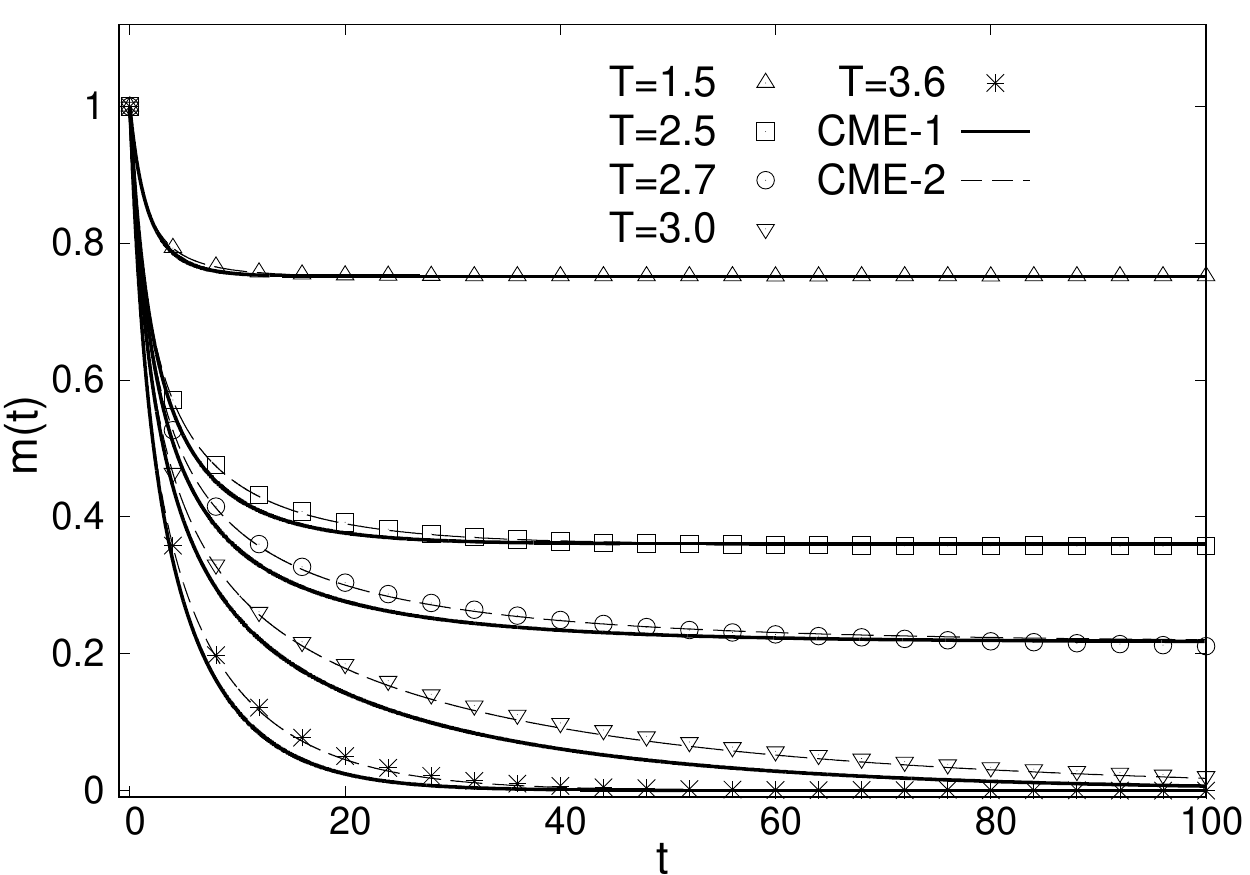}
\includegraphics[keepaspectratio=true,width=0.45\textwidth]{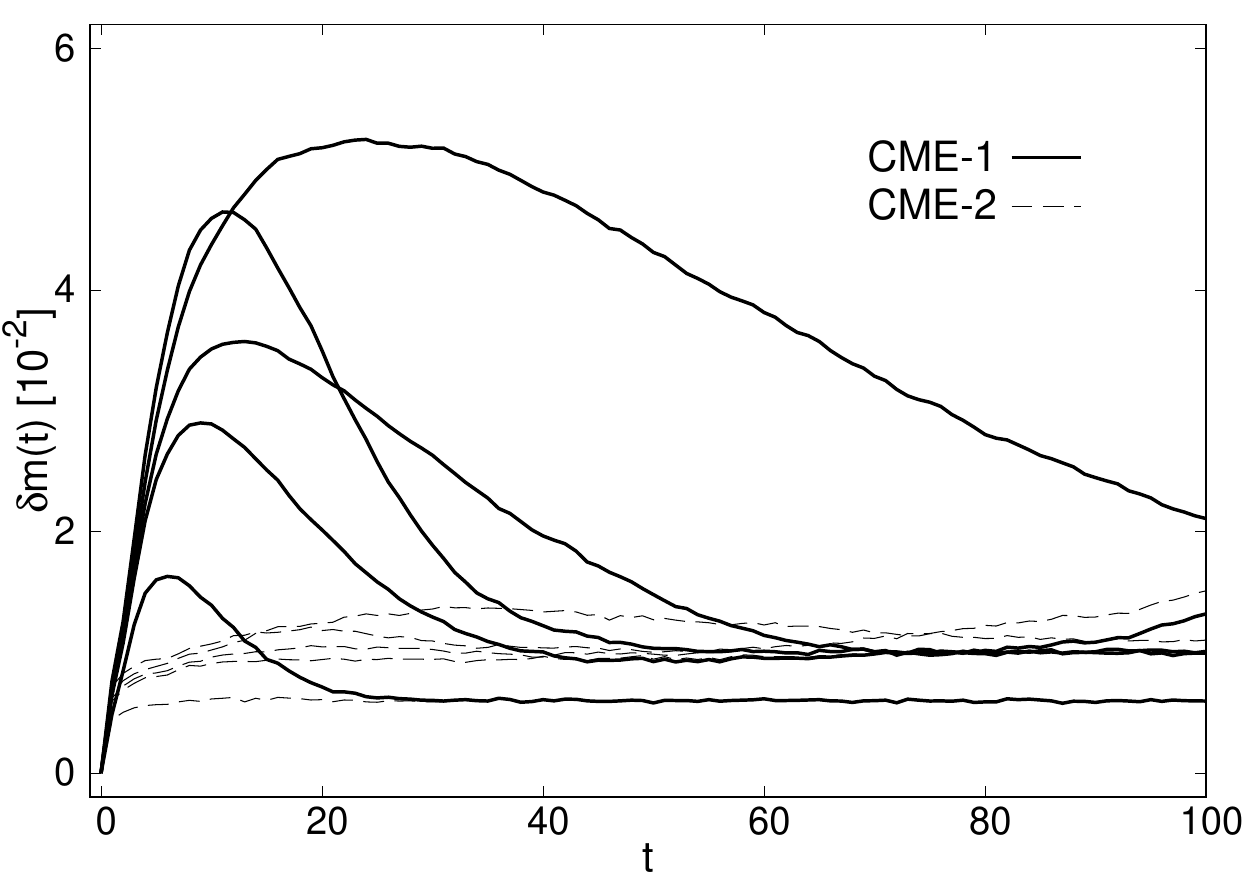}

\includegraphics[keepaspectratio=true,width=0.45\textwidth]{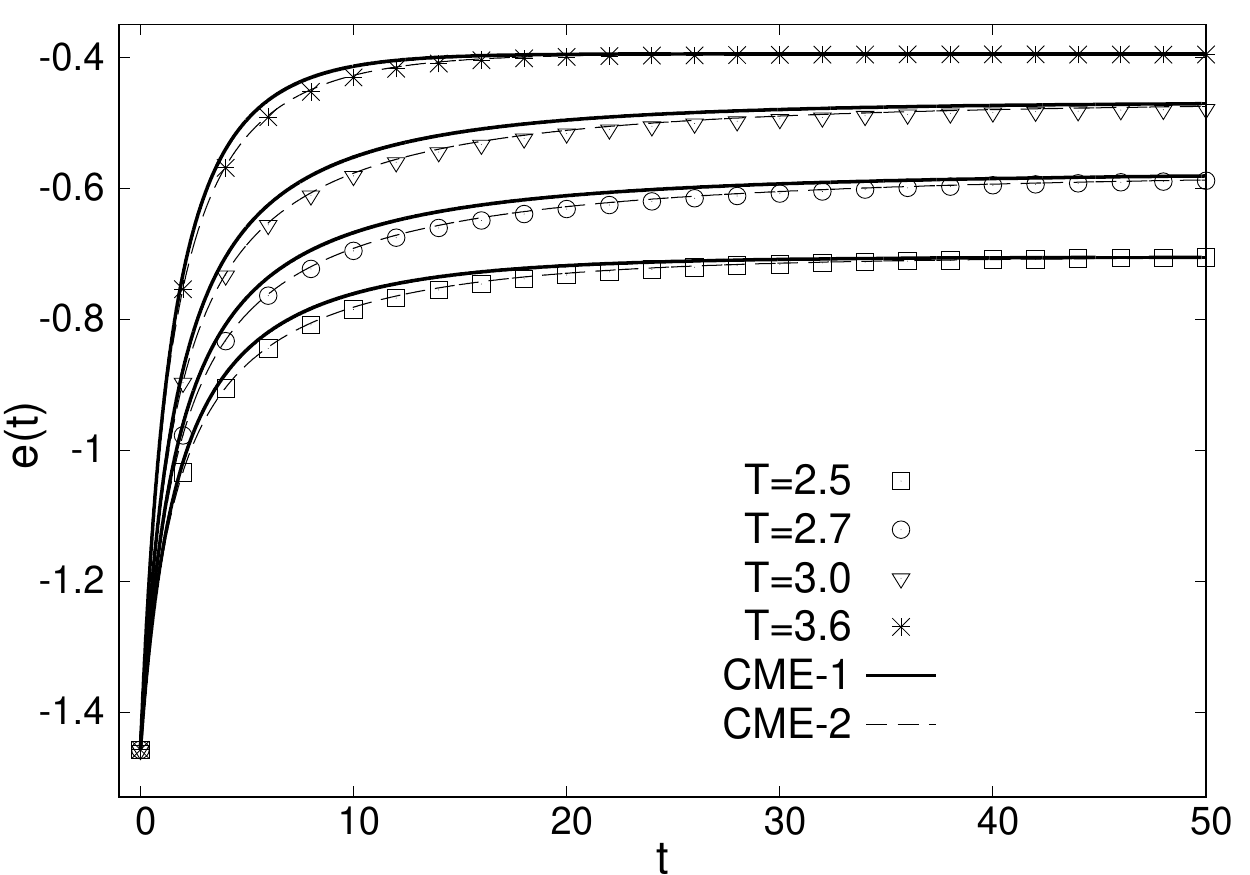}
\includegraphics[keepaspectratio=true,width=0.45\textwidth]{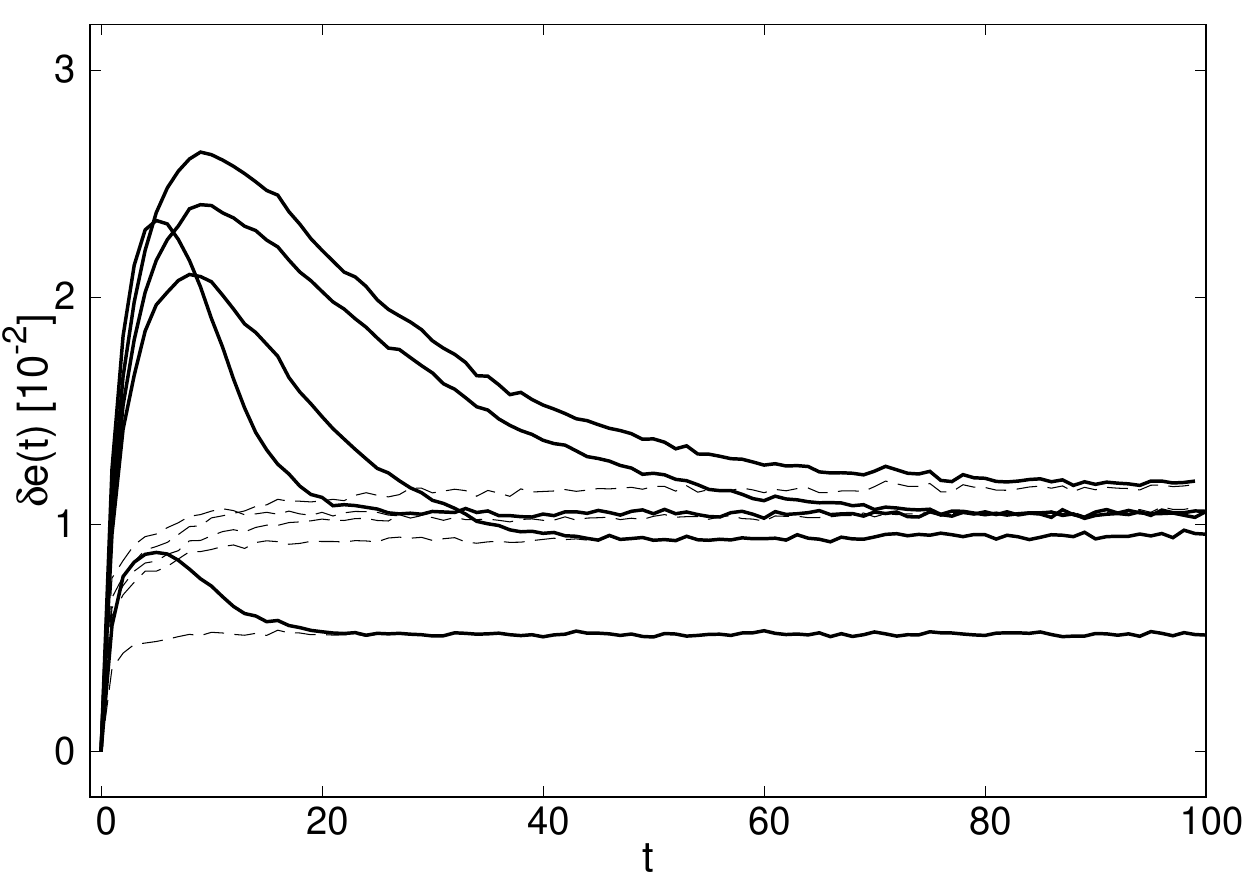}
\caption{Dynamics of the Ising ferromagnet according to Kinetic Monte Carlo simulations (points), and the integration of equations in levels \textit{CME-1} (thick lines) and \textit{CME-2} (dashed lines). In all cases an initially fully-magnetized system evolves in time in contact with a heat bath at a given temperature. Calculations were done for Erdos-Renyi graphs of size $N=5000$ with mean connectivity $c=3$. \textbf{Top-left panel:} Time evolution of system's magnetization (see equation (\ref{eq:mag})). \textbf{Top-right panel:} Time evolution of the error (\ref{eq:local_error_mag}) between local magnetizations predicted by \textit{CME-1, 2} and the results of Kinetic Monte Carlo simulations. \textbf{Bottom-left panel:} Time evolution of system's energy density (see equation (\ref{eq:ener})). \textbf{Bottom-right panel:} Time evolution of the error (\ref{eq:local_error_ener}) written for the expected values of local energy terms.}
\label{fig:CME2_Ferro}
\end{figure}

Figure (\ref{fig:CME2_Ferro}) shows results of these simulations alongside the semi-analytical output of our dynamic cavity method in the first (\textit{CME-1}) and second (\textit{CME-2}) levels of approximation. Top-left panel contains the time evolution of system's magnetization (equation (\ref{eq:mag})), while top-right panel illustrates the corresponding error for the local magnetization (equation (\ref{eq:local_error_mag})) between a theoretical (TH) approach and the simulation (MC) :

\begin{eqnarray}
m(t) &=& \frac{1}{N} \sum_{i=1}^{N} \sum_{\sigma_i} \sigma_i \, P^{t}(\sigma_i) \equiv \frac{1}{N} \sum_{i=1}^{N} m_i(t) \label{eq:mag} \\
\delta m^{\text{TH}}(t) &=& \sqrt{\frac{1}{N} \sum_{i=1}^{N} \big( m_{i}^{\text{TH}}(t) - m_{i}^{\text{MC}} (t) \big)^{2}}
\label{eq:local_error_mag}
\end{eqnarray}

The behavior depicted in top-left panel is typical of ferromagnets. A transition between two regimes occurs at some $T_C$ (when the average connectivity is $c=3$ we have $T_C \approx 2.89$). Steady-state magnetization is zero for high temperatures ($T > T_C$), and at low temperatures ($T < T_C$) we have non-zero magnetization for all times.

Both levels of approximation, \textit{CME-1} and \emph{CME-2}, give a good qualitative and quantitative description of the magnetization obtained from the simulations, especially at the steady-state. The results of \textit{CME-2} are particularly accurate, and this happens in the steady-state and also at the transient. As top-right panel of (\ref{fig:CME2_Ferro}) shows, the equations in the level \textit{CME-2} reproduce local magnetizations $m_{i}^{\text{MC}}(t)$ also quite well, with errors of order $10^{-3}-10^{-2}$.

Bottom-left panel of (\ref{fig:CME2_Ferro}) shows the time evolution of ferromagnet's energy density:

\begin{equation}
e(t) = -\frac{1}{2N} \sum_{i \neq j} \sum_{\sigma_i} \sum_{\sigma_j} P^{t}(\sigma_i, \sigma_j) J_{ij} \sigma_i \sigma_j = \frac{1}{2} \sum_{i \neq j} e_{ij}(t)
\label{eq:ener}
\end{equation}
which is a measure of how many unsatisfied interactions are in the system. Again, our dynamic cavity method gives very accurate results, even for the expected values of local energy terms $e_{ij}(t)$. Bottom-right panel of (\ref{fig:CME2_Ferro}) shows the time evolution of local error:

\begin{equation}
\delta e^{\text{TH}}(t) = \sqrt{\frac{1}{2M} \sum_{i \neq j} \big( e_{ij}^{\text{TH}}(t) - e_{ij}^{\text{MC}} (t) \big)^{2}}
\label{eq:local_error_ener}
\end{equation}
where $M$ is the number of connected pairs $(i,j)$. This error remains of order $10^{-3}-10^{-2}$ for all times.

Another way of choosing the couplings is to draw each one from the bi-modal distribution $d(J_{ij}) = 1/2 \delta(J_{ij}-1) + 1/2 \delta(J_{ij}+1)$. In this model, also defined over Erdos-Renyi random graphs, we have a transition to a spin-glass phase at a finite temperature $T_{SG}$. As correlations play an important role when temperature decreases, we know that the approximations we made in \ref{subsec:closure} can't be as accurate as with the ferromagnet.

\begin{figure}[htb]
\centering
\includegraphics[keepaspectratio=true,width=0.45\textwidth]{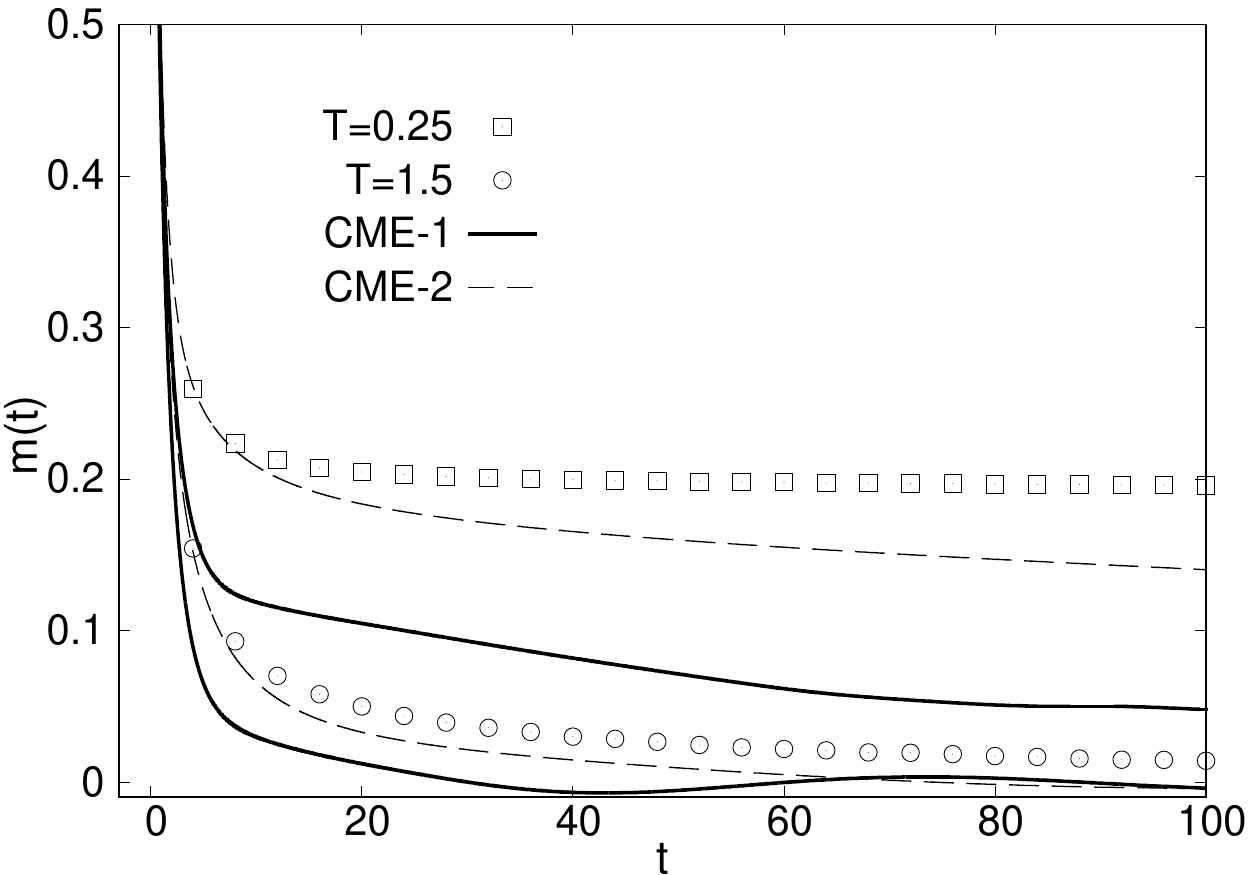}
\includegraphics[keepaspectratio=true,width=0.45\textwidth]{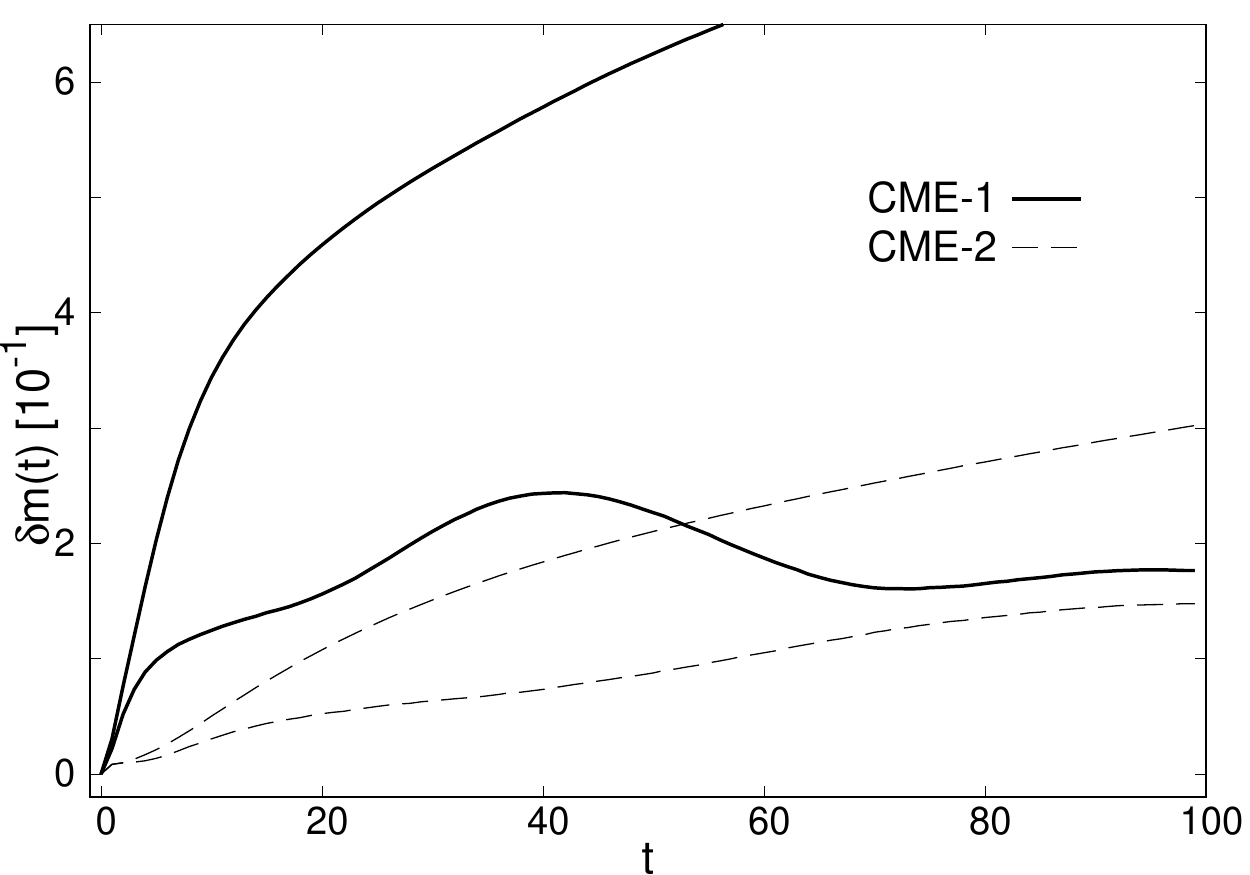}

\includegraphics[keepaspectratio=true,width=0.45\textwidth]{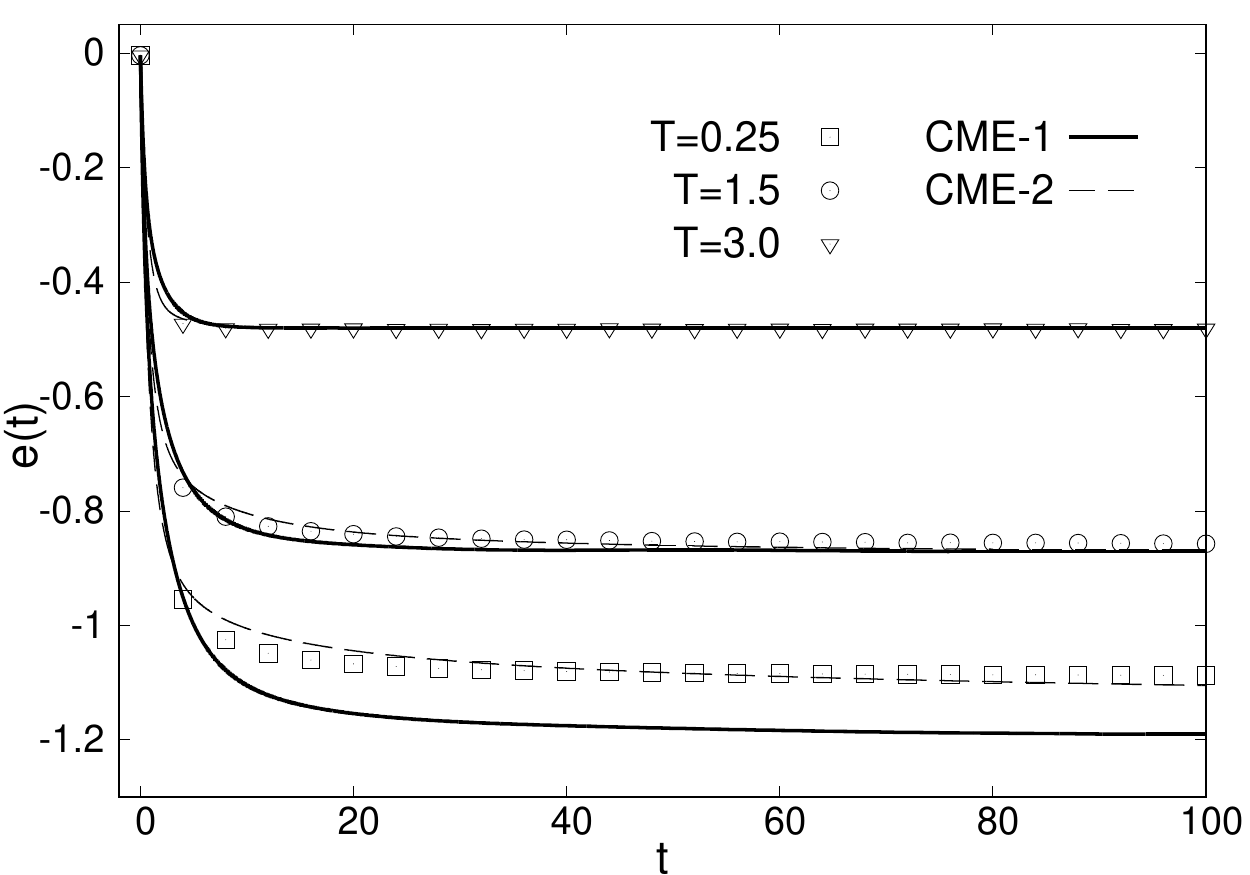}
\includegraphics[keepaspectratio=true,width=0.45\textwidth]{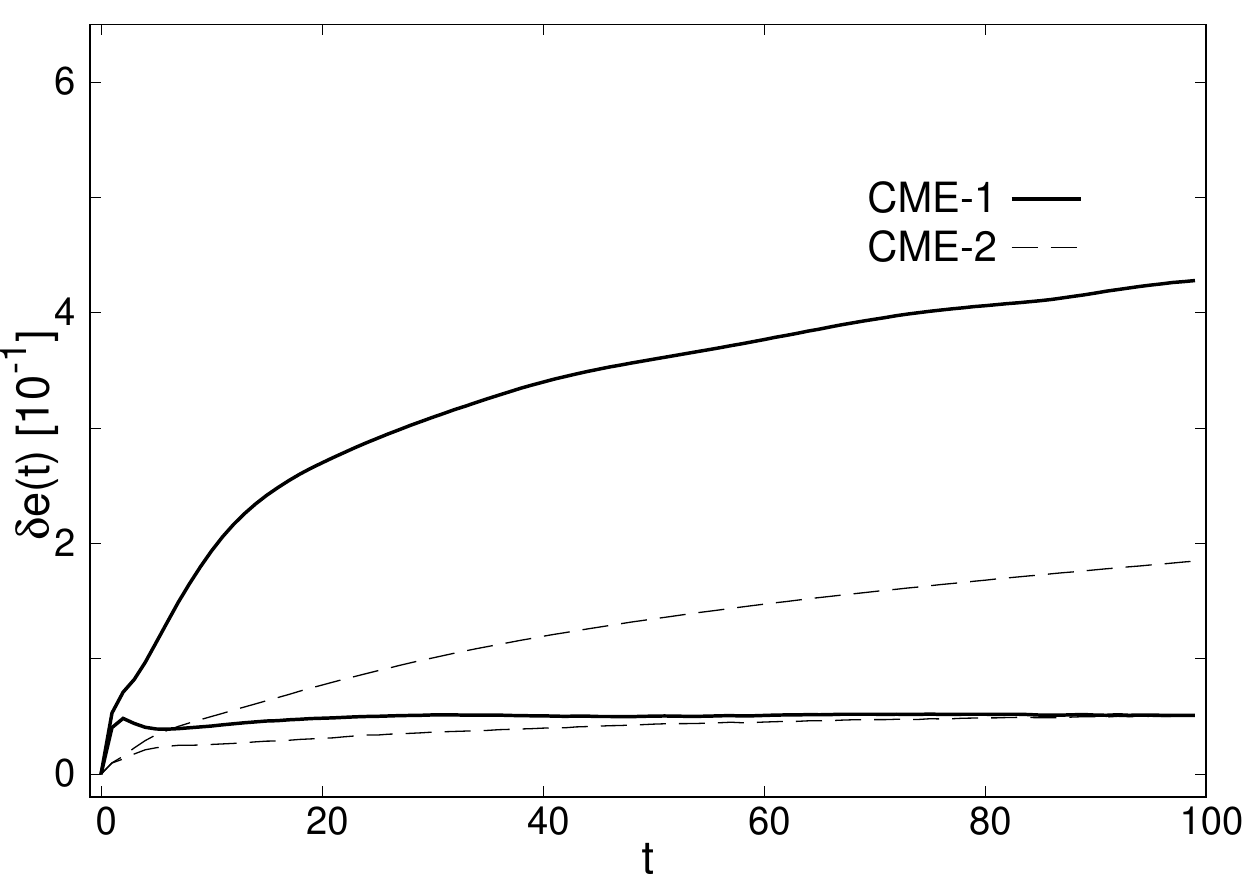}
\caption{Dynamics of the Viana-Bray spin-glass according to Kinetic Monte Carlo simulations (points), and the integration of equations in levels \emph{CME-1} (thick lines) and \emph{CME-2} (dashed lines). In all cases an initially fully-magnetized system evolves in time in contact with a heat bath at a given temperature. Calculations were done for Erdos-Renyi graphs of size $N=1000$ with mean connectivity $c=3$. \textbf{Top-left panel:} Time evolution of system's magnetization (see equation (\ref{eq:mag})). \textbf{Top-right panel:} Time evolution of the error (\ref{eq:local_error_mag}) between local magnetizations predicted by \textit{CME-1, 2} and the results of Kinetic Monte Carlo simulations. \textbf{Bottom-left panel:} Time evolution of system's energy density (see equation (\ref{eq:ener})). \textbf{Bottom-right panel:} Time evolution of the error (\ref{eq:local_error_ener}) written for the expected values of local energy terms.}
\label{fig:CME2_VB}
\end{figure}

Figure (\ref{fig:CME2_VB}) shows results for this model in a graph with $c=3$, where $T_{SG} \approx 1.52$. As expected, below $T_{SG}$ the description is not as successful as for the ferromagnetic case. However, it still holds that \textit{CME-2} performs better than \textit{CME-1}. In the top-right panel, we see that for $0 \leq t \leq 100$ the errors are on a scale of $10^{-1}$, which is one order higher than what we saw in the ferromagnet. Furthermore, even with the approximation \textit{CME-2}, the error seems to monotonically increase with time.

However, our theoretical description of the simulations is noticeably improved just by going up one level in our hierarchical approximations. Not only errors for local magnetizations are appreciably smaller with the level \textit{CME-2} than with \textit{CME-1}. If we look at the bottom panels of the figure we see how this second level of approximation already gives a good description of the average $e(t)$ even at very low temperatures like $T=0.25 \ll T_{SG}$. However, this result might be misleading because the errors for the expected values of local energy terms are still of order $10^{-1}$.

Results in figures (\ref{fig:CME2_Ferro}) and (\ref{fig:CME2_VB}) indicate that the performance of the approach used in \cite{CME-PRE, CME-Pspin, CME-PRL, CME-AvAndBP} (in \textit{CME-1}) is significantly improved just by using the next level of approximation. It is important to say that the integration of equations in the second level, although very accurate, does not take high efforts considering the capabilities of present-day personal computers, not to mention high-performance computers.

\subsection{Comparison with other theoretical approach} \label{subsec:other_theo}
The hierarchical system of approximations introduced for the study of stochastic local search algorithms \cite{SemerjianwalkSAT2003, BarthelwalkSAT2003} and systematized in \cite{Semerjian_dyn_2004} combines theoretical simplicity with numerical accuracy in a variety of traditional models from statistical mechanics. Here, we will compare our theoretical approach with the Dynamical Independent Neighbor Approximation (DINA in what follows), which constitutes the second level of the scheme presented in \cite{Semerjian_dyn_2004}.

In the case of an Ising ferromagnet with Glauber rates (\ref{eq:Glauber_rule_sat_unsat}) defined over random regular graphs, where every node has $K$ neighbors, the DINA works with the exact equation:

\begin{eqnarray}
\frac{d \hat{P}^{t}(\sigma, u)}{dt} &=& -r(K, u) \, \hat{P}^{t}(\sigma, u) + r(K, K-u) \, \hat{P}^{t}(-\sigma, K-u) - \nonumber \\
& & - \sum_{u'=0}^{K-1} (K-u') \, r(K, u') \, \hat{P}^{t}(\sigma, u') \Big[ \hat{P}^{t}(u \mid \sigma, \sigma, u') - \hat{P}^{t}(u-1 \mid \sigma, \sigma, u') \Big] -
\label{eq:DINA_exact} \\
& & - \sum_{u'=0}^{K-1} (u' \! + \! 1) \, r(K, u' + 1) \, \hat{P}^{t}(\sigma, u' + 1) \Big[ \hat{P}^{t}(u - 1 \mid \sigma, -\sigma, u') - \hat{P}^{t}(u \mid \sigma, -\sigma, u') \Big] \nonumber
\end{eqnarray}
together with the closure relations \cite{Semerjian_dyn_2004}:
\begin{eqnarray}
\hat{P}^{t}(\hat{u} \mid \sigma, \sigma, \hat{u}') &\approx& \frac{(K-\hat{u}) \hat{P}^{t}(\sigma, \hat{u})}{\sum_{u'} (K-u') \, \hat{P}^{t}(\sigma, u')} \label{eq:closure_DINA_SAT} \\
P^{t}(\hat{u} \mid \sigma, -\sigma, \hat{u}') &\approx& \frac{(\hat{u} + 1) P^{t}(\sigma, \hat{u}+1)}{\sum_{u'} u' \, \hat{P}^{t}(\sigma, u')}\label{eq:closure_DINA_UNSAT}
\end{eqnarray}

Here, $\hat{P}^{t}(\sigma, u)$ is the probability density of having a node with spin $\sigma$ and $u$ unsatisfied interactions with its neighbors, where the variable $u$ is an integer between zero and $K$. This can be written in terms of the densities $P^{t}(\sigma_i, \sigma_{\partial i})$ as:

\begin{equation}
\hat{P}^{t}(\sigma, u) = \lim_{N \rightarrow \infty} \ll \frac{1}{N} \sum_{i=1}^{N} \sum_{\sigma_i} \sum_{\sigma_{\partial i}} P_{\xi_{K}(N)}^{t}(\sigma_i, \sigma_{\partial i}) \, \delta_{\sigma, \sigma_i} \:\: \delta_{(\sum_{k \in \partial i} \sigma_k), K - 2u} \gg_{\xi_{K}(N)} \label{eq:av_def}
\end{equation}
where $\xi_{K}(N)$ is the ensemble of random regular graphs with connectivity $K$ and size $N$, and $\delta_{x, y}$ is a Kronecker delta evaluated at $(x,y)$. The symbol $\ll \cdot \gg_{\xi_{K}(N)}$ represents an average over this ensemble, with proper probabilistic weights.

We can similarly define the probability density $\hat{P}^{t}(\sigma, \sigma', \hat{u}, \hat{u}')$ of having a connected pair $(\sigma, \sigma')$ with $\hat{u}$ and $\hat{u}'$ unsatisfied interactions with their other neighbors, respectively. In this case, $\hat{u}, \hat{u}'$ are integers between zero and $K-1$. The magnitude $\hat{P}^{t}(\hat{u} \mid \sigma, \sigma', \hat{u}')$ that we have in (\ref{eq:DINA_exact}) is the conditional probability density:

\begin{equation}
\hat{P}^{t}(\hat{u} \mid \sigma, \sigma', \hat{u}') = \frac{\hat{P}^{t}(\sigma, \sigma', \hat{u}, \hat{u}')}{\sum_{u'} \hat{P}^{t}(\sigma, \sigma', \hat{u}, u')}
\label{eq:cond_probs_DINA}
\end{equation}

The closure relation used in DINA can be derived starting from the assumptions of the dynamical replica approach under replica symmetry \cite{Semerjian_dyn_2004}. This explains why DINA gives very accurate results in models where detailed balance holds. In this case, it is more likely that the microscopic probability distribution function is a constant within a subspace with a finite number of order parameters. Indeed, we know that in those models the condition is satisfied at least at equilibrium. 

\begin{figure}[htb]
\centering
\includegraphics[keepaspectratio=true,width=0.55\textwidth]{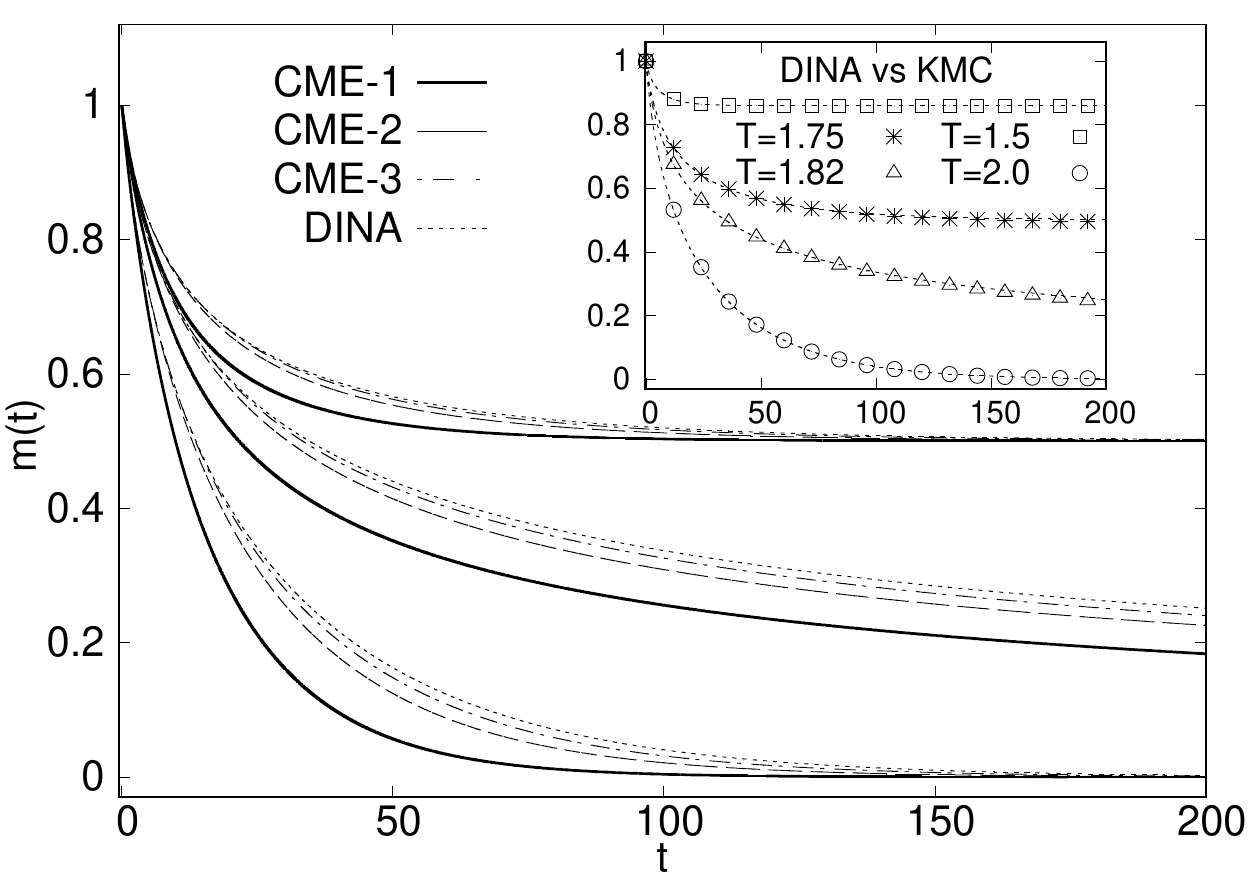}
\caption{Dynamics of the Ising ferromagnet according to Kinetic Monte Carlo simulations (points), the integration of the DINA (thin-dashed lines) and equations in the levels \emph{CME-1} (thick lines), \\ \emph{CME-2} (dashed lines) and \textit{CME-3} (dashed-dotted lines). In all cases an initially fully-magnetized system evolves in time in contact with a heat bath at a given temperature. From top to bottom, we have $T=1.5, 1.75, 1.82, 2.0$. KMC's calculations were done for a random regular graph of size $N=10000$ with connectivity $K=3$. The main panel and the inserted graphic show the time evolution of system's magnetization (see equation (\ref{eq:mag})). The first one compares the different levels of approximation of our dynamic cavity method with the DINA, and the latter compares the DINA with the results of Kinetic Monte Carlo simulations.}
\label{fig:CME_vs_DINA}
\end{figure}

As can be seen, (\ref{eq:DINA_exact}) is an average over the graphs ensemble of the differential equation for $P^{t}(\sigma_i, \sigma_{\partial i})$, which has the form (\ref{eq:P_derivative_tree_gen}). Therefore, closure relations (\ref{eq:closure_DINA_SAT}) and (\ref{eq:closure_DINA_UNSAT}) are equivalent to our closure relation (\ref{eq:closure_P_CME2}) in the sense that both are approximations for the same kind of conditional probabilities. Actually, they both drop the information about the number of unsatisfied relations $\hat{u}'$. An important difference is in the fact that the first ones are written for the average case probability densities, while the latter is written for the single-instance.

However, for our dynamic cavity method all sites become equivalent at random regular graphs with homogeneous initial conditions. In that case, all probability densities defined over connected sets with the same structure follow the same differential equations. We can drop information about the local structure of the graph. As we show at Appendix \ref{app:avRGfc}, we can forget the irrelevant details of, let's say, $P^{t}(\sigma_i, \sigma_{\partial i})$. All the combinations of $\sigma_i$, $\sigma_{\partial i}$ with the same value of $\sigma_i$ and the same number of unsatisfied interactions between this spin and its neighborhood will have the same probability density $P^{t}(\sigma=\sigma_i, u = \sum_{k \in \partial i} \delta_{\sigma_k, -\sigma_i})$.

We are easily able to write average case equations (see Appendix \ref{app:avRGfc}) for several levels of approximation, as the density (\ref{eq:av_def}) can be directly written as $\hat{P}^{t}(\sigma, u) = \binom{K}{u} P(\sigma, u)$. Here we present one of the equations we used to get the time dependence of the densities at (\ref{eq:av_def}):

\begin{eqnarray}
\frac{d \hat{P}^{t}(\sigma, u)}{dt} &=& -r(K, u) \, \hat{P}^{t}(\sigma, u) + r(K, K-u) \, \hat{P}^{t}(-\sigma, K-u) - \nonumber \\
& & -
(K - u) \: \big\langle \, r(K, u') \, \big\rangle_{\sigma, \sigma} \, \hat{P}^{t}(\sigma, u) + (u + 1) \: \big\langle \, r(K, u') \, \big\rangle_{-\sigma, \sigma} \, \hat{P}^{t}(\sigma, u + 1) - \nonumber \\
& & - u \: \big\langle \, r(K, u') \, \big\rangle_{-\sigma, \sigma} \, \hat{P}^{t}(\sigma, u) + (K - u + 1) \: \big\langle \, r(K, u') \, \big\rangle_{\sigma, \sigma} \, \hat{P}^{t}(\sigma, \hat{u} - 1) \label{eq:ME2avRGfc}
\end{eqnarray}
where we have assumed that $1 \leq u \leq K-1$ and used the averages $\big\langle \, r(K, u') \, \big\rangle_{\sigma, \sigma'} = \sum_{u' = 0}^{K-1} r(K, u') \, \hat{p}^{t}(u' \mid \sigma \: \big| \big| \: \sigma' )$. The equations for $u=0,K$ are easily obtained from (\ref{eq:ME2avRGfc}) by suppressing its second or third line, respectively.

Equation (\ref{eq:ME2avRGfc}) is analogous to (\ref{eq:DINA_exact}), not only because they both give the time derivative of the same quantities, but also because they have similar structures. The main difference is that we have made the approximation:

\begin{equation}
\hat{P}^{t}(\sigma, u) \, \hat{P}^{t}(u' \mid \sigma', \sigma, u) \approx \hat{P}^{t}(\sigma, u) \, \hat{p}^{t}(u' \mid \sigma', \sigma) \approx \hat{P}^{t}(\sigma, u) \, \hat{p}^{t}(u' \mid \sigma' \: \big| \big| \: \sigma) \label{eq:cav_closure_av}
\end{equation}

The closure (\ref{eq:cav_closure_av}) works as a substitution drawn from our dynamic cavity method for the conditional probabilities $\hat{P}^{t}(u' \mid \sigma', \sigma, u)$. The densities $\hat{p}^{t}(\hat{u} \mid \sigma \: \big| \big| \: \sigma')$ can be obtained from our cavity treatment (see Appendix \ref{app:avRGfc}). On the other hand, closures (\ref{eq:closure_DINA_SAT}) and (\ref{eq:closure_DINA_UNSAT}) of DINA are the substitutions for $\hat{P}^{t}(u' \mid \sigma', \sigma, u)$ that correspond to the assumptions of DRT.

Main panel of figure (\ref{fig:CME_vs_DINA}) shows a comparison between approximations in the first, second and third levels of our dynamic cavity method, and the DINA. Each level is defined by the approximation we use to obtain the cavity probability densities $\hat{p}^{t}(\hat{u} \mid \sigma \: \big| \big| \: \sigma')$. As the level increases, our method predicts a time dependence of the magnetization which is closer to what is obtained from DINA. This means that our cavity closure on the densities $\hat{p}^{t}(\hat{u} \mid \sigma \: \big| \big| \: \sigma')$ becomes similar to the DRT-like closures (\ref{eq:closure_DINA_SAT}) and (\ref{eq:closure_DINA_UNSAT}), which work very well in this case.


\subsection{Susceptible-infectious-susceptible model for epidemics}\label{subsec:SIS}

The propagation of an epidemic is a relevant issue present in various scenarios. Problems like the dissemination of a disease within a population \cite{pastor2015epidemic} or the spreading of a computer virus or rumors over a network \cite{pastor2001epidemic} are studied by numerous scientists all over the world. We do not have to explain the significance that global COVID-19 pandemics has brought nowadays to the development of theoretical tools for understanding epidemic outbreaks \cite{cuevas2020agent}.

There are a vast variety of models for epidemic processes. Since the seminal work by Kermack and McKendric \cite{kermack1927contribution} about the susceptible-infectious-recovered (SIR) model, we are used to seeing compartment models whose main idea is to divide the population into several groups. Each one of those groups is assumed to be homogeneous, in the sense that all individuals interact with rules that depend on which group they belong to. However, this does not necessarily mean that all of the members of a compartment are equivalent. The dynamics, for example, can be defined over a specific contact network. Here, we will concentrate on the susceptible-infectious-susceptible (SIS) model defined over random regular graphs.

\begin{figure}
\includegraphics[width=0.3\textwidth, angle=0]{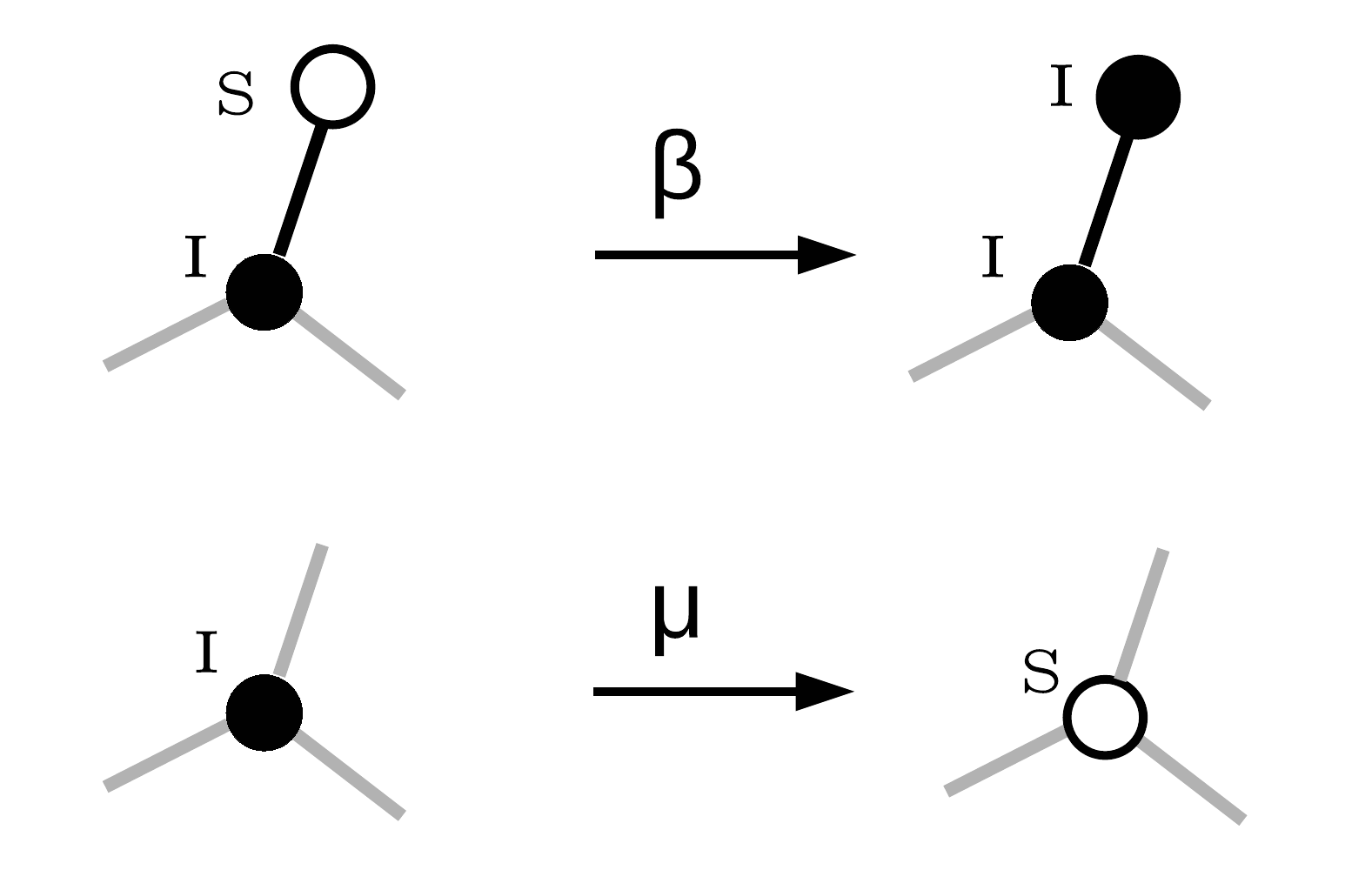} \caption{Allowed transitions in SIS compartment model on networks. \label{fig:SISdiagram}}
\end{figure}

The standard SIS uses two compartments (states) $x_i = 0 \equiv \mbox{susceptible}$, or $x_i = 1 \equiv \mbox{infectious}$ and is the simplest model for recurrent transmissible diseases. The epidemic is thus a continuous-time stochastic process with only two admitted transitions. An infectious node transmits the disease to each one of its neighbors with rate $\beta$, and recovers with rate $\mu$, as represented in Figure \ref{fig:SISdiagram}. The ratio $\lambda=\beta/\mu$, known as spreading rate, is commonly used as the control parameter of the model.

Such stochastic process can be described by a master equation like (\ref{eq:Gen_Mas_Eq}), where the relevant quantities are probability densities over discrete variables: $P^{t}(\vec{x})$. This allows us to apply our dynamic cavity method here, and, as said before, we only have to adapt the rates:

\begin{equation}
r_i(x_i, x_{\partial i}) = \beta \, \delta_{x_i, 0} \sum_{k \in \partial i} x_{k} + \mu \, \delta_{x_i, 1}
\label{eq:rates_SIS}
\end{equation}

The probability per time unit that $x_i$ changes its value, $r_i(x_i, x_{\partial i})$, is equal to $\mu$ when node $i$ is infectious ($x_i=1$), and is equal to $\beta$ times the number of infected neighbors of $i$ when the node is occupied by a susceptible individual ($x_i=0$). This corresponds with our description of the model.

The dynamics of SIS on random networks has motivated abundant literature. One of the most successful theoretical approaches in this scenario is known as quenched mean-field (QMF) theory \cite{vanmieghem2009QMF, castellano2010epid}, which allows calculating epidemics threshold by complementing the master equation with some suitable factorization. The most accurate version of QMF is Pair Quenched Mean-Field (PQMF), or pair-based mean-field. This approximation considers pair correlations and has been intensively investigated lately \cite{silva19PQMF, silva20PQMF} with very good results.

\begin{figure}[htb]
\centering
\includegraphics[keepaspectratio=true,width=0.45\textwidth]{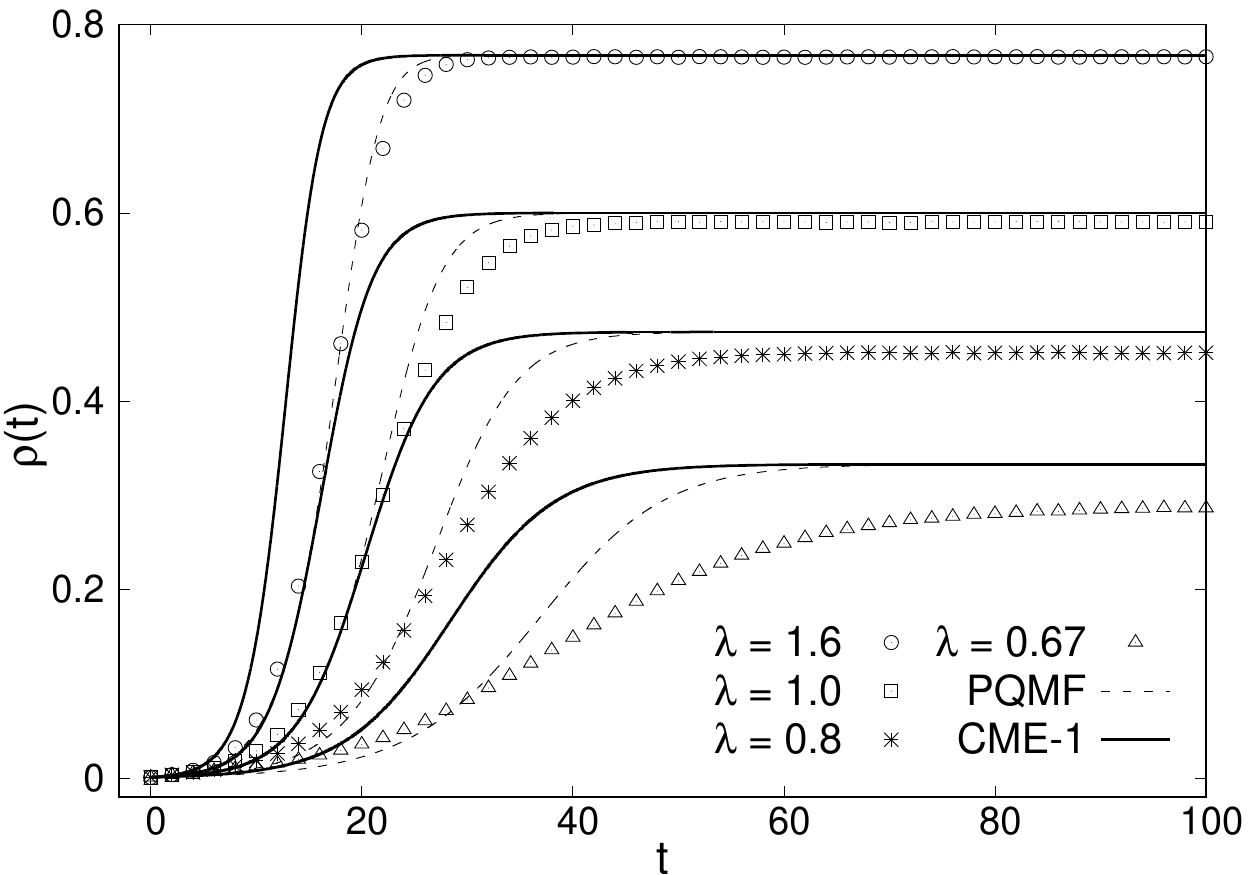}
\includegraphics[keepaspectratio=true,width=0.45\textwidth]{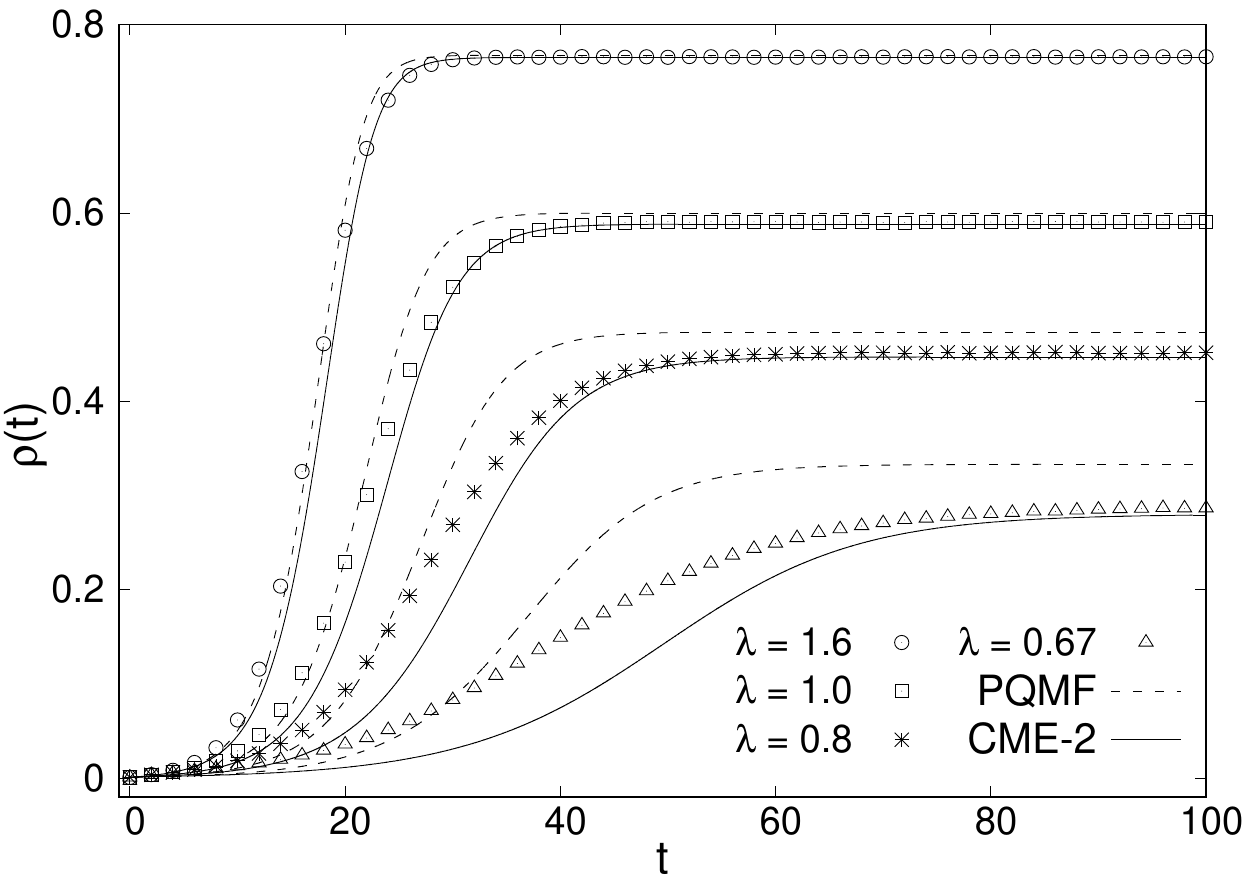}

\hspace{7pt}
\includegraphics[keepaspectratio=true,width=0.43\textwidth]{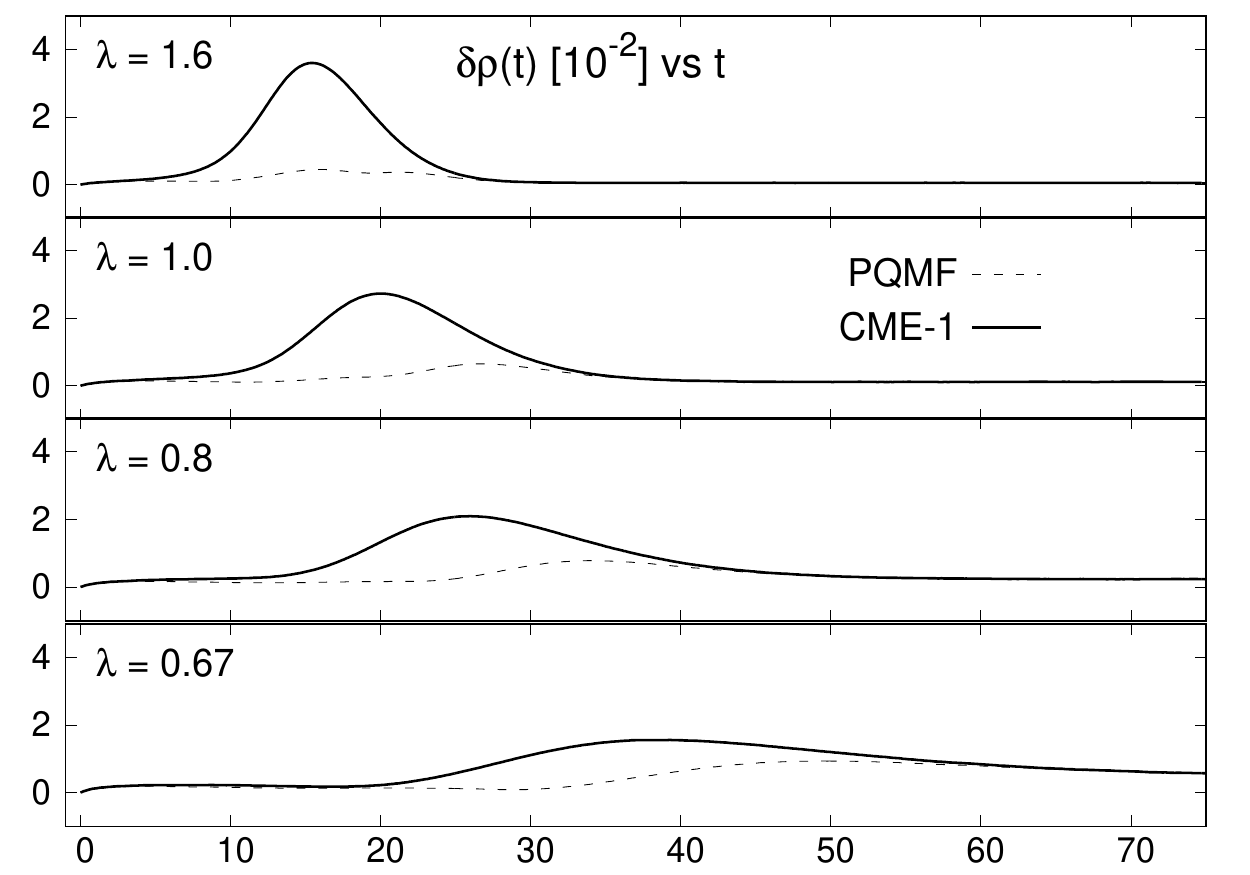}
\includegraphics[keepaspectratio=true,width=0.44\textwidth]{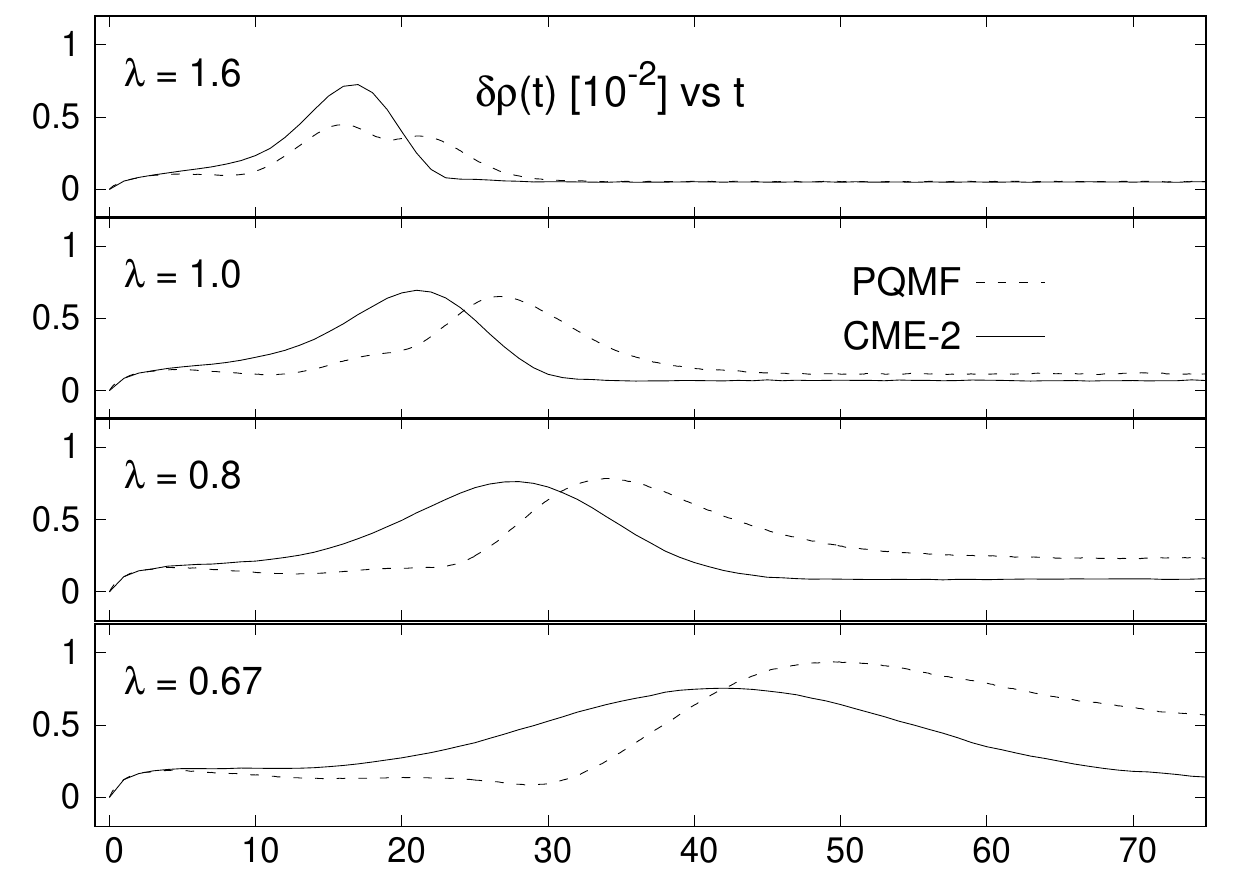}
\caption{Dynamics of the SIS model for epidemics according to Kinetic Monte Carlo simulations (points), the integration of PQMF (dashed lines) and equations in the first (\textit{CME-1}, thick lines) and second (\textit{CME-2}, thin lines) levels of our dynamic cavity method. Calculations were done for $\beta=0.4$ and $\mu=0.25, 0.4, 0.5, 0.6$ (that in the figure correspond to $\lambda=1.6, 1.0, 0.8, 0.67$). In all cases the infection started in one node of a random regular graph of size $N=1000$ with connectivity $K=3$. \textbf{Top panels panel:} Time evolution of system's average $\rho(t) = \frac{1}{N}\sum_{i} \rho_{i}(t)$. \textbf{Bottom panels:} Time evolution of the local error (computed analogously to \ref{eq:local_error_mag}) between these semi-analytical methods and Kinetic Monte Carlo simulations.}
\label{fig:CME12_vs_PQMF}
\end{figure}

The PQMF equations read:

\begin{eqnarray}
\frac{d \rho_i}{d t} &=& - \mu \rho_i + \beta \sum_{j \in \partial i} (\rho_j - \psi_{ij}) \label{eq:site_PQMF}
\end{eqnarray}
and
\begin{eqnarray}
\frac{d \psi_{ij}}{d t} &=& -2 (\mu + \beta) \psi_{ij} + \beta(\rho_i + \rho_j) + \nonumber \\
& & + \beta \frac{\rho_j - \psi_{ij}}{1 - \rho_i} \sum_{k \in \partial i \setminus j} (\rho_k - \psi_{ik}) + \beta \frac{\rho_i - \psi_{ij}}{1 - \rho_j} \sum_{k \in \partial j \setminus i} (\rho_k - \psi_{jk})
\label{eq:pair_PQMF}
\end{eqnarray}
where $\rho_i(t) = \sum_{x_i} x_i \, P^{t}(x_i)$ and $\psi_{ij}(t) = \sum_{x_i} \sum_{x_j} x_i \, x_j \, P^{t}(x_i, x_j)$

An extensive comparison with equations in the first level of approximation of our dynamic cavity method (see (\ref{eq:CME}) and (\ref{eq:ME})) has been already carried out at \cite{CME-Epid-1}. As it is shown there, PQMF gives more accurate predictions for the system's observables. Left panels of figure (\ref{fig:CME12_vs_PQMF}) illustrate this for the particular case of an epidemic outbreak that begins with a single infectious individual in a random regular graph of size $N=1000$ and connectivity $K=3$. Although the steady-state seems to be similarly predicted, PQMF performs indisputably better at the transient regime than the equations in \textit{CME-1}. The corresponding maximum local errors, depicted in the bottom-left panel, are approximately four times bigger for the latter than for PQMF.

Right panels of the figure (\ref{fig:CME12_vs_PQMF}) compares Kinetic Monte Carlo simulations with PQMF's results and the numeric integration of our dynamic cavity equations at the second level of approximation. Equations (\ref{eq:CME2})-(\ref{eq:closure_P_CME2}) give a very good description of the simulated epidemic outbreak. The bottom-right panel shows that its maximum local errors are of the same order as PQMF's, and for $\lambda \leq 1.0$ is clear that our cavity method gives even preciser predictions for the steady-state probability densities. This is also observable in the top-right panel of the figure, which illustrates the time dependence of the average of $\rho_i$ over all sites. There we can also see that PQMF does not capture the transient regime appreciably better than our new approach.

Thus, the second level of our dynamic cavity method outperforms PQMF at least in describing the propagation of epidemics through a random regular graph. Nevertheless, we expect the same to happen in other families of random networks, like Erdos-Renyi graphs. As PQMF is currently the state-of-the-art among mean-field approximations for the SIS dynamics \cite{silva19PQMF, silva20PQMF}, this is a very relevant result.

\section{Conclusions} \label{sec:disc}

We derived, using the Theory of Random Point Processes, a hierarchical scheme of Cavity Master Equations directly applicable to the continuous-time dynamics of systems with discrete variables. We described carefully the approximations made at each level of the scheme and explained their significance, pointing to the system's correlation length as the parameter defining the accuracy of the approximation. Our scheme clarifies some of the contents and approximations made on the recent literature about what it has been called Cavity Master Equation \cite{CME-PRE, CME-Pspin, CME-PRL, CME-AvAndBP}.

We performed numerical tests at different levels of the approximation in spin and epidemic models. For spin models, the accuracy of the technique is comparable with the ones obtained with well-known successful methods like Dynamic Independent Neighbor Approximation and Dynamical Replica Theory, and improves as we increase the level of approximation. Studying the susceptible-infectious-susceptible model for epidemics we show that our equations give better predictions for the stationary state of the epidemics than the widely used Pair Quenched Mean-Field Approximation, and, depending on the parameters, can also perform better in the transient regime.

\section*{Acknowledgements}

This work was supported by the project \emph{Mathematical Modelling of Epidemics, from PNCB CITMA, Republic of Cuba}




\bibliographystyle{unsrt}
\bibliography{ref_cvme}

\appendix

\section{Connecting cavity messages with instantaneous magnitudes}\label{app:relation_mu_lambda}

For completeness, let's reproduce the derivation of equation (\ref{eq:lambda_mu_app_full}) that appears in \cite{CME-PRE}.

If we marginalize (\ref{eqn:marginal_prob_dist_mess}) on $X_{\partial i \setminus \{i,j\}}$ and combine the result with (\ref{eqn:pair_prob_dist_mess}) we get a cavity message passing equation:

\begin{equation}
\mu_{i \rightarrow (ij)}^{t}(X_i \mid X_j) = \sum_{\{X_k\},k\in\partial i\setminus j} \Phi_i^{t} (X_i|X_{\partial i}) \prod_{k\in\partial i\setminus j} \mu_{k \rightarrow (ki)}^{t}(X_k \mid X_i)
\label{eq:update_bethe}
\end{equation}

Where $X_i(t)$ is the history of spin $i$ to time $t$. To simplify the notation we will sometimes write $\mu_{i \rightarrow (ij)}^{t}$ for the cavity conditional message.

We have learned that probability densities in the random point processes formulation can be written as (see equation (\ref{eq:mu_param})):

\begin{equation}
\mu_{i \rightarrow (ij)}^{t}(X_i \mid X_j) = \prod_{l_i=1}^{s_i} \lambda_{i \rightarrow (ij)}(X_i, X_j, t_{l_i}) \exp\{ - \int_{t_0}^{t} \lambda_{i \rightarrow (ij)}(X_i, X_j, \tau) d\tau \}
\label{eq:cavity_mess_param}
\end{equation}

Changing indexes accordingly, we can use the same parametrization for the other cavity messages in the update equation.

The interaction term $\Phi_i^{t}(X_i|X_{\partial i})$ can be interpreted as the probability density of $X_i$ conditioned on the histories of spins
in $\partial i$:

\begin{equation}
\Phi_i^{t}(X_i|X_{\partial i}) = \prod_{l_i=1}^{s_i} r_{i}(\sigma_i(t_{l_i}),\sigma_{\partial i}(t_{l_i}))
\exp\{ - \int_{t_0}^{t} r_{i}(\sigma_i(\tau),\sigma_{\partial i}(\tau)) d\tau \}
\end{equation}

Here, $r_i$ is the real jumping rate of $i$. Under a Markov assumption this is an instantaneous quantity, meaning that at time $\tau$ it depends only on the values of $\sigma_{i}(\tau) , \sigma_{\partial i}(\tau)$.

The trace on the right hand side of (\ref{eq:update_bethe}) can be written in more detail. Let $F$ be the argument of the sum:

\begin{equation}
\sum_{\{X_k\},k\in\partial i\setminus j}^{[t_0,t]} F(X_i,X_{\partial i},t) = \sum_{\{s_k\},k\in\partial i\setminus j}
\left[ \prod_{k=1}^{d} \int_{t_0}^{t}dt_1^{k}\int_{t_1^{k}}^{t}dt_2^{k}\ldots \int_{t_{s_k-1}^{k}}^{t}dt_{s_k}^{k}\right]
F(X_i,X_{\partial i},t)
\end{equation}
In all equations, $s_k$ will be the number of jumps of the history of spin $k$.

If we write (\ref{eq:update_bethe}) for every pair $(i,j)$ in the network we get a system of coupled equations. Having these functions, we could
describe all the dynamics of the system. However (\ref{eq:update_bethe}) is a very involved expression and we need to transform it to a more tractable one. With that objective, we will differentiate both sides of (\ref{eq:update_bethe}).

Differentiation in this context should be handled carefully since increasing $t$ means we are changing the sample space itself. Therefore, it is safer to use the definition of differentiation for both sides of equation (\ref{eq:update_bethe}). We will compute the limit:

\begin{equation}
\lim_{\Delta t\rightarrow 0} \dfrac{\mu_{i\rightarrow (ij)}^{t+\Delta t}(X_i \mid X_j) - \mu_{i\rightarrow (ij)}^{t}(X_i \mid X_j)}{\Delta t}
\end{equation}

A very important question arises at this point. What is the relation of $(X_i(t+\Delta t),X_j(t+\Delta t))$ and $(X_i(t),X_j(t))$? Or in other
words, what happens in the interval $(t,t+\Delta t)$? The answer is important because expressions for $\mu_{i\rightarrow (ij)}^{t+\Delta t}$ are different
whether we consider there can be jumps in the small $\Delta t$ interval or not. The first thing that makes sense to impose is that histories
must agree up to time $t$. In $(t,t+\Delta t)$ we can have several combinations.

An implicit assumption throughout all this theory is that on an infinitesimal interval only two things can happen to a spin; it can
stay on its current state or make one and only one jump to the opposite orientation. Two or more jumps are not allowed. Considering this we have four
cases to analyze:

\begin{itemize}
\item There are $s_i,s_j$ jumps in $(t_0,t)$ and neither $i$ nor $j$ jumps in $(t,t + \Delta t)$. This occurs with a probability
$(1-\lambda_i \Delta t)(1-\lambda_j \Delta t)$.
\item There are $s_i,s_j$ jumps in $(t_0,t)$ and $i$ XOR $j$ jumps in $(t,t + \Delta t)$. This occurs with a probability
$(1-\lambda_i \Delta t)(\lambda_j \Delta t)$ or $(1-\lambda_j \Delta t)(\lambda_i \Delta t)$. These are two cases in one.
\item There are $s_i,s_j$ jumps in $(t_0,t)$ and both $i$ and $j$ jumps in $(t,t + \Delta t)$. This has a probability of $\lambda_j \lambda_i \Delta t ^2$

\end{itemize}

When $\Delta t$ goes to zero, from the previous analysis we conclude that the derivative should be computed, with probability 1, using the first option, where histories for $i$ and
$j$ have no jumps in the interval of length $\Delta t$.

To differentiate the left hand side of (\ref{eq:update_bethe}) we can use the parametrization (\ref{eq:cavity_mess_param}):

\begin{eqnarray}
\nonumber
\mu_{i \rightarrow (ij)}^{t + \Delta t} &=& \prod_{l_i=1}^{s_i} \lasubsim{i}{j}{t_{l_i}} \exp\{ - \int_{t_0}^{t + \Delta t} \lasubsim{i}{j}{\tau} d\tau \} \\
\nonumber
\mu_{i \rightarrow (ij)}^{t + \Delta t} &=& [1-\lasubsim{i}{j}{t} \Delta t] \prod_{l_i=1}^{s_i} \lasubsim{i}{j}{t_{l_i}} \exp\{ - \int_{t_0}^{t} \lasubsim{i}{j}{\tau} d\tau \} + o(\Delta t) \\
\mu_{i \rightarrow (ij)}^{t + \Delta t} &=& [1-\lasubsim{i}{j}{t} \Delta t]\:\mu_{i \rightarrow (ij)}^{t} + o(\Delta t)
\end{eqnarray}

Then, with probability 1, the time derivative of the left hand side of equation (\ref{eq:update_bethe}) is equal to $
-\lasubsim{i}{j}{t}\:\mu_{i \rightarrow (ij)}^{t} $

To calculate the derivative of the right hand side of (\ref{eq:update_bethe}) we should compute:

\begin{equation}
\lim_{\Delta t \rightarrow 0} \dfrac{ \sum_{\{X_k\},k\in\partial i\setminus j}^{[t_0,t + \Delta t]} F(X_i,X_{\partial i},t + \Delta t) - \mu_{i \rightarrow (ij)}^{t} }{\Delta t} \label{eq:RHS_der}
\end{equation}
\
Let's focus on the first term of the numerator in the previous expression. It can be expanded to the first order in $\Delta t$. It is important to remember that $\Delta t$ appears
in the integration limits as well as in the integrand $F$. In addition, we should keep in mind that all jumps for $X_i$ and $X_j$ must occur before $t$. This restriction, however, does not
apply to the histories $X_k $ for $k$ in $\partial i \setminus j$.

The expansion of (\ref{eq:RHS_der}) can be explained as follows. First, let's remember that $F$ is the joint
probability of $X_i$ and $\{X_k\}$ with $k\in \partial i \setminus j$, conditioned on $X_j$. All histories in the term of interest are in the interval $[t_0,t+\Delta t]$.
The expression:

\begin{equation}
\displaystyle \sum_{\{X_k\},k\in\partial i\setminus j}^{[t_0,t + \Delta t]} F(X_i,X_{\partial i},t + \Delta t)
\end{equation}

is the marginalization of the mentioned joint probability distribution. The previous sum, to order $\Delta t$, has two main contributions. One comes from summing over $\{X_k\}$ with all $X_k$ having no jumps in $[t,t+\Delta t]$:

\begin{equation}
A = \sum_{\{X_k\},k\in\partial i\setminus j}^{[t_0,t]} F(X_i,X_{\partial i},t) \Big\lbrace 1 - \Big[ \sum_k \lasubsim{k}{i}{t} + r_i(t) \Big] \Delta t \Big\rbrace
\end{equation}

The other considers all the possibilities of having one of the $X_k$ with a jump in the interval of
length $\Delta t$:

\begin{equation}
B = \sum_{k} \displaystyle \sum_{\{X_k\},k\in\partial i\setminus j}^{[t_0,t]} F(X_i,X_{\partial i},t) \lasubsim{k}{i}{t}\Delta t
\end{equation}

Then:

\begin{equation}
\displaystyle \sum_{\{X_k\},k\in\partial i\setminus j}^{[t_0,t + \Delta t]} F(X_i,X_{\partial i},t + \Delta t) = A + B + o(\Delta t)
\end{equation}

Putting all together we see that B cancels out with the $\lambda$ part of A, and the remaining term of order 1 is $\musub{i}{j}{t}$, which
cancels when inserted in the limit expression.

The final outcome of this differentiation process is:

\begin{equation}
\lambda_{i \rightarrow (ij)} [X_i,X_j, t]\;\;\mu_{i \rightarrow (ij)}^{t}(X_i \mid X_j) = \displaystyle \sum_{\{X_k\},k\in\partial i\setminus j}^{[t_0,t]}
r_{i}[\sigma_i(t),\sigma_j(t),\sigma_{\partial i \setminus j}(t)] F(X_i,X_{\partial i},t) \label{eq:lambda_mu_app_F}
\end{equation}

We can now marginalize the right-hand side of the above equation over all the histories of the spins $k \in \partial i\setminus j$ by keeping the configuration of these spins at the last time $t$ fixed.
The result reads

\begin{equation}
\lambda_{i \rightarrow (ij)} (X_i,X_j, t)
\;\mu_{i \rightarrow (ij)}^{t}(X_i \mid X_j) = \displaystyle \sum_{\sigma_{\partial i\setminus j}(t)} \!\! r_{i}(\sigma_i(t),\sigma_j(t),\sigma_{\partial i \setminus j}(t)) \, p^{t}(\sigma_{\partial i \setminus j}, X_i|X_j)
\label{eq:lambda_mu_app}
\end{equation}
where we introduced the function $p$ as the marginalization of the function $F$ over all the spin histories of the neighbors of $i$ except $j$, with the configuration at the final time fixed. Note that the notation $\sigma_{\partial i \setminus j}(t)$ is equivalent to $\lbrace \sigma_k (t) \rbrace k \in \partial i \setminus j$ and that
in $p$ above appears again explicitly the conditional nature of
the probability distribution $F$.

Equation (\ref{eq:lambda_mu_app}) represents the differential dynamic message-passing update equation obtained by differentiating (\ref{eq:update_bethe}) in
time. It connects the derivative of the dynamic message $\mu_{i \rightarrow (ij)}^{t}$, and so the jumping rate $\lambda_{i \rightarrow (ij)}(t)$ of spin $i$ in the cavity used to
parametrize the message in (\ref{eq:cavity_mess_param}), with the transition rate of the
same spin $r_{i}(\sigma_i(t),\sigma_j(t),\sigma_{\partial i \setminus j}(t))$ at time $t$.

\section{Average dynamic cavity equations in random regular graphs}\label{app:avRGfc}

Whenever we have initial conditions independent of the site, and homogeneous node connectivity, the equations of our dynamic cavity method acquire a spatial symmetry that will allow us to reduce them to a few average case equations. Let's show this through an example.

We will start with equation (\ref{eq:CME2}).

\begin{eqnarray}
\frac{d p^{t}(\sigma_{\partial i \setminus j}, \sigma_i \: \big| \big| \: \sigma_j)}{dt} &=& - r_{i}(\sigma_i, \sigma_{\partial i}) \, p^{t}(\sigma_{\partial i \setminus j}, \sigma_i \: \big| \big| \: \sigma_j) + r_{i}(-\sigma_i, \sigma_{\partial i}) \, p^{t}(\sigma_{\partial i \setminus j}, -\sigma_i \: \big| \big| \: \sigma_j) \nonumber \\
& & -
\sum_{l \in \partial i \setminus j} \sum_{\sigma_{\partial l \setminus i} } \Big\lbrace r_{l}(\sigma_l, \sigma_{\partial l}) \, p^{t}(\sigma_{\partial l \setminus i} \mid \sigma_l \: \big| \big| \: \sigma_i ) \, p^{t}(\sigma_{\partial i \setminus j}, \sigma_i \: \big| \big| \: \sigma_j) - \label{eq:CME2_app} \\
& & \:\:\:\:\:\:\:\:\:\:\:\:\:\:\:\:\:\:- r_{l}(-\sigma_l, \sigma_{\partial l}) \, p^{t}(\sigma_{\partial l \setminus i} \mid -\sigma_l \: \big| \big| \: \sigma_i ) \, p^{t}(F_{l}[\sigma_{\partial i \setminus j}], \sigma_i \: \big| \big| \: \sigma_j) \Big\rbrace \nonumber
\end{eqnarray}

In a random regular graph with connectivity $K$, a spatial symmetry results from choosing an initial condition for $p^{t}(\sigma_{\partial i \setminus j}, \sigma_i \: \big| \big| \: \sigma_j)$ which does not depends on $i$ and $j$. In that case, the equation governing the time evolution of $p^{t}(\sigma_{\partial i \setminus j}, \sigma_i \: \big| \big| \: \sigma_j)$ wont depend on the values of $i, j$ for any time. We can write a single equation for all sites:

\begin{eqnarray}
\frac{d p^{t}(\hat{u}, \sigma \: \big| \big| \: \sigma')}{dt} &=& - r(K, \hat{u} + \delta_{\sigma,-\sigma'}) \, p^{t}(\hat{u}, \sigma \: \big| \big| \: \sigma') + r(K, K - \hat{u} - \delta_{\sigma,-\sigma'}) \, p^{t}(K-\hat{u}, -\sigma \: \big| \big| \: \sigma') - \nonumber \\
& & \!\!\!\!\!\!\!\!\!\!\!\!\!\!\!\!\!\!\!\!\!\!\!\! -
(K - 1 - \hat{u}) \sum_{u'=0}^{K-1} \binom{K-1}{u'} r(K, u') \, \Big\lbrace p^{t}(u' \mid \sigma \: \big| \big| \sigma) \, p^{t}(\hat{u}, \sigma \: \big| \big| \: \sigma') - \nonumber \\
& & \:\:\:\:\:\:\:\:\:\:\:\:\:\:\:\:\:\:\:\:\:\:\:\:\:\:\:\:\:\:\:\:\:\:\:\:\:\:\:\:\:\:\:\:\:\:\:\:\:\:\:\:\:\:\:\:\:\:\: - p^{t}(u' \mid -\sigma \: \big| \big| \sigma) \, p^{t}(\hat{u} + 1, \sigma \: \big| \big| \: \sigma') \Big\rbrace - \nonumber \\
& & \!\!\!\!\!\!\!\!\!\!\!\!\!\!\!\!\!\!\!\!\!\!\!\! - \hat{u} \sum_{u'=0}^{K-1} \binom{K-1}{u'} r(K, u') \, \Big\lbrace p^{t}(u' \mid -\sigma \: \big| \big| \sigma) \, p^{t}(\hat{u}, \sigma \: \big| \big| \: \sigma') - \nonumber \\
& & \:\:\:\:\:\:\:\:\:\:\:\:\:\:\:\:\:\:\:\:\:\:\:\:\:\:\:\:\:\:\:\:\:\:\:\:\:\:\:\:\:\:\:\:\:\:\:\:\:\:\:\:\:\:\:\:\:\:\: - p^{t}(u' \mid \sigma \: \big| \big| \sigma) \, p^{t}(\hat{u} - 1, \sigma \: \big| \big| \: \sigma') \Big\rbrace\label{eq:CME2_single_RGfc}
\end{eqnarray}

In (\ref{eq:CME2_single_RGfc}), regardless of the site where they are defined, the probability densities depend only on two connected spin variables, $\sigma$ and $\sigma'$, and on the number $\hat{u}=0,\ldots, K-1$ of unsatisfied interactions that $\sigma$ has with its neighbors other than $\sigma'$. The sum in the fourth and fifth lines takes into account that $\hat{u}$ unsatisfied interactions, all of which contribute equally to the derivative. The same happens with second and third lines and the contribution of the remaining $(K-1-\hat{u})$ satisfied interactions.

So far, we have not defined new probability densities, we just re-denoted $p^{t}(\sigma_{\partial i \setminus j}, \sigma_i \: \big| \big| \: \sigma_j)$ into $p^{t}(\hat{u}, \sigma \: \big| \big| \: \sigma')$ after realizing that the values of sites $i$ and $j$ where irrelevant due to spatial symmetry, and that it was not important to keep track of all the combinations of $\sigma_{\partial i \setminus j}$, we needed only to record the number of unsatisfied $\hat{u}$ interactions between $\sigma_{\partial i \setminus j}$ and $\sigma_i$. Now, we will introduce the total densities $\hat{p}^{t}(\hat{u}, \sigma \: \big| \big| \: \sigma') = \binom{K-1}{\hat{u}} p^{t}(\hat{u}, \sigma \: \big| \big| \: \sigma')$, noticing that these are analogous to the average probability densities in (\ref{eq:av_def}). Actually, exactly as before, we have:

\begin{eqnarray}
\hat{p}^{t}(\hat{u}, \sigma \: \big| \big| \: \sigma') = \lim_{N \rightarrow \infty} \ll \frac{1}{N K} \sum_{i=1}^{N} \sum_{j \in \partial i} \sum_{\sigma_i} \sum_{\sigma_j} \sum_{\sigma_{\partial i} \setminus j} p_{\xi_{K}(N)}^{t}(\sigma_{\partial i \setminus j}, \sigma_i \: \big| \big| \: \sigma_j) \times \nonumber \\
\times \, \delta_{\sigma, \sigma_i} \delta_{\sigma', \sigma_j} \:\: \delta_{(\sum_{k \in \partial i \setminus j} \sigma_k), K - 1 - 2\hat{u}} \gg_{\xi_{K}(N)} \label{eq:av_p_def}
\end{eqnarray}
where $\xi_{K}(N)$ is the ensemble of random regular graphs with connectivity $K$ and size $N$. The symbol $\ll \cdot \gg_{\xi_{K}(N)}$ represents an average over this ensemble, with proper probabilistic weights.

The differential equation for this average probability densities is:

\begin{eqnarray}
\frac{d \hat{p}^{t}(\hat{u}, \sigma \: \big| \big| \: \sigma')}{dt} &=& - r(K, \hat{u} + \delta_{\sigma,-\sigma'}) \, \hat{p}^{t}(\hat{u}, \sigma \: \big| \big| \: \sigma') + r(K, K - \hat{u} - \delta_{\sigma,-\sigma'}) \, \hat{p}^{t}(K-\hat{u}, -\sigma \: \big| \big| \: \sigma') - \nonumber \\
& & \!\!\!\!\!\!\!\!\!\!\!\!\!\!\!\!\!\!\!\!\!\!\!\! -
(K - 1 - \hat{u}) \: \big\langle \, r(K, u') \, \big\rangle_{\sigma, \sigma} \, \hat{p}^{t}(\hat{u}, \sigma \: \big| \big| \: \sigma') + (\hat{u} + 1) \: \big\langle \, r(K, u') \, \big\rangle_{-\sigma, \sigma} \, \hat{p}^{t}(\hat{u} + 1, \sigma \: \big| \big| \: \sigma') - \nonumber \\
& & \!\!\!\!\!\!\!\!\!\!\!\!\!\!\!\!\!\!\!\!\!\!\!\! - \hat{u} \: \big\langle \, r(K, u') \, \big\rangle_{-\sigma, \sigma} \, \hat{p}^{t}(\hat{u}, \sigma \: \big| \big| \: \sigma') + (K - \hat{u}) \: \big\langle \, r(K, u') \, \big\rangle_{\sigma, \sigma} \, \hat{p}^{t}(\hat{u} - 1, \sigma \: \big| \big| \: \sigma') \label{eq:CME2avRGfc}
\end{eqnarray}
where we have written the averages $\big\langle \, r(K, u') \, \big\rangle_{\sigma, \sigma'} = \sum_{u' = 0}^{K-1} r(K, u') \, \hat{p}^{t}(u' \mid \sigma \: \big| \big| \: \sigma' )$. The conditional cavity probability densities in (\ref{eq:CME2avRGfc}) are:

\begin{equation}
\hat{p}^{t}(\hat{u} \mid \sigma \: \big| \big| \: \sigma' ) = \frac{\hat{p}^{t}(\hat{u} , \sigma \: \big| \big| \: \sigma' )}{\sum_{u'} \hat{p}^{t}(u' , \sigma \: \big| \big| \: \sigma' )} \label{eq:closure_CME2avRGfc}
\end{equation}

Equations (\ref{eq:CME2avRGfc}) and (\ref{eq:closure_CME2avRGfc}) form a closed system that can be numerically integrated. This must be complemented with:

\begin{eqnarray}
\frac{d \hat{P}^{t}(\sigma, u)}{dt} &=& -r(K, u) \, \hat{P}^{t}(\sigma, u) + r(K, K-u) \, \hat{P}^{t}(-\sigma, K-u) - \nonumber \\
& & -
(K - u) \: \big\langle \, r(K, u') \, \big\rangle_{\sigma, \sigma} \, \hat{P}^{t}(\sigma, u) + (u + 1) \: \big\langle \, r(K, u') \, \big\rangle_{-\sigma, \sigma} \, \hat{P}^{t}(\sigma, u + 1) - \nonumber \\
& & - u \: \big\langle \, r(K, u') \, \big\rangle_{-\sigma, \sigma} \, \hat{P}^{t}(\sigma, u) + (K - u + 1) \: \big\langle \, r(K, u') \, \big\rangle_{\sigma, \sigma} \, \hat{P}^{t}(\sigma, \hat{u} - 1) \label{eq:ME2avRGfc_app}
\end{eqnarray}
which can be derived similarly.

To finish writing the equations of the levels of approximation shown in main text's figure (\ref{fig:CME_vs_DINA}), we still need to address first and third levels: \textit{CME-1} and \textit{CME-3}. In the first case we start with equations (\ref{eq:CME}) and (\ref{eq:ME})

\begin{eqnarray}
\frac{d p^{t}(\sigma_i \: \big| \big| \: \sigma_j)}{dt} &=& - \sum_{\sigma_{\partial i \setminus j} } \Big\lbrace r_{i}(\sigma_i, \sigma_{\partial i}) \Big[ \prod_{k \in \partial i \setminus j} p^{t}(\sigma_k \: \big| \big| \: \sigma_i) \Big] p^{t}(\sigma_i \: \big| \big| \: \sigma_j) - \nonumber \\
& & \:\:\:\:\:\:\:\:\:\:\:\:\:\:\:\:\:\:- r_{i}(-\sigma_i, \sigma_{\partial i}) \Big[ \prod_{k \in \partial i \setminus j} p^{t}(\sigma_k \: \big| \big| -\sigma_i) \Big] p^{t}(-\sigma_i \: \big| \big| \: \sigma_j) \Big\rbrace
\label{eq:CME_app} \\
\frac{d P^{t}(\sigma_i)}{dt} &=& -
\sum_{\sigma_{\partial i } } \Big\lbrace r_{i}(\sigma_i, \sigma_{\partial i}) \, \Big[ \prod_{k \in \partial i} p^{t}(\sigma_k \: \big| \big| \sigma_i) \Big] \, P^{t}(\sigma_i) - \label{eq:ME_app} \\
& & \:\:\:\:\:\:\:\:\:\:\:\:\:\:\:\:\:\:- r_{i}(-\sigma_i, \sigma_{\partial i}) \, \Big[ \prod_{k \in \partial i} p^{t}(\sigma_k \: \big| \big| -\sigma_i) \Big] \, P^{t}(-\sigma_i) \Big\rbrace \nonumber
\end{eqnarray}

It is easy so see that the average case versions of this equations at random regular graphs with homogeneous initial conditions are:

\begin{eqnarray}
\frac{d \hat{p}^{t}(\sigma \: \big| \big| \: \sigma')}{dt} &=& - \sum_{u'=0}^{K-1} \Big\lbrace r(K, u' + \delta_{\sigma, -\sigma'}) \big[\hat{p}(-\sigma \mid \sigma) \big]^{u'} \, \big[\hat{p}(\sigma \mid \sigma) \big]^{K-1-u'} \hat{p}^{t}(\sigma \: \big| \big| \: \sigma') - \nonumber \\
& & \:\:\:\:\:\:- r(K, u' + \delta_{\sigma, \sigma'}) \big[\hat{p}(\sigma \mid -\sigma) \big]^{u'} \, \big[\hat{p}(-\sigma \mid -\sigma) \big]^{K-1-u'} \hat{p}^{t}(-\sigma \: \big| \big| \: \sigma') \Big\rbrace
\label{eq:CMEav_app} \\
\frac{d \hat{P}^{t}(\sigma)}{dt} &=& -
\sum_{u'=0}^{K} \Big\lbrace r(K, u') \, \big[\hat{p}(-\sigma \mid \sigma) \big]^{u'} \, \big[\hat{p}(\sigma \mid \sigma) \big]^{K-u'} \, \hat{P}^{t}(\sigma) - \label{eq:MEav_app} \\
& & \:\:\:\:\:\:\:\:\:\:\:\:\:\:\:\:\:\:- r(K, u') \, \big[\hat{p}(\sigma \mid -\sigma) \big]^{u'} \, \big[\hat{p}(-\sigma \mid -\sigma) \big]^{K-u'} \, \hat{P}^{t}(-\sigma) \Big\rbrace \nonumber
\end{eqnarray}

Equations (\ref{eq:CMEav_app}) and (\ref{eq:MEav_app}) also form a closed system and its numerical integration is shown in figure (\ref{fig:CME_vs_DINA}) of main text. The equations in the third level of approximation, \textit{CME-3}, whose results appear also in that figure, require some extra work.

\begin{figure}[htb]
\centering
\includegraphics[keepaspectratio=true,width=0.45\textwidth]{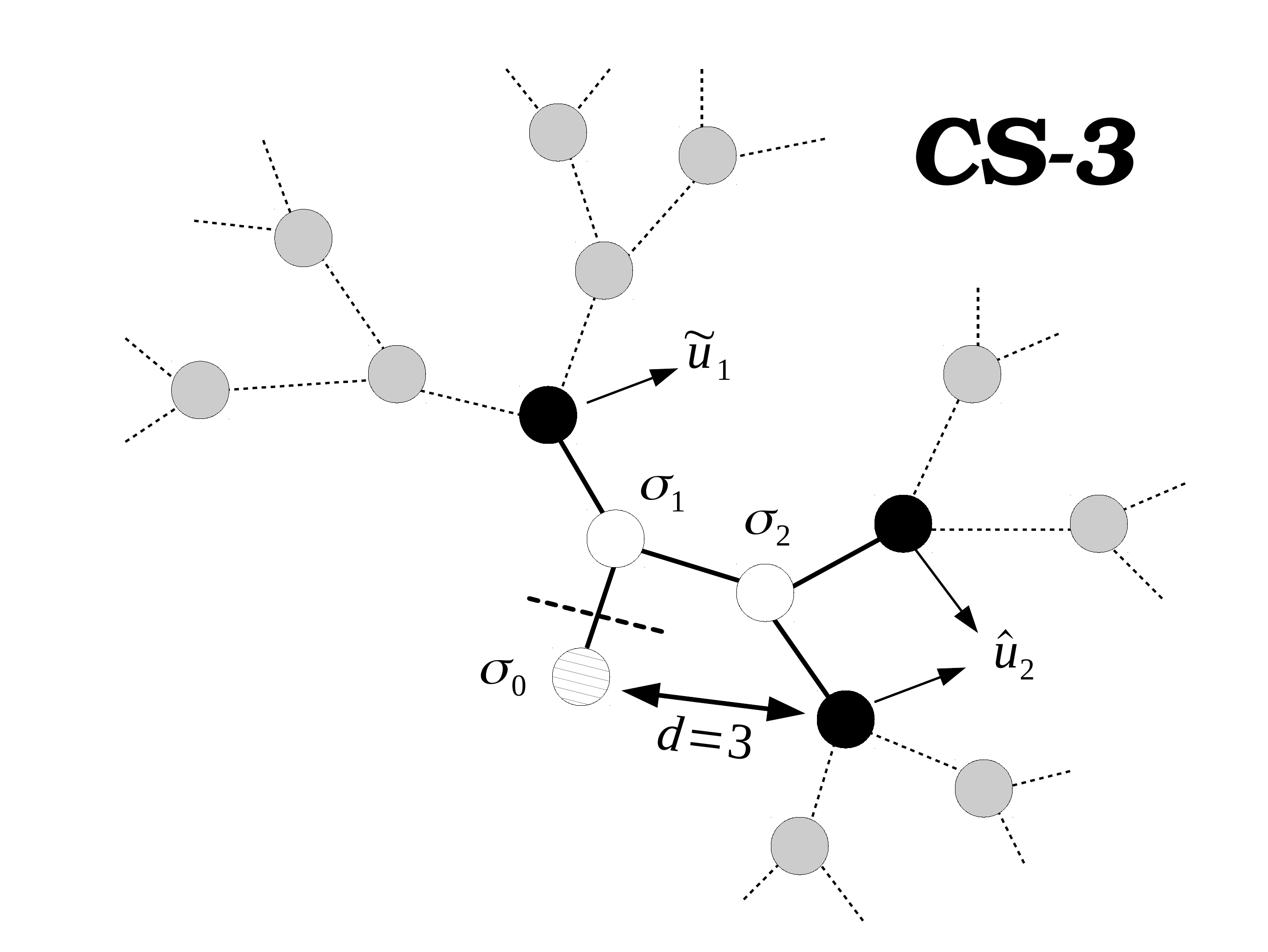}
\caption{The cavity probability density $p^{t}(\hat{u}_2, \sigma_2, \tilde{u}_1, \sigma_1 \: \big| \big| \: \sigma_0)$ is defined over a connected set in \textit{CS-3}. This figure illustrates meaning of each one of the variables $\hat{u}_2$, $\sigma_2$, $\tilde{u}_1$, $\sigma_1$ and $\sigma_0$.}
\label{fig:illustration_CS-3_app}
\end{figure}

Our probability densities are now $p^{t}(\hat{u}_2, \sigma_2, \tilde{u}_1, \sigma_1 \: \big| \big| \: \sigma_0)$, which are defined over the connected sets illustrated in bottom-left panel of (\ref{fig:illustration_CS-N}). The spins $\sigma_0$, $\sigma_1$ and $\sigma_2$ are marked in figure (\ref{fig:illustration_CS-3_app}). The integer $\hat{u}_2$, which goes from zero to $K-1$, represents the number of unsatisfied interactions between $\sigma_2$ and its neighbors, without counting $\sigma_1$. Finally, the integer $\tilde{u}_1$, which goes from zero to $K-2$, is the number of unsatisfied interactions between $\sigma_1$ and its neighbors, without counting $\sigma_0$ and $\sigma_2$.

As we learned with (\ref{eq:p_derivative_tree_gen_markov}), the exact equations for this densities are:

\begin{eqnarray}
\frac{d p^{t}(\hat{u}_2, \sigma_2, \tilde{u}_1, \sigma_1 \: \big| \big| \: \sigma_0)}{dt} &=& - r(K, \tilde{u}_1 + \delta_{\sigma_0,-\sigma_1} + \delta_{\sigma_2, -\sigma_1}) \, p^{t}(\hat{u}_2, \sigma_2, \tilde{u}_1, \sigma_1 \: \big| \big| \: \sigma_0) + \label{eq:CME3_single_RGfc} \\
& & \!\!\!\!\!\!\!\!\!\!\!\!\!\!\!\!\!\!\!\!\!\!\!\! + r(K, K - \tilde{u}_1 - \delta_{\sigma_0,-\sigma_1} - \delta_{\sigma_2,-\sigma_1}) \, p^{t}(\hat{u}_2, \sigma_2, K - 2 - \tilde{u}_1, -\sigma_1 \: \big| \big| \: \sigma_0) - \nonumber \\
& & \!\!\!\!\!\!\!\!\!\!\!\!\!\!\!\!\!\!\!\!\!\!\!\! - r(K, \hat{u}_2 + \delta_{\sigma_1,-\sigma_2}) \, p^{t}(\hat{u}_2, \sigma_2, \tilde{u}_1, \sigma_1 \: \big| \big| \: \sigma_0) + \nonumber \\
& & \!\!\!\!\!\!\!\!\!\!\!\!\!\!\!\!\!\!\!\!\!\!\!\! + r(K, K - \hat{u}_2 - \delta_{\sigma_1,-\sigma_2}) \, p^{t}(K - 1 - \hat{u}_2, -\sigma_2, \tilde{u}_1, \sigma_1 \: \big| \big| \: \sigma_0) - \nonumber \\
& & \!\!\!\!\!\!\!\!\!\!\!\!\!\!\!\!\!\!\!\!\!\!\!\!\!\!\!\!\!\!\!\!\!\!\!\!\!\!\!\!\!\!\!\!\!\!\!\!\!\!\!\!\!\!\!\!\!\!\!\! -
(K - 2 - \tilde{u}_{1}) \sum_{u'_1=0}^{K-1} \binom{K-1}{u'_1} r(K, u'_1) \, \Big\lbrace p^{t}(u'_1 \mid \hat{u}_2, \sigma_2, \tilde{u}_1, \sigma_1 \: \big| \big| \sigma_0) \, p^{t}(\hat{u}_2, \sigma_2, \tilde{u}_1, \sigma_1 \: \big| \big| \: \sigma_0) - \nonumber \\
& & - p^{t}(u'_1 \mid \hat{u}_2, \sigma_2, \tilde{u}_1 + 1, \sigma_1 \: \big| \big| \sigma_0) \, p^{t}(\hat{u}_2, \sigma_2, \tilde{u}_1 + 1, \sigma_1 \: \big| \big| \: \sigma_0) \Big\rbrace - \nonumber \\
& & \!\!\!\!\!\!\!\!\!\!\!\!\!\!\!\!\!\!\!\!\!\!\!\!\!\!\!\!\!\!\!\!\!\!\!\!\!\!\!\!\!\!\!\!\!\!\!\!\!\!\!\!\!\!\!\!\!\!\!\! -
\tilde{u}_{1} \sum_{u'_1=0}^{K-1} \binom{K-1}{u'_1} r(K, u'_1) \, \Big\lbrace p^{t}(u'_1 \mid \hat{u}_2, \sigma_2, \tilde{u}_1, \sigma_1 \: \big| \big| \sigma_0) \, p^{t}(\hat{u}_2, \sigma_2, \tilde{u}_1, \sigma_1 \: \big| \big| \: \sigma_0) - \nonumber \\
& & - p^{t}(u'_1 \mid \hat{u}_2, \sigma_2, \tilde{u}_1 - 1, \sigma_1 \: \big| \big| \sigma_0) \, p^{t}(\hat{u}_2, \sigma_2, \tilde{u}_1 - 1, \sigma_1 \: \big| \big| \: \sigma_0) \Big\rbrace \nonumber - \nonumber \\
& & \!\!\!\!\!\!\!\!\!\!\!\!\!\!\!\!\!\!\!\!\!\!\!\!\!\!\!\!\!\!\!\!\!\!\!\!\!\!\!\!\!\!\!\!\!\!\!\!\!\!\!\!\!\!\!\!\!\!\!\! -
(K - 1 - \hat{u}_{2}) \sum_{u'_2=0}^{K-1} \binom{K-1}{u'_2} r(K, u'_2) \, \Big\lbrace p^{t}(u'_2 \mid \hat{u}_2, \sigma_2, \tilde{u}_1, \sigma_1 \: \big| \big| \sigma_0) \, p^{t}(\hat{u}_2, \sigma_2, \tilde{u}_1, \sigma_1 \: \big| \big| \: \sigma_0) - \nonumber \\
& & - p^{t}(u'_2 \mid \hat{u}_2 + 1, \sigma_2, \tilde{u}_1, \sigma_1 \: \big| \big| \sigma_0) \, p^{t}(\hat{u}_2 + 1, \sigma_2, \tilde{u}_1, \sigma_1 \: \big| \big| \: \sigma_0) \Big\rbrace - \nonumber \\
& & \!\!\!\!\!\!\!\!\!\!\!\!\!\!\!\!\!\!\!\!\!\!\!\!\!\!\!\!\!\!\!\!\!\!\!\!\!\!\!\!\!\!\!\!\!\!\!\!\!\!\!\!\!\!\!\!\!\!\!\! -
\hat{u}_{2} \sum_{u'_2=0}^{K-1} \binom{K-1}{u'_2} r(K, u'_2) \, \Big\lbrace p^{t}(u'_2 \mid \hat{u}_2, \sigma_2, \tilde{u}_1, \sigma_1 \: \big| \big| \sigma_0) \, p^{t}(\hat{u}_2, \sigma_2, \tilde{u}_1, \sigma_1 \: \big| \big| \: \sigma_0) - \nonumber \\
& & - p^{t}(u'_2 \mid \hat{u}_2 - 1, \sigma_2, \tilde{u}_1, \sigma_1 \: \big| \big| \sigma_0) \, p^{t}(\hat{u}_2 - 1, \sigma_2, \tilde{u}_1, \sigma_1 \: \big| \big| \: \sigma_0) \Big\rbrace \nonumber
\end{eqnarray}

Unlike what we have seen before, our variables are not defined over regular connected sets \textit{rCS-3} (see figure (\ref{fig:illustration_CS-3_app})). We need a slightly different closure, or more specifically, two new closures. These are necessary because the time derivative of densities in \textit{CS-3} that appears in (\ref{eq:CME3_single_RGfc}) involves probability densities defined over connected sets in \textit{CS-4}.

The first approximation targets the conditional cavity probability densities that appear in the lines from the fifth to the eight of (\ref{eq:CME3_single_RGfc}) and reduces them to densities of the form $p^{t}(\hat{u}_2, \sigma_2, \tilde{u}_1, \sigma_1 \: \big| \big| \: \sigma_0)$:

\begin{eqnarray}
p_{\text{SAT}}^{t}(u'_1 \mid \hat{u}_2, \sigma_2, \tilde{u}_1, \sigma_1 \: \big| \big| \sigma_0) &\approx& \frac{p^{t}(u'_1, \sigma_1, \bar{u}_1, \sigma_1 \: \big| \big| \sigma_0)}{\sum_{u'}p^{t}(u', \sigma_1, \bar{u}_1, \sigma_1 \: \big| \big| \sigma_0)}
\label{eq:closure_CME3_u1_SAT} \\
p_{\text{UNSAT}}^{t}(u'_1 \mid \hat{u}_2, \sigma_2, \tilde{u}_1, \sigma_1 \: \big| \big| \sigma_0) &\approx& \frac{p^{t}(u'_1, -\sigma_1, \bar{u}_1, \sigma_1 \: \big| \big| \sigma_0)}{\sum_{u'}p^{t}(u', -\sigma_1, \bar{u}_1, \sigma_1 \: \big| \big| \sigma_0)}
\label{eq:closure_CME3_u1_UNSAT}
\end{eqnarray}

\begin{figure}[htb]
\centering
\includegraphics[keepaspectratio=true,width=0.45\textwidth]{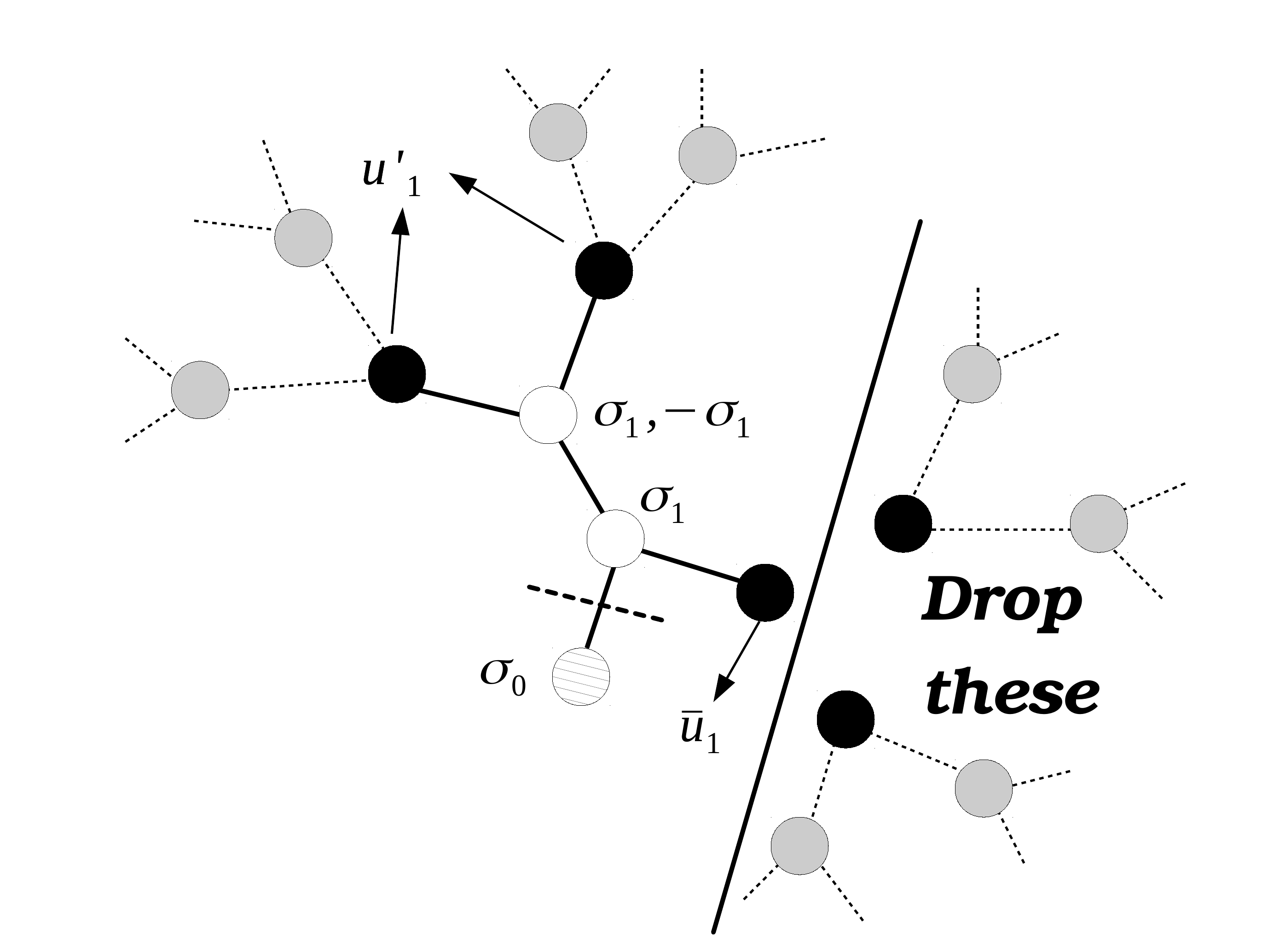}
\includegraphics[keepaspectratio=true,width=0.45\textwidth]{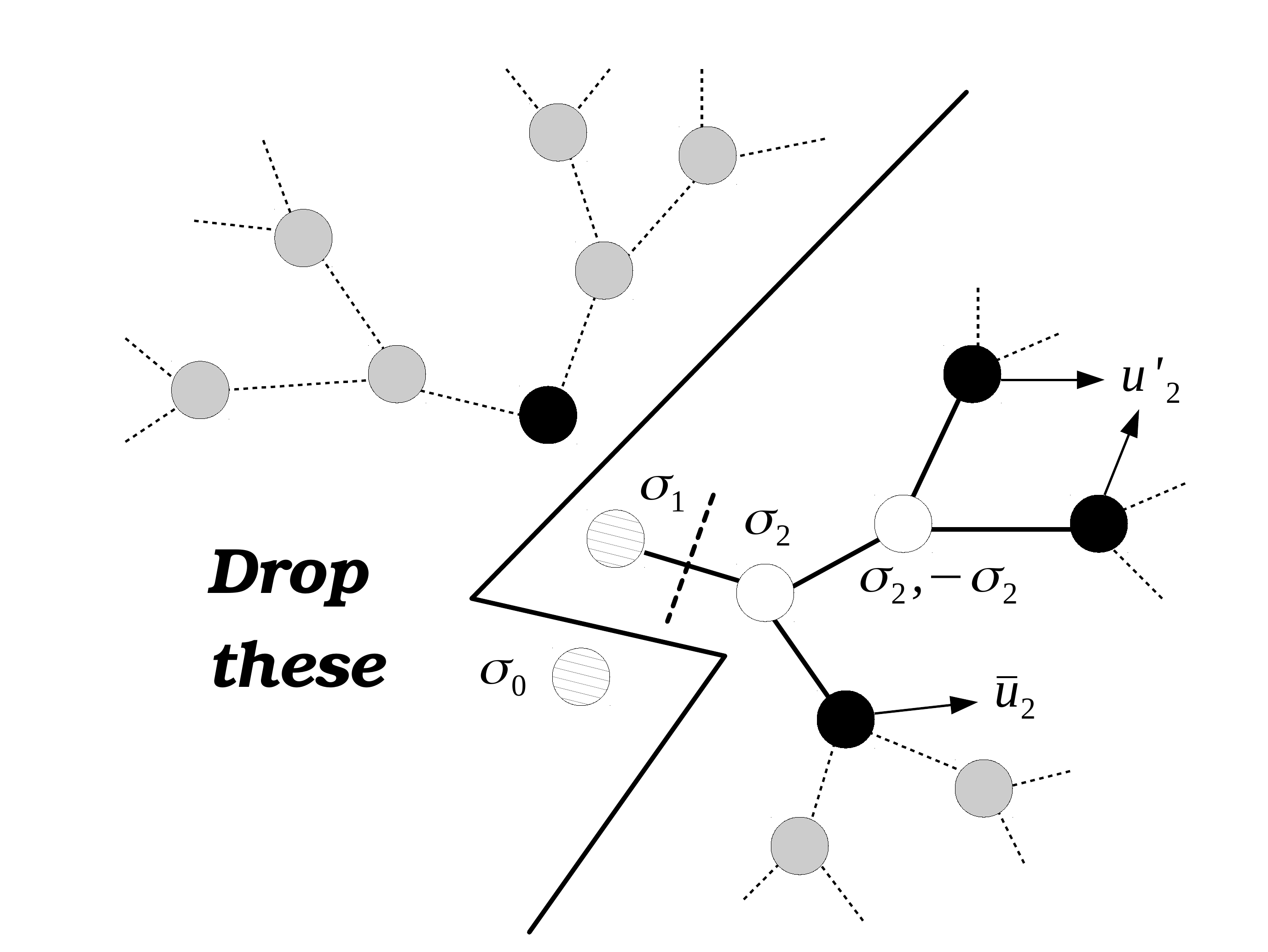}
\caption{Illustration of the approximations (\ref{eq:closure_CME3_u1_SAT}), (\ref{eq:closure_CME3_u1_UNSAT}), (\ref{eq:closure_CME3_u2_SAT}) and (\ref{eq:closure_CME3_u2_UNSAT}) applied to conditional cavity probability densities. These are necessary because the time derivative (\ref{eq:CME3_single_RGfc}) involves probability densities defined over connected sets in \textit{CS-4}. In each panel, the nodes not colored in gray belong to such connected sets, and we have to drop the dependency on some of them in order to give a closure to (\ref{eq:CME3_single_RGfc}).}
\label{fig:illustration_CS-3_app_closures}
\end{figure}

The left panel of figure (\ref{fig:illustration_CS-3_app_closures}) illustrates equations (\ref{eq:closure_CME3_u1_SAT}) and (\ref{eq:closure_CME3_u1_UNSAT}). There, the integer $u'_1=0, \ldots, K-1$ represents the number of unsatisfied interactions related to a specific neighbor of $\sigma_1$ which can either have the same value of $\sigma_1$ (equation (\ref{eq:closure_CME3_u1_SAT})) or the opposite value $-\sigma_1$ (equation (\ref{eq:closure_CME3_u1_UNSAT})). On the other hand, $\bar{u}_1=0, \ldots, K-2$ is the number of unsatisfied interactions between $\sigma_1$ and its neighbors other than $\sigma_0$ and the one mentioned in the previous sentence (this means that $\tilde{u}_1$ includes the interaction with $\sigma_2$). Notice that we have dropped the dependency on the number of unsatisfied relations $\hat{u}_2$ because the spins involved in that interactions (except $\sigma_2$) are at a distance $d=4 > 3$ from the ones involved in $u'_1$.

The second approximation targets the conditional cavity probability densities that appear in the lines from ninth to twelfth of (\ref{eq:CME3_single_RGfc}):

\begin{eqnarray}
p_{\text{SAT}}^{t}(u'_2 \mid \hat{u}_2, \sigma_2, \tilde{u}_1, \sigma_1 \: \big| \big| \sigma_0) &\approx& \frac{p^{t}(u'_2, \sigma_2, \bar{u}_2, \sigma_2 \: \big| \big| \sigma_1)}{\sum_{u'}p^{t}(u', \sigma_2, \bar{u}_2, \sigma_2 \: \big| \big| \sigma_1)}
\label{eq:closure_CME3_u2_SAT} \\
p_{\text{UNSAT}}^{t}(u'_1 \mid \hat{u}_2, \sigma_2, \tilde{u}_1, \sigma_1 \: \big| \big| \sigma_0) &\approx& \frac{p^{t}(u'_2, -\sigma_2, \bar{u}_2, \sigma_2 \: \big| \big| \sigma_1)}{\sum_{u'}p^{t}(u', -\sigma_2, \bar{u}_2, \sigma_2 \: \big| \big| \sigma_1)}
\label{eq:closure_CME3_u2_UNSAT}
\end{eqnarray}

Right panel of figure (\ref{fig:illustration_CS-3_app_closures}) illustrates the meaning of this equations. We won't give more details because they are very similar to what we said about (\ref{eq:closure_CME3_u2_SAT}) and (\ref{eq:closure_CME3_u2_UNSAT}).

From this point, it is easy to derive equations for the averages $\hat{p}^{t}(\hat{u}_2, \sigma_2, \tilde{u}_1, \sigma_1 \: \big| \big| \: \sigma_0)$, which together with the averages of closure relations (\ref{eq:closure_CME3_u1_SAT}), (\ref{eq:closure_CME3_u1_UNSAT}), (\ref{eq:closure_CME3_u2_SAT}) and (\ref{eq:closure_CME3_u2_UNSAT}) form a system that we can numerically integrate to obtain the results in figure (\ref{fig:CME_vs_DINA}).

\end{document}